\documentclass[12pt]{article}
\usepackage{epsf,epsfig}
\usepackage{cite}
\usepackage{amssymb}
\usepackage[dvips]{color}

\makeatletter
\renewcommand\section{\@startsection {section}{1}{\z@}%
                                   {-3.5ex \@plus -1ex \@minus -.2ex}
                                   {2.3ex \@plus.2ex}%
                                   {\normalfont\large\bfseries}}
\renewcommand\subsection{\@startsection{subsection}{2}{\z@}%
                                     {-3.25ex\@plus -1ex \@minus -.2ex}%
                                     {1.5ex \@plus .2ex}%
                                     {\normalfont\bfseries}}
\makeatother

\let\non\nonumber

\newcommand{\bea}{\begin{eqnarray}}
\newcommand{\eea}{\end{eqnarray}}
\newcommand{\be}{\begin{equation}}
\newcommand{\ee}{\end{equation}}

\newcommand{\detail}[1]{{~\\~\underline{\it #1}} ~}

\newcommand{\Z}{{\mathbb Z}}
\newcommand{\R}{{\mathbb R}}
\newcommand{\M}{{\cal M}}
\newcommand{\E}{{\cal E}}
\newcommand{\V}{{\cal V}}
\newcommand{\cO}{{\cal O}}

\newcommand{\RV}{Ro{\v c}ek-Verlinde}

\newcommand{\lrp}{{\buildrel \leftrightarrow \over \partial}}

\newcommand{\G}{\Gamma}
\newcommand{\e}{\epsilon}

\newcommand{\rr}{\rightarrow}
\newcommand{\m}{\mu}
\newcommand{\n}{\nu}
\newcommand{\p}{\partial}

\def\slash#1{\setbox0=\hbox{$#1$}#1\hskip-\wd0\hbox to\wd0{\hss\sl/\/\hss}}
\newcommand{\mP}{\mathbb P}
\newcommand{\mC}{\mathbb C}

\newcommand{\C}[1]{$(\ref{#1})$}

\newcommand{\rt}{{\sqrt 2}}

\newcommand{\half}{{1\over 2}}

\newcommand{\IC}{{\mathbb C}}

\newcommand{\IP}{{\mathbb P}}

\newcommand{\QP}{Q_+}
\newcommand{\QPB}{\bar{Q}_+}

\renewcommand{\a}{\alpha}

\newcommand{\s}{\sigma}
\renewcommand{\l}{\lambda}
\newcommand{\lb}{\bar{\lambda}}

\newcommand{\U}{\Upsilon}
\renewcommand{\L}{\Lambda}
\newcommand{\LB}{\bar{\Lambda}}

\renewcommand{\t}{\theta}
\newcommand{\tb}{\bar{\theta}}
\newcommand{\tp}{{\theta^+}}
\newcommand{\tpb}{{\bar{\theta}^+}}

\newcommand{\D}{{\rm D}}
\newcommand{\DB}{\bar{\rm D}}

\newcommand{\F}{{\cal F}}
\newcommand{\GG}{{\cal G}}
\newcommand{\N}{{\cal N}}
\renewcommand{\S}{{\cal S}}

\newcommand{\A}{{\cal A}}
\newcommand{\Sm}{{\Sigma}}
\newcommand{\TT}{$(2,2)$}
\newcommand{\ZT}{$(0,2)$}

\newcommand{\nn}{\nonumber}

\begin{document}
\begin{titlepage}

\begin{center}

\today
\hfill                  hep-th/0309226

\hfill EFI-03-33, HUTP-03/A052

\vskip 2 cm {\Large \bf {(0,2) Duality}} \vskip 1.25 cm { Allan
Adams$^{a,b}$\footnote{ allan@schwinger.harvard.edu}, Anirban
Basu$^c$\footnote{ basu@theory.uchicago.edu} and Savdeep
Sethi$^c$\footnote{ sethi@theory.uchicago.edu}
}\\
{\vskip 0.5cm $^a$ Jefferson Physical Laboratory,
Harvard University, Cambridge, MA 02138, USA \\}
{\vskip 0.5cm $^b$  Department of Theoretical Physics,
Tata Institute of Fundamental Research, Mumbai 400005, India \\}
{\vskip 0.5cm $^c$ Enrico Fermi Institute, University of Chicago,
  Chicago, IL 60637,  USA \\}

\end{center}

\vskip 2 cm

%
%
\begin{abstract}
\baselineskip=18pt

We construct dual descriptions of \ZT\ gauged linear sigma models. In
some cases, the dual is a \ZT\ Landau-Ginzburg theory, while in other
cases, it is a non-linear sigma model. The duality map defines an
analogue of mirror symmetry for \ZT\ theories. Using the dual
description, we determine the instanton corrected chiral ring for some
illustrative examples. This ring defines a \ZT\ generalization of the quantum
cohomology ring of \TT\ theories.

\end{abstract}
\end{titlepage}
\pagestyle{plain}
\baselineskip=19pt

%
%
\tableofcontents
\begin{center}
-----------------------------------------------------------------------------
\end{center}

%
%
%
%
\section{Introduction}

Mirror symmetry is one of the more spectacular predictions of string
theory~\cite{Greene:1990ud}. Strings propagating on topologically
distinct spaces can give rise to the same effective space-time
physics. This duality is best understood for theories that can be
constructed as \TT\ gauged linear sigma models
(GLSM)~\cite{Witten:1993yc}.

In the space of perturbative heterotic string compactifications,
\TT\ world-sheet theories are quite special. The more general
supersymmetric string compactification only requires \ZT\
supersymmetry. To describe a geometric heterotic  string
compactification (without fluxes), we need to specify a K\"ahler
space, $\M$, with tangent bundle $T\M$ and a holomorphic bundle,
$\V$, satisfying the conditions 
\bea\label{conditionone} & {\rm
c}_1(T\M) =0, \quad {\rm c}_1(\V) = 0 \,\, ({\rm mod} \,\, 2)
\\ \label{conditiontwo}
& {\rm ch}_2 (\V) = {\rm ch}_2(T\M). 
\eea 
The assumption of
world-sheet \TT\ supersymmetry corresponds to the choice, 
\be
\V = T\M. 
\ee 
The moduli space of $\M$ locally decomposes into
K\"ahler and complex structure deformations which are exchanged
under the mirror map.

It is natural to ask whether a generalization of mirror symmetry exists
for the larger class of \ZT\ theories. At the outset, there is
a potential problem; namely, specifying an $(\M, \V)$ obeying
\C{conditionone}\ and \C{conditiontwo}\ does not guarantee the
existence of a corresponding superconformal field theory. Except
under special conditions~\cite{Distler:1987wm, Distler:1988ee}, we
expect world-sheet instantons to destabilize most \ZT\ non-linear
sigma models. Fortunately, this potential problem vanishes for
\ZT\ theories that can be realized as linear sigma
models~\cite{Silverstein:1995re, Basu:2003bq, Beasley:2003fx}; this
vanishing can be quite non-trivial, as shown
in~\cite{Berglund:1995yu}, because individual instantons can give
non-zero contributions. However, the net contribution to the space-time
superpotential is zero.

The next basic issue is defining a non-perturbative duality. The
moduli space
for a geometric \ZT\ superconformal field theory consists of
K\"ahler and complex structure deformations together with
deformations of the gauge bundle. We could imagine many different
dualities permuting these three kinds of moduli.  
A natural
extension of \TT\ mirror symmetry was studied in a special
class of solvable \ZT\ models by Blumenhagen et.
al.~\cite{Blumenhagen:1997vu}. Some mirror pairs related by
quotient actions (as in the original \TT\
construction~\cite{Greene:1990ud}) were described
in~\cite{Blumenhagen:1997tv, Blumenhagen:1997pp}. This notion of 
mirror symmetry
intimately involves a superpotential in both the original
and dual descriptions. In related work, a description of
equivariant sheaves and their relevence to \ZT\ mirror symmetry
appears in~\cite{Knutson:1998yt}, while an extension of the
monomial divisor mirror map~\cite{Aspinwall:1993rj}\ to a class of
\ZT\ theories appears in~\cite{Sharpe:1998wa}. Note that unlike
the \TT\ case, we believe that \ZT\ mirror symmetry should map 
certain instanton sums on $\M$ to instanton sums on the mirror.
Generically, both sides of the duality receive non-perturbative
corrections.

Our aim in this effort is to define a non-perturbative $(0,2)$ duality for
theories that can be constructed from gauged linear sigma models.
We generalize an approach
used recently by Hori and Vafa  to construct
$(2,2)$ mirror pairs~\cite{Hori:2000kt}.
Their approach suggests an
equivalence between
certain \TT\ gauged linear sigma models and
\TT\ Landau-Ginzburg theories. This equivalence is derived using a
generalization of world-sheet abelian duality, and is closely related
to an earlier attempt at deriving mirror
symmetry~\cite{Morrison:1996yh}. The  manifold associated
with the gauged linear sigma model is a toric variety with
non-negative first Chern class. The basic approach used
in~\cite{Hori:2000kt}\ is to dualize the torus action which is
implemented via an abelian gauge symmetry in the GLSM. This
dualization exchanges charged fields for uncharged fields. However,
because the circle action is not free, a superpotential is
generated by instantons. The dual description is therefore a
Landau-Ginzburg theory.

We construct an analogue of abelian duality for \ZT\
theories. Applied to a GLSM, this duality generates a non-perturbative
dual. There are some important points to note: in this analysis, we
dualize models without a superpotential. In a sense, the parameter
space that is natural for us is orthogonal to the one studied  
in~\cite{Blumenhagen:1997vu, Blumenhagen:1997tv, 
Blumenhagen:1997pp}. To connect the two
notions of duality will require understanding the dualization process
in the presence of a superpotential. It seems to us that this problem
can be addressed (at least for special models).

We consider both conformal and non-conformal models. For
non-conformal models, we can relax condition \C{conditionone}\ and
permit the weaker
constraint
\be
c_1(T\M) > 0.
\ee
Using the dual description, we can determine the exact chiral ring of
the original theory, including instanton corrections. We use this ring
to define a generalization of the quantum cohomology ring of a \TT\
model~\cite{Witten:1988xj, Witten:1990ig}. 
In some particularly nice illustrative examples based on
$\mP^1\times \mP^1$, we determine this instanton corrected ring precisely. 

The structure of our dual theory depends sensitively on whether ${\rm
rk}(\V) \geq {\rm rk} (T\M)$ or whether  ${\rm
rk}(\V) < {\rm rk} (T\M)$.  In the former case, the low-energy dual
theory is generically 
a \ZT\ Landau-Ginzburg model with isolated supersymmetric vacua. In
the latter case, the dual theory is typically a \ZT\ non-linear sigma
model. The sigma model metric is singular on certain loci where
the accompanying dilaton also diverges. It is worth noting that the
duality maps the canonical K\"ahler moduli of a GLSM to superpotential
terms in the dual description.   

In section 2, we establish our \ZT\ superspace, superfield, and
component field conventions. Section $3$ contains a derivation of the
perturbative duals to both ungauged and gauged \ZT\ models. In
section $4$, we determine the exact form of non-perturbative
corrections to the dual superpotential. The vacuum
structure, and the nature of instanton corrections to \ZT\ theories
are described in section $5$. Finally, in section $6$ we present
an analysis of some illustrative examples, together with an explanation of how
the different dual descriptions emerge depending on the relation
between ${\rm rk} (\V)$ and ${\rm rk} (T\M)$. 

We should mention a few of the future directions that seem worth
exploring to us. Restricting our dualization results to the case of
$(0,4)$ theories should help classify heterotic compactifications on a
$K3$ surface, extending the classification given
in~\cite{deBoer:2001px}\ for tori. The analogue of the quantum cohomology ring
that we have described should be computable in a large class of
examples (conformal and non-conformal), perhaps with the help of localization
techniques~\cite{Kontsevich:1994na}. In section~\ref{reldef}, we found
a nice example of a bundle
degeneration which is easily generalizable. The two-dimensional field
theory should provide a resolution of the singularity which seems
worth comparing with the cases studied in~\cite{Distler:1996tj}.
There should be some space-time duality argument, generalizing the SYZ
construction~\cite{Strominger:1996it}\ for \TT\ theories. In a related
vein, the heterotic instanton corrections we consider are related, via
S-duality, to type I D-string corrections. The relation between the
open and closed string instanton moduli spaces is likely to be
fascinating (see, for example,~\cite{Katz:2001vm}). Lastly, we would
like to know how much we can learn about
the Yukawa couplings of generic \ZT\ heterotic theories, and perhaps about
superpotentials for vector bundle moduli (studied recently, for
example, in~\cite{Buchbinder:2002ic}).

\section{The Structure of \ZT\ Theories}
In this section we review some basic facts, and fix our notation for
\ZT\ supersymmetric
field theories in $1+1$ dimensions.

\subsection{\ZT\ Supersymmetry}

Chiral \ZT\ supersymmetry is generated by two supercharges, $\QP$ and
$\QPB=\QP^\dagger$,
the bosonic generators $H$, $P$ and $M$ of translations and rotations,
and the generator $F_+$ of a $U(1)$ R-symmetry.
The algebra itself is
\bea
\QP^2 =\QPB^2 =0 ~~&~&~~ \{\QP , \QPB \}=2(H-P)  \nn\\
 \left[ M, \QP \right]=-\QP   ~~&~&~~ \left[ M,\QPB \right] =-\QPB   \nn\\
 \left[ F_+,\QP \right]=-\QP  ~~&~&~~ \left[F_+ , \QPB \right]= +\QPB
 \nn.
\eea

Much of what follows is simplified by the use of superspace.
Let the \ZT\ superspace coordinates be $(y^+, y^-,\t^+,\tb^+)$,
where $y^\pm = (y^0\pm y^1)$.  Spinor conventions are as in Wess 
\& Bagger~\cite{Wess:1992cp}.
The superderivatives are
\bea
\D_+ =   {\p \over \p \tp}  -i\tpb \p_+
~~~~~~&&~~~~~~
\DB_+ = - {\p \over \p \tpb} +i\tp  \p_+  \\
\{\D_+,\D_+\} = \{\DB_+,\DB_+\}=0
~~~~~~&&~~~~~~
\{\DB_+,\D_+\} =2i\p_+.
\eea
Unconstrained superfields are arbitrary functions of $(y^+, y^-,
\theta^+, \tb^+)$. In general, we will work with superfields
constrained in different ways. For this reason, it is worth noting
that $\DB_+$ annihilates the combinations
$z^+=y^+-i\tp\tpb$, $z^-=y^-$, and $\theta^+$.

\subsubsection{The \ZT\ Gauge Multiplet}

To construct gauge theories, we need to extend our superspace
derivatives, $\D_+$ and $\DB_+$, to gauge covariant superderivatives.
The gauge covariant superderivatives ${\cal{D}}_+,
{\bar{\cal{D}}}_+$ acting on charge $1$ fields, and
$\cal{D_{\alpha}}$ ($\alpha=1,2$) satisfy the algebra
\be {\cal{D}}_+^2 = {\bar{\cal{D}}}_+^2 =0, \qquad  \{\ {\cal{D}}_+,
    {\bar{\cal{D}}}_+ \}\ =2i ({\cal{D}}_0+{\cal{D}}_1). \ee
The first two equations imply that ${\cal{D}}_+ = e^{-\Psi} D_+
e^{\Psi}$ and ${\bar{\cal{D}}}_+ = e^{\bar{\Psi}} {\bar{D}}_+ e^{-\bar{\Psi}}$
where $\Psi$ is a superfield taking values in the Lie algebra of the
gauge group. We will restrict to abelian theories in our
discussion. In  Wess-Zumino gauge, the component expansion of $\Psi$ gives
$$\Psi = \theta^+ {\bar{\theta}}^+ (A_0 + A_1) (y^{\alpha}), $$
while
\bea  {\cal{D}}_0 + {\cal{D}}_1 &=& \partial_0 + \partial_1 +i(A_0 +
A_1), \\
 {\cal{D}}_+ &=&\frac{\partial}{\partial \theta^+} -i {\bar{\theta}}^+
({\cal{D}}_0 + {\cal{D}}_1), \\
 {\bar{\cal{D}}}_+ &=& -\frac{\partial}{\partial {\bar{\theta}}^+} +i
\theta^+  ({\cal{D}}_0 + {\cal{D}}_1), \\
 {\cal{D}}_0 - {\cal{D}}_1 &=& \partial_0 -\partial_1 +i V. \eea
The vector superfield $V$ is given by,
\be V= A_0 - A_1 -2i \theta^+ {\bar{\lambda}}_- -2i {\bar{\theta}}^+
\lambda_- +2 \theta^+ {\bar{\theta}}^+ D. \ee
We see that the $A_-$ component of the gauge-field has two real
gaugino partners, while $A_+$ does not.
Under a gauge transformation with
gauge parameter $\L$ satisfying a chiral constraint $\DB_+ \L =0$, the two
 gauge-fields $V$ and $\Psi$ transform as follows
 \bea
&&\delta_\L V = \p_-(\L + \LB), \nn\\
&&\delta_\L \Psi = i (\L-\LB) .\nn
 \eea
Finally, the natural field strength is an uncharged fermionic
chiral superfield, \be \U= [{\bar{\cal{D}}}_+
  ,{\cal{D}}_0   - {\cal{D}}_1] =\DB_+(\p_-\Psi + i V) =
-2\{\l_-(z)-i\tp(D-iF_{01})\}, \ee for which the natural action is
\be S_{\Upsilon} = {1\over 8e^2}\int\!{\rm{d^2}}y\, d^2\t ~\bar{\U}\U
=\frac{1}{e^2}
\int{\rm{d^2}}y\,
\left\{\frac{1}{2}F_{01}^2  +i {\bar{\lambda}}_- (\partial_0 +
\partial_1)\lambda_-  +\frac{1}{2} D^2  \right\}. \ee Since
$\U$ is a chiral fermion, we can also add an FI term of the form
\be \label{defineFI} S_{\rm FI} = {t\over 4}\int\!{\rm{d^2}}y\,
d\tp~ \U \vert_{{\bar{\theta}}^+ =0} + {\rm h.c.} \ee where
$t=ir+{\t\over 2\pi}$  is the complexified FI parameter.

\subsubsection{Chiral Multiplets}

An uncharged chiral superfield is one which satisfies $\DB_+\Phi^0=0$.
Chiral superfields are therefore
naturally expanded in the $z$ coordinates
$z^+=y^+-i\tp\tpb$, $z^-=y^-$, and $\theta^+$.
Bosonic chiral superfields contain the components fields,
 \bea
 \Phi^0&=&\phi(z) +\rt\theta^+\psi_+(z)\\
 &=& \phi(y) +\rt\theta^+\psi_+(y) -i\tp\tpb\p_+\phi(y).
\non \eea

The action for a chiral boson is
 \be S_{\Phi^0} = -{i\over 2}\int\ \,{\rm{d^2}}y\, {\rm{d^2}}\t~ \bar{\Phi^0}\p_-\Phi^0. \ee
With the definition $\Phi^0 = e^{-Q \Psi} \Phi$, we note that $\Phi$ satisfies the
covariant chirality constraint  ${\bar{\cal{D}}}_+  \Phi =0$ for a
 field with $U(1)$ charge $Q$. In
 components,
 \be
 \Phi = \phi(y) +\rt\theta^+\psi_+(y) -i\tp\tpb (D_0+D_1)\phi(y),
 \ee
where $D_\a = \partial_\a + iQ  A_\a$. The corresponding gauge invariant
 Lagrangian is given by,
\bea
 S_{\Phi} &=&  - \frac{i}{2} \int \,{\rm{d^2}}y\, {\rm{d^2}} \theta \
 {\bar{\Phi}} ({\cal{D}}_0 -{\cal{D}}_1)\Phi, \\ &=&
 \int {\rm{d^2}}y\,  \Big\{ - \vert D_{\alpha} \phi\vert^2 +
    i{\bar{\psi}}_{+}(D_0 - D_1) \psi_{+} -i Q {\sqrt 2}
    {\bar{\phi}} \lambda_- \psi_{+} \cr & & +i Q {\sqrt 2} \phi
    {\bar{\psi}}_{+} {\bar{\lambda}}_-  + Q D \vert \phi \vert^2\Big\}. \non     \eea


\subsubsection{Fermi Multiplets}

In addition to bosonic chiral multiplets, there are also fermionic
multiplets which, for uncharged fields, satisfy the condition
\be \DB_+ \G^0 = \sqrt{2} E^0 \ee where $E^0$ satisfies $$ \DB_+ E^0 =0. $$
A component expansion gives the terms \be \G^0 = \chi_- -\rt\tp G
- i\tp \tpb \p_+ \chi_- - \sqrt{2} \tpb E^0. \ee
Note that fermi multiplets have negative chirality.

To satisfy the covariant chirality condition, we again define $\G^0 =
e^{-Q \Psi} \G$, and $E^0 = e^{-Q \Psi} E$ so that
\be \label{almostchiral} {\bar{\cal{D}}}_+ \G = \sqrt{2} E. \ee
The choice of $E$
plays an important role in our discussion for reasons that we will
describe later. We follow \cite{Witten:1993yc}\ and
assume that $E$ is a holomorphic function of chiral superfields
$\Phi_i$.
The action for $\G$ is given by
 \bea \label{LF} S_{\G} &=& -{1\over 2}\int\!{\rm{d^2}}y\, d^2\t
 ~\bar{\G} \G \\
&=& \int {\rm{d^2}}y\, \left\{i
    {\bar{\chi}}_{-}  (D_0 + D_1) \chi_{-} + \vert G \vert^2 - \vert E
    \vert^2  - \left( \bar{\chi}_- {\partial E \over \partial
      \phi_i} \psi_{+i} + \bar{\psi}_{+i}
{\partial \bar{E} \over \partial \bar{\phi}_i} \chi_- \right) \right\}. \non \eea

A special case of \C{LF}\ of particular importance to us; namely, where
$E = \Sigma  {\cal E}(\Phi_i)$ and
$\Sigma$ is an uncharged chiral superfield with component expansion
\be \label{defSm} \Sigma = \s +\rt  \tp\lb_+ -i\tp {\bar\theta}^+\partial_+\s.\ee
Then the action for $\G$ is given by
 \bea  S_{\G} &=&
  \int {\rm{d^2}}y\, \Bigg\{ i
    {\bar{\chi}}_{-}  (D_0 + D_1) \chi_{-} + \vert G \vert^2 - \vert
    \s {\cal E}
    \vert^2
-  \left( \s \bar{\chi}_- {\partial  {\cal E}\over \partial
      \phi_i} \psi_{+i} + \bar{\s} \bar{\psi}_{+i}
{\partial \bar{{\cal E}} \over \partial \bar{\phi}_i} \chi_- \right) \\
& & - \left({\cal E}  \bar{\chi}_- {\bar\lambda}_+ +\bar{{\cal E}} \lambda_+ \chi_-
\right) \Bigg\}. \non \eea

\subsubsection{\ZT\ Superpotentials}

In general, we can also add superpotential terms. These
terms depend on Fermi superfields, $\G_a$, and holomorphic
functions, $J^a$,
of the chiral superfields
\bea \label{superpotential}
S_{\cal W} &=& -{1\over \sqrt{2}}\int\!{\rm{d^2}}y\,d\tp~ \G_a
J^a \vert_{{\bar\theta}^+=0} - {\rm
  h.c.},\\ &=& -\int{\rm{d^2}}y\, \left\{G_a J^a(\phi_i) + \chi_{-a}
\psi_{+i}  {\partial J^a \over
  \partial \phi_i} \right\}- {\rm
  h.c.}.
\non \eea
Since $\G_a$ is not an honest chiral superfield but satisfies
\C{almostchiral}, we need to impose the condition
\be E \cdot J=0 \ee
to ensure that the superpotential is chiral. Lastly, note that gauge invariance
requires
$$Q_{\G_a}=-Q_{J^a}.$$

\subsection{$(2,2)$ Supersymmetry}
\label{ttsusy}
A special class of \ZT\ theories have enhanced $(2,2)$
supersymmetry. To describe these theories, we enlarge our superspace
by adding two fermionic coordinates,
$(y^+, y^-,\t^+,\tb^+, \t^-, \tb^-)$, and we introduce additional
supercovariant derivatives
\bea  D_- &=& \frac{\partial}{\partial \theta^-} - i{\bar{\theta}}^-
\partial_-, \\
{\bar{D}}_- &=& - \frac{\partial}{\partial {\bar{\theta}}^-} + i\theta^-
\partial_-.
\eea
We normalize integrals over all the fermionic coordinates of
superspace with the convention that
\be \int d^4 \theta ~\theta^+ {\bar{\theta}}^+ \theta^-
 {\bar{\theta}}^- =1. \ee
Unlike the \ZT\ case, there are two kinds of chiral
multiplet. Conventional chiral multiplets, $\Phi$, satisfy the
conditions
\be
\DB_+ \Phi = \DB_- \Phi =0,
\ee
while twisted chiral multiplets, $Y$, satisfy the conditions
\be
\DB_+ Y = \D_- Y =0.
\ee
Both kinds of multiplet can be reduced to \ZT\
multiplets.  An uncharged $(2,2)$ chiral
multiplet gives a \ZT\ chiral and Fermi multiplet,
\be
\Phi^{(0,2)} = \Phi \vert_{\t^- = \tb^-=0}, \qquad \G^{(0,2)} =
    {1\over \sqrt{2}} \D_- \Phi\vert_{\t^- = \tb^-=0}.
\ee
Similarly, a twisted chiral multiplet (which is always uncharged) also
gives a chiral and
Fermi multiplet,
\be
Y^{(0,2)} = Y \vert_{\t^- = \tb^-=0}, \qquad F^{(0,2)} =
   - {1\over \sqrt{2}} \DB_- Y\vert_{\t^- = \tb^-=0}.
\ee
There is also a $(2,2)$ vector superfield, $V$, whose field strength
is a twisted chiral multiplet (often denoted $\Sm$). On reduction to
\ZT, we obtain a chiral multiplet, $\Sm^{(0,2)}$, and a vector
multiplet, $V^{(0,2)}$,
as follows:
\be
\tb^+ \Sm^{(0,2)} = - {1\over \sqrt{2}} \D_- V\vert_{\t^- = \tb^-=0},
\qquad  V^{(0,2)} - i \p_-
\Psi^{(0,2)} = - \DB_- \D_- V\vert_{\t^- = \tb^-=0}.
\ee
Lastly, we note that a $(2,2)$ chiral multiplet with $U(1)$ charge $Q$
reduces to a charged \ZT\ chiral multiplet, $\Phi^{(0,2)}$, and a charged
Fermi multiplet, $\G^{(0,2)}$, with a particular non-vanishing $E$ so that
$$  {\bar{\cal{D}}}_+ \G^{(0,2)} = \sqrt{2} E $$
where $E$ is given by~\cite{Witten:1993yc}
\be E = {\sqrt 2} Q \Sigma^{(0,2)} \Phi^{(0,2)}.
\ee

%
%

\section{Duality in \ZT\ Models}\label{derivation}
\subsection{Duality in Free \ZT\ Theories}

The essential magic of mirror symmetry is that a priori distinct target spaces
may lead to identical string spectra. A simple example of a mirror symmetry is
T-duality, which identifies the spectrum of strings on tori of radii $R$ and $1/R$.
Since the world-sheet theory on a torus is exactly solvable, T-duality of tori, unlike
general mirror symmetry, is easily derived directly in the world-sheet theory.
In this section, we recall the standard prescription for deriving
such dualities, following \RV\ (RV)~\cite{Rocek:1992ps}. We then
apply this prescription to \ZT\ models; this will play an essential role in
our dualization of \ZT\ GLSMs in the following sections.
We begin by reviewing the dualization procedure for free \ZT\ theories before addressing the
more interesting case of \ZT\ GLSMs.

\subsubsection{T-duality as Abelian Duality}

T-duality identifies the momentum (winding) modes on a circle of radius $R$ with the
winding (momentum) modes on a circle of radius $1/R$.  As such, it may
be implemented via a Legendre transformation in a way we now recall.
The theory must admit a $U(1)$ isometry, and the simplest
example is a free scalar on a circle of radius $R$ with action
\be \label{trivial} S = {R^2 \over 4\pi} \int d^2 y~ (\partial\phi)^2 . \ee
To dualize the shift symmetry of $\phi$,  we introduce a Lagrange
multiplier $1$-form, $B$, with modified action
\be \label{modaction} S = {1\over 4\pi R^2} \int  B\wedge *B
      -{i\over 2\pi}     \int  \phi \,  dB. \ee
Path-integrating out $B$ in Euclidean space amounts to
solving the $B$ equation of motion
giving $$ B=-i R^2 * d\phi.$$
When plugged into the action, we recover our original theory~\C{trivial}.

To obtain the dual description, we instead integrate out $\phi$. This
enforces the condition that $B$ be closed,
$$ d B = 0. $$
Locally, we can express $B$ in the form
$$ B= d \theta $$
where $\theta$ is not necessarily single-valued. The only caveat to
this argument is that $\phi$ is periodic so for the action
\C{modaction}\ to be well-defined, we require that $B$ be an integral
class.

The dual action is therefore
\be S = {1\over 4\pi R^2} \int d^2 y~  (\partial\theta)^2. \ee
We note that $\theta$ must be periodic with radius $1/R$. This is most
easily seen by
comparing the spectra of the original and dual descriptions.
A momentum mode for $\phi$ can only correspond to a
solitonic excitation for $\theta$ implying that $\theta$ is compact.

\subsubsection{Dualization in \ZT\ Superspace}

The above reasoning can be extended to \ZT\ superspace.  To dualize a \ZT\
chiral multiplet, $Y$, we consider the action
\be {\cal S}_{ch} = -\frac{1}{4}\int d^2 y~ d^2 \theta ~(R^2 AB + A(Y+\bar{Y})
      -iB\partial_-(Y-\bar{Y}) ),\ee
where $A$ and $B$ are unconstrained real superfields without kinetic
terms.
Integrating out these non-dynamical real superfields gives the relations
\bea A &=& \frac{i}{R^2}\partial_-(Y-\bar{Y}), \\
     B &=& -\frac{1}{R^2}(Y+\bar{Y}).
\eea
Inserted back into the action, these relations give, up to total
derivatives, the standard action for $Y$
\be {\cal S}_{ch} =
-\frac{i}{2R^2}\int  d^2 y~d^2\theta~ \bar{Y} \partial_- Y .\ee

To obtain the dual description, we instead integrate out the chiral
 superfield, $Y$, which gives the relation,
\be \DB_+ (A+i \partial_-B)=0,\ee
allowing us to write $A = i\partial_-(\Phi-\bar{\Phi})$ and $B =
(\Phi+\bar{\Phi})$,
where $\Phi$ is a bosonic chiral superfield.\footnote{The general solution for
$B$ includes an arbitrary superfield anihilated by $\p_-$, i.e.,
$B=\Phi + \bar{\Phi} + {\cal S}$, where $\p_-{\cal S}=0$.  Plugging
 this solution into the action and integrating
by parts reveals that all terms involving ${\cal S}$ vanish. We can
 therefore safely neglect any such $\cal S$.}
The resulting dual action is
\be {\cal S}_{ch} = -\frac{i R^2}{2} \int  d^2 y~d^2 \theta ~ \bar{\Phi} \partial_- \Phi.
\ee
The duality map is therefore
\be \label{mapbos} (Y+\bar{Y}) = -R^2(\Phi+\bar{\Phi}) \qquad\quad
\p_-(Y-\bar{Y}) = R^2\p_-(\Phi-\bar{\Phi}) .\ee

We can also dualize a chiral Fermi multiplet in a similar way.  Let
$F$ be a chiral Fermi multiplet satisfying $\DB_+ F=0$, and let $\N$ be
an unconstrained Fermi superfield.
To induce dual descriptions, we consider the following first-order action
\be {\cal S}_f= \int\!  d^2 y~d^2\t \left\{  -{R^2\over 2}\bar{\N}\N ~ -\half(F\bar{\N}
+\N\bar{F}) \right\}. \ee
Integrating out $\bar{\N}$ gives the relation
\be \N = {1\over R^2} F, \ee
which when substituted into the action gives
\be {\cal S}_f= {1\over 2 R^2}\int\!  d^2 y~d^2\t ~ \bar{F} F.\ee
Integrating out $F$ instead gives the relation
\be \label{chiralconstraint} \DB_+ \bar{\N} =0 \ee
which gives a dual action
\be {\cal S}_f= {R^2\over 2}\int\! d^2 y~ d^2\t ~ \bar{\Gamma}\Gamma\ee
where the chiral superfield $\Gamma = \bar{\N}$ and $\bar{\N}$ satisfies the chirality
constraint \C{chiralconstraint}. The corresponding duality map is
\be \label{mapferm} {\bar{\Gamma}} = {1\over R^2} F .\ee
{}From this map, we see that the action on fermions is no more than a
rescaling at the level of free-fields.
We should point out that this map becomes more complicated when we
include interactions with chiral bosons (via the $\DB_+\G= \sqrt{2}
E(\Phi)$ coupling), as we shall see in detail in section~\ref{generalzt}. 

\detail{The Duality Map in Components}\\
It is useful to restate these dualities in component form.
Start with a supersymmetric sigma model on a
cylinder of radius $R$ with action,
\be S ~=
R^2 \int d^2 y ~\left(- \partial_\alpha \bar{\phi}~ \partial^{\alpha} \phi
         +i~{\bar{\psi}}_+ \partial_- \psi_+ \right), \ee
where
\be \phi = \rho + i \varphi \ee
is the lowest component of a \ZT\ bosonic chiral multiplet,
$\Phi=\phi(z)+\rt\tp\psi_+$,
whose imaginary part is periodic
$$ \varphi \sim \varphi + 2 \pi. $$
Dualizing this isometry amounts to dualizing $\varphi$ and $\psi_+$.
Starting with the bosonic fields, we see that the resulting dual
metric is
\be  ds^2=R^2d\rho^2 + \frac{1}{R^2}d\vartheta^2 = \frac{1}{R^2}(R^4d\rho^2 + d\vartheta^2), \ee
where
$$ \vartheta \sim \vartheta + 2\pi. $$
This suggests that the natural dual coordinate is $\eta = R^2 \rho - i \vartheta$. Written
in terms of $\eta$, the full dual lagrangian takes the simple form
\be S =  \frac{1}{R^2} \int d^2 y
    ~ \left( - \partial_{\alpha} \bar{\eta} \partial^{\alpha} \eta
    ~ +i\bar{{\widetilde{\psi}}}_+ \p_- {\widetilde{\psi}}_+ \right).\ee
If we have correctly identified our dual fields, their supervariations should again close
with the correct normalizations.  By dualizing the original susy variations, we find
\be \delta \psi_+ = {\sqrt 2} i \partial_+ \phi {\bar{\epsilon}}_-
~~~\Rightarrow~~~
    \delta {\widetilde{\psi}}_+ = {\sqrt 2} i \partial_+ \eta {\bar{\epsilon}}_- ,\ee
where ${\widetilde{\psi}}_+ = R^2 \psi_+$, so our dualization is consistent
with supersymmetry and the dual
fields also fill out a \ZT\ chiral multiplet
$$Y = \eta(z) + {\sqrt 2} \theta^+ {\widetilde{\psi}}_+.$$

The case of a free fermionic \ZT\ chiral supermultiplet, $\G = \chi_-(z) + \rt\tp g$,
 can also be expressed in components.
The initial action is
\be S ~=  {i}R^2 \int d^2 y ~\left(\bar{\chi}_- \p_+ \chi_- ~+~ |g|^2 \right) \ee
and the dual action is simply
\be S ~=   {i\over R^2} \int d^2 y ~ \left(\bar{\psi}_- \p_+ \psi_-
~+~ |f|^2 \right), \ee
where $\psi_- = R^2 \bar{\chi}_-$, and we can assemble $\psi_-$ and $f$ into
a chiral Fermi superfield $F=\psi_-(z)+\rt\tp f$.

\subsection{Duality in \ZT\ Gauge Theories}

We next consider the dualization of \ZT\ gauged linear sigma
models. The dualization for $(2,2)$ gauged linear sigma models has been carried
out in~\cite{Hori:2000kt}. Since a $(2,2)$ model is a special case of
a $(0,2)$ model, we can reduce the duality map of~\cite{Hori:2000kt}\
to a map on $(0,2)$ fields. This gives us a particular case of a \ZT\ duality. Next we
generalize this duality to arbitrary \ZT\ theories.

It is important to keep in mind that the $U(1)$ action we wish to
dualize is no longer free. This is
an issue that we will ignore for the moment. The way this
issue emerges in the dual description is via the generation of a
non-perturbative superpotential to which we turn in
section~\ref{derivesup}. This is, perhaps, the most critical aspect of the
dualization procedure.


\detail{Warm-up: \TT\ Duality in \ZT\ Superspace} 

We begin by expressing the results of Hori and
Vafa~\cite{Hori:2000kt}\ in \ZT\ language. This gives a special case
of a more general \ZT\ duality map.   
 The simplest such
theory is that of a single chiral \TT\ multiplet with charge $Q$
coupled to a \TT\ vector multiplet. 
The
slightly involved rewriting of the original $(2,2)$ theory in \ZT\
language is performed in Appendix~\ref{reduction}.  

When reduced to
\ZT\ language, the end result is a
\ZT\ gauge theory with a chiral multiplet, $\Phi$, a chiral
Fermi multiplet, $\G$, both with $U(1)$ charge $Q$. In addition, the
\TT\ vector multiplet reduces to a \ZT\ vector multiplet
with field strength $\U$,
and an uncharged chiral multiplet, $\Sm$. The Lagrangian for these
fields is given by
\bea \label{originalLag}
L &=&  - {i\over 2} \int d^2 \theta ~{\bar{\Phi}} ({\cal{D}}_0 - {\cal{D}}_1 ) \Phi
 -{1\over 2} \int d^2 \theta ~\bar{\Gamma} \Gamma + {i\over 2 e^2} \int d^2 \theta
 ~{\bar{\Sm}} \p_- \Sm  + {1\over 8e^2}\int d^2\t ~\bar{\U}\U \non \\ & & + \left\{
{ t\over 4}\int d\tp~ \U
\vert_{{\bar{\theta}}^+ =0} + {\rm h.c.} \right\} \eea
where $e$ is the gauge coupling, and $t=ir+{\t\over 2\pi}$
is the complexified FI parameter. The Fermi superfield satisfies
${\bar{\cal{D}}}_+ \Gamma= {\sqrt 2} E$ with $E$ given by~\cite{Witten:1993yc}
\be
E = {\sqrt 2} Q \Sigma \Phi.
\ee

Dualizing an isometry means exchanging the roles of the generator of
the isometry and its canonical conjugate. This means that under this
generalized world-sheet T-duality, a charged field maps to
an uncharged field. The dual variables, a chiral superfield $Y$ and
chiral Fermi superfield $F$,  are therefore neutral.

The dual action is again obtained by reducing the $(2,2)$ result
in Appendix~\ref{reduction}. The result is, \bea \widetilde{L} &=&
\frac{i}{8} \int d^2 \theta ~\left[\frac{Y-\bar{Y}}{Y + \bar{Y}}
\partial_- (Y + \bar{Y})
 +2i\frac{\bar{F} F}{Y + \bar{Y}} \right]  - \left( \frac{Q}{2}
\int d \theta^+  [ \Sigma F + \frac{i}{2} Y \Upsilon ]
+ {\rm h.c.} \right) \non \\ & & + {i\over 2 e^2} \int d^2 \theta
 ~{\bar{\Sm}} \p_- \Sm  + {1\over 8e^2}\int d^2\t ~\bar{\U}\U
\non + \left\{
{ t\over 4}\int d\tp~ \U
\vert_{{\bar{\theta}}^+ =0} + {\rm h.c.} \right\},  \eea
where ${\DB}_+ Y = {\DB}_+ F = 0$. The $(2,2)$ duality map can also be
expressed in \ZT\ language as described in
Appendix~\ref{reduction}. The map becomes,
\bea \bar\Phi \Phi &=& \frac{1}{2} (Y + \bar{Y}), \\
 \label{dec} -i \bar{\Phi} ({\buildrel \leftrightarrow \over
   \partial_-} +iQ V) \Phi + \bar\Gamma \Gamma
&=& \frac{i}{4} \partial_- (Y-\bar{Y}), \\
\frac{1}{2} \bar{F} &=& \bar\Phi \Gamma,\eea
where $$ \bar{\Phi} {\buildrel \leftrightarrow \over
   \partial_-}  \Phi = {1\over 2} \left(  \bar{\Phi}\p_- \Phi - \Phi
\p_- \bar{\Phi} \right). $$ Since it is rather important, we must
emphasize that $Y_i$ is not a conventional $\mathbb{C}$-valued
field. Rather, \be \label{Yperiod} {\rm Im}(Y_i) \sim {\rm
Im}(Y_i) + 2 \pi, \qquad {\rm Re}(Y_i) \geq 0. \ee

One must interpret this duality map (and the $(2,2)$ map) with
great care. As an equivalence between superfields, the map does
not make sense. The component expansions on both sides of the
equivalence do not agree. However, we will only use the relations
between the lowest components when we need explicit relations.
Those relations and the dualization procedure itself (as an
equivalence between theories) do make sense.


\subsubsection{Dualizing \ZT\ Chiral Multiplets}

In string compactifications, chiral multiplets describe the
geometry of our target space, while chiral Fermi multiplets  
define a vector bundle over this
space. Our current
task is to dualize charged chiral and Fermi multiplets.
We begin by considering just chiral multiplets with no coupled Fermi
multiplets. 

We need a starting action along the lines described
earlier: let us start with the candidate action
\be \label{sch}
 {\cal S}_{ch}= \int  d^2 y ~d^2\theta~ \left\{ -\frac{i}{2}e^{2(\Psi + B)} (iV+iA) ~
                 -i \F~\!\DB_+(\p_-B + iA) + {\rm h.c.} \right\}
\ee
where $\F$ is a neutral unconstrained fermionic superfield, while $A$ and $B$ are
unconstrained real superfields.

Integrating out the unconstrained Lagrange multiplier field $\F$ yields the constraint
\be \DB_+ (\p_-B + iA)=0 , \ee
the general solution of which is\footnote{A note of caution is in order. In general we should
write $2B = \Pi +\bar{\Pi} + 2\S_R$, where $\S_R$ is a real bosonic superfield
annihilated by $\p_-$. However, a real bosonic \ZT\ superfield can always be written
as the real part of a complex chiral superfield, $2\S_R=(\S+\bar{\S})$; both have four
independent real components. Absorbing $\S$ into $\Pi$ gives the Lagrangian written above
up to a shift $V\to V+c$, where $c$ is a constant c-number; since this may be absorbed by
a gauge transformation, (\ref{ADMM1}) is indeed the most general solution of the constraint.}
\be \label{ADMM1} 2B=\Pi +\bar{\Pi}   ~~~~~~~ 2iA=\p_-(\Pi-\bar{\Pi})
\ee
where $\Pi$ is a chiral superfield.
Plugging this back into the action gives, after some reordering,
\be  {\cal S}_{ch}=-\frac{i}{2}\int  d^2 y ~d^2\theta~ e^{\Psi+\bar{\Pi}} ~(\p_- +iV)~\!
e^{\Psi+\Pi}. \ee
We can make the kinetic term canonical by changing variables to the covariantly chiral field
$\Phi=e^{\Psi + \Pi}$, in terms of which the action reads
\be  {\cal S}_{ch}= -\frac{i}{2}\int  d^2 y ~d^2\theta~ \bar{\Phi}
~({\mathcal{D}}_0 -{\mathcal{D}}_1)~\!\Phi  .\ee

Integrating out instead the auxiliary gauge fields $A$ and $B$ requires first
integrating the constraint terms by parts.  Defining
${1\over 4} Y=\DB_+\F$, the auxiliary field variations give
\bea
\label{Aeom}
\delta_{A} \,  & \Rightarrow & \, \half e^{2(\Psi + B)} - {1\over 4}(Y+\bar{Y})=0, \\
\delta_{B} \, & \Rightarrow & \, -ie^{2(\Psi + B)}(iV+iA) -{i\over 4} \p_-
(Y-\bar{Y})=0. \eea
 Solving these gives
\be \label{ADMM2} 2 B = -2\Psi+ \ln(\frac{Y+\bar{Y}}{2}) ~~~~~~ iA=-iV-{\p_-(Y-\bar{Y})\over 2(Y+\bar{Y})}. \ee
Plugging back into the action and simplifying gives
\be \label{kinetic}{\cal S}_{ch} =  {i\over 8}\int  d^2 y ~d^2\theta~
    {(Y-\bar{Y})~\!\p_-(Y+\bar{Y})\over
(Y+\bar{Y})}
                 - {i\over 4}\int  d^2 y ~d\tp Y\Upsilon + {\rm h.c.}
\ee
Comparing \C{ADMM1} and \C{ADMM2}, we see that the duality map is
 \be \label{mirror1}  \bar{\Phi}\Phi ~=~\half (Y+\bar{Y}),~~~~
 \bar{\Phi} ( {\buildrel \leftrightarrow \over \partial_-} +i V) \Phi
 =-{1\over 4}\partial_- (Y-\bar{Y}).\ee
On comparing with \C{dec}, we see that the fermion bilinear has
dropped out as we intuitively expect for this special case with no
coupling to the left-moving fermions.

\subsubsection{Dualizing \ZT\ Fermi Multiplets} \label{FermiDual}

We can similarly dualize Fermi supermultiplets. The first order Lagrangian is
 \be \label{sf}
{\cal S}_f = ~\int\! d^2 y ~d^2\t ~\left\{-{1\over 2}~ \bar{\N} \N
                 + \S \left( {\bar{\cal D}}_+\N - \sqrt{2} E \right) ~
 - \bar{\S}\left(
{\cal D}_+\bar{\N} + \sqrt{2} \bar{E} \right) \right\},
 \ee
where $\N$ is an unconstrained Fermi superfield, $\S$ is an
unconstrained bosonic superfield,  and $E$ is a bosonic (covariantly) chiral
 multiplet. Both $\N$ and $E$ have charge $Q$ while $S$ has charge $-Q$.
Integrating out $\S$
gives the equation of motion
$${\bar{\cal D}}_+\N= \sqrt{2} E,$$
which is solved by $\N=\G$,
where $\G$ is a chiral
Fermi superfield in the general sense of \C{almostchiral}.
The corresponding Lagrangian is just
 \be {\cal S}_f = ~-\half\int\! d^2 y ~d^2\t ~\bar{\Gamma}\Gamma. \ee
Solving the $\N$ equation of motion instead gives the relation
\be \label{smallrel} {\bar{\cal D}}_+ \S = - {1\over 2} \bar{\N} \ee
Let us set $ \bar{\N} = \GG$ so \C{smallrel}\ implies that
$${\bar{\cal D}}_+ \GG=0.$$
Substituting gives the action,
\be
{\cal S}_f = \int{\! d^2 y ~d^2\t ~ \left\{- {1\over 2} \bar{\GG}\GG - \sqrt{2} \S E -
  \sqrt{2} \bar{\S} \bar{E} \right\}.}
\ee
We now write,
$$  \int\! d^2 y ~d^2\t ~ \sqrt{2} \S E = -\int  d^2 y ~d \t^+
\sqrt{2}\left(\bar{\cal D}_+ \S \right) E =  \int  d^2 y ~d\t^+ {1\over \sqrt{2}} \GG E
$$
since $\bar{\cal D}_+ E=0.$ Note that $\GG$ has charge $-Q$ while $E$
has charge $Q$. Let us define a neutral superfield $F = \GG E$. The
reason to do this is so that (in nice cases) we can express the action
in terms of the
dual chiral fields, $Y$. In terms of $F$, the action takes the form
\be
{\cal S}_f = - {1\over 2}\int\! d^2 y ~d^2\t ~ {\bar{F} F \over {\bar E} E}
  - \left\{ \int\! d^2 y ~d\t^+ {1\over \sqrt{2}} F + {\rm h.c.} \right\}.
\ee
Let us define
\be |E|^2 =  {Y_E + \bar{Y}_E \over 2} \ee
so
\be \label{nfermact}
{\cal S}_f = - \int\! d^2 y ~d^2\t ~ {\bar{F} F \over  Y_E + \bar{Y}_E}
  - \left\{ \int\! d^2 y ~d\t^+ {1\over \sqrt{2}} F + {\rm h.c.} \right\}.
\ee The duality map for Fermi superfields is then given by \be
\label{fermiduality} F = E \, \bar{\G}. \ee 
In nice cases, we can find explicit expressions
for $|E|^2$ using the duality map \C{mirror1}; for example, if $E$
is a monomial.

An important special case, related to the discussion around eq. \C{defSm}, is when
$E = \Sm \cal{E}$, where $\Sm$ is an uncharged chiral boson. In this case we can
rescale $\GG$ by $\cal{E}$ rather than $E$ to get \be
\label{specialcase} {\cal S}_f = -\int\! d^2 y ~d^2\t ~ {\bar{F} F
\over  {\bar {\cal E}}{\cal E} }
 - \left\{\int\! d^2 y ~d\t^+ {1\over \sqrt{2}} \Sm F + {\rm h.c.} \right\}.
\ee

\subsubsection{Dualizing General \ZT\ Models}
\label{generalzt}
Things get more interesting when we dualize chiral multiplets, $\Phi_i$, coupled to
Fermi multiplets, $\G_a$, via constraints of the form 
$${\bar {\cal{D}} }_+ \G_a=\sqrt{2}\Sm\E_a(\Phi_i).$$  
In a situation like this, we can perform our previous dualization
procedure but we can only explicitly solve for the dual action when
$\E$ is a monomial. 

Start with the sum of first order actions 
$$ S = S_{ch} +S_{f}$$
where $S_{ch}$ is given in \C{sch}\ and $S_f$ is given in \C{sf}. 
We permit $\E$ to be an arbitrary (generally non-local) function of
$A,B$ and $\Psi$. As before, integrating out $\S$ and $\F$ gives an
action, 
$$  {S}= -\frac{i}{2}\int  d^2 y ~d^2\theta~ \bar{\Phi}
~({\mathcal{D}}_0 -{\mathcal{D}}_1)~\!\Phi 
~-\half\int\! d^2 y ~d^2\t ~\bar{\Gamma}\Gamma,
$$
where ${\bar {\cal{D}} }_+\G=\sqrt{2}\Sm\E(\Phi)$. 

To get the dual description, we integrate out $A, B$ and $\N$. From
integrating out $\N$, we get $|\E|^2$ in the kinetic term for the
fermions as in \C{specialcase}. In
general, the $A$ and $B$ equations of motion are complicated
(non-local) functions of $A$ and $B$. For the particular case,
\be
|\E|^2 = e^{-2N (\Psi+B)},
\ee
the $A$ equation of motion is unchanged from \C{Aeom}, but the $B$
equation of motion gives
\be
A = -V+{i\over 2} {\p_- (Y-\bar{Y}) \over Y+\bar{Y}} - N \left( {2\over
  Y+\bar{Y}}\right)^{N+1} { \bar{F} F \over |\Sm|^2}. 
\ee
In the original theory, this corresponds to the case $\E = \Phi^N$. 

The corresponding dual action is given by,
\bea S &=&  \int  d^2 y ~d^2\theta~
   \left[ {i\over 8} {(Y-\bar{Y})~\!\p_-(Y+\bar{Y})\over (Y+\bar{Y})}
    - {2^{N-1}\bar{F} F\over (Y+\bar{Y})^N} \right] \cr & & - 
\int\! d^2 y ~d\t^+ \left[ {1\over \sqrt{2}} \Sm F 
                 - {i\over 4} Y\Upsilon \right]+ {\rm h.c.},
\eea
so the action takes the same form we found before. 
What has changed is the duality map, which now reads
\be \label{gen_mirror1}  \bar{\Phi}\Phi ~=~\half (Y+\bar{Y}),~~~~
                     \bar{\Phi} ( \lrp_- +i V) \Phi -i N\bar{\G}\G
                        =-{1\over 4}\partial_- (Y-\bar{Y}).
\ee
On comparing with \C{mirror1}, we note the appearance of a fermion bilinear; for the special
case $N=1$, this reproduces the \TT\ result \C{dec}, as expected.

Unfortunately, things rapidly become difficult once we consider general functions
$\E(\Phi)$, because the $A, B$ equations of motion involve complicated
functions of $A$ and $B$. So the action (and duality map) cannot, in
general, be written in closed form. There are really two issues: the
first is that we
cannot express $|\E|^2$
in terms of $Y$ and $\bar{Y}$. However, this only affects the kinetic terms for the
dual Fermi multiplets, but not any holomorphic quantities. 
The second issue is finding the exact duality map. Fortunately, the
correction to the naive
duality map always involves terms with two or more fermions. This kind
of correction will play no role in our subsequent computations, so we
can safely ignore it.

\section{The Exact Dual Superpotential} \label{derivesup}

\subsection{Lagrangians and Conventions}

We have derived the perturbative superpotential of the dual theory. It is
easy to extend the analysis of the previous sections to theories with
several superfields carrying arbitrary charges. Let us consider a theory
containing chiral superfields, $\Phi_i$, with charges $Q_i$ and Fermi
superfields, $\Gamma_a$, with charges $Q_a$. We shall always assume that the
charges satisfy the gauge anomaly cancellation condition required for
a consistent quantum field theory
\be
\sum_i Q_i^2 = \sum_a Q_a^2.
\ee
This condition is equivalent in the infra-red to the geometric
constraint given in \C{conditiontwo}.

When dualizing \ZT\ models, we are faced with the natural
question: which fields should we dualize? To answer this question,
we need to consider different choices for $E$. The first 
choice we might consider is $E=0$, but this is problematic because (in general)
there is no natural way to construct a neutral dual Fermi
superfield. In section~\ref{equalandopp}, we will describe a particular model in
which there is a natural choice.

One possible way to proceed for $E=0$ is to dualize the chiral
superfields leaving the Fermi fields untouched. This seems reasonable
because chiral and the Fermi superfields
interact only indirectly via their coupling to gauge-fields. 
In this situation,
the fields map as follows from the original to the dual
description
$$ ( \Phi_i, \G_a) \rightarrow (Y_i, \G_a). $$
The chiral superfields $Y_i$ are uncharged, while the
Fermi superfields $\Gamma_a$ are charged.
The difficulty we seem to encounter is with the superpotential. 
Under a partial
dualization where the theory is described in terms of $(Y_i,
\G_a)$, it is hard to even define what is meant by a
superpotential. There is clearly no perturbative superpotential of the
form appearing in \C{nfermact}\ because of gauge invariance. It is
also unclear how to take into account instanton effects in the
original theory; it seems likely that these non-perturbative effects
result in a non-local dual theory. For these reasons, for the most
part we restrict
to $E\neq 0$.  

When the charged
chiral and Fermi superfields interact with each other via
$E_a \neq 0$, we must dualize both the chiral and Fermi
superfields
$$ ( \Phi_i, \G_a) \rightarrow (Y_i, F_a) $$
where both $Y_i$ and $F_a$ are neutral.

We give the Lagrangians for the dual theory for two classes
of $E_a$. Omitted is the kinetic term for the vector
multiplet with field strength $\U$
given, for example, in~\C{originalLag}.
{}For $E_a = f_a (\Phi_i)$,
\be \label{NP2} \widetilde{L} =  \frac{i}{8} \sum_i \int d^2 \theta
~ \frac{Y_i - {\bar{Y}}_i}{Y_i + {\bar{Y}}_i}
\partial_- (Y_i + {\bar{Y}}_i)
-\sum_a \int d^2 \theta~\frac{{\bar{F}}_a F_a}{Y_{f_a} + {\bar{Y}}_{f_a}}
 + (\int d \theta^+ ~\widetilde{W} + {\rm h.c.}),\ee
where, \be \widetilde{W} = -\frac{i \Upsilon}{4}(\sum_i Q_i Y_i +
it) - \frac{1}{\sqrt 2} \sum_a  F_a \ee and \be \frac{1}{2}
(Y_{f_a} + {\bar{Y}}_{f_a}) =  \vert f_a (\phi_i) \vert^2.\ee
{}For the second case, $E_a = \Sigma g_a (\Phi_i)$,
and $\Sm$ is a neutral chiral superfield with canonical kinetic
terms. Rescaling as in \C{specialcase}\ gives the Lagrangian \be
\label{NP3} \widetilde{L} =  \frac{i}{8} \sum_i \int d^2 \theta ~
\frac{Y_i - {\bar{Y}}_i}{Y_i + {\bar{Y}}_i}
\partial_- (Y_i + {\bar{Y}}_i)
-\sum_a \int d^2 \theta~\frac{{\bar{F}}_a F_a}{Y_{g_a} + {\bar{Y}}_{g_a}}
 + (\int d \theta^+ ~\widetilde{W} + {\rm h.c.}),\ee
where,
\be \widetilde{W} = -\frac{i \Upsilon}{4}(\sum_i Q_i Y_i + it)
- \frac{\Sigma}{\sqrt 2} \sum_a F_a, \ee
and,
\be \frac{1}{2} (Y_{g_a} + {\bar{Y}}_{g_a}) =  \vert g_a (\phi_i) \vert^2.\ee

The dual superpotential is exact in perturbation theory because of
perturbative  non-renormalization
theorems~\cite{Dine:1986zy, Dine:1987bq}. However, there can be
non-perturbative corrections. Our aim
is to determine the exact form of the dual superpotential taking into
account the non-perturbative effects generated by vortex
instantons in the original theory~\cite{vortex, Nielsen:1973cs}. We should
note, however, that the superpotential of the original theory
does not receive non-perturbative corrections as recently shown
in~\cite{Basu:2003bq}.

Before proceeding further, we state our field
expansion conventions and some relevant formulae that we need both here and in
later discussion.
In the original theory, the charged chiral superfields, $\Phi_i$,
satisfy ${\bar{\cal{D}}}_+ \Phi_i =0$, and have the component field
expansion
\be \Phi_i =  \phi_i + {\sqrt 2}\theta ^+ \psi_{+i}
-i \theta ^+ {\bar{\theta}}^+ D_+ \phi_i  .\ee
The charged Fermi superfields, $\Gamma_a$, satisfy
${\bar{\cal{D}}}_+ \Gamma_a = {\sqrt 2} E_a$, and have the component
field expansion
\be \Gamma_a = \chi_{-a} - {\sqrt 2}\theta ^+ G_a
-i \theta ^+ {\bar{\theta}}^+ D_+ \chi_{-a} -
{\sqrt 2} {\bar{\theta}}^+ E_a .
\ee
In the dual theory, the neutral chiral superfields,
$Y_i$, satisfy ${\bar{D}}_+ Y_i =0$, and have the component field expansion
\be Y_i =  y_i + {\sqrt 2}\theta ^+ {\bar{\xi}}_{+i}
-i \theta ^+ {\bar{\theta}}^+ \partial_+ y_i ,  \ee
while the neutral Fermi
superfields, $F_a$, satisfy ${\bar{D}}_+ F_a =0$, and have the
component field expansion
\be F_a = \eta_{-a}  - {\sqrt 2}\theta ^+ H_a
-i \theta ^+ {\bar{\theta}}^+ \partial_+ \eta_{-a} .
\ee

{}Finally let us state some general results obtained from the duality
maps that
we derived in section~\ref{derivation}. We will need these formulae
for studying non-perturbative corrections to the dual superpotential,
and later for verifying various dual descriptions.
We define
\bea \phi_i &=& \rho_i e^{i \varphi_i}, \cr
     y_i &=& \varrho_i -i \vartheta_i. \eea
{}From~\C{mirror1},  we find from the first relation that
\bea \label{M1}
\varrho_i &=& \rho_i^2 , \cr
{\bar{\xi}}_{+i} &=& 2 {\bar{\phi}}_i \psi_{+i}, \cr
 \xi_{+i} &=& 2 \phi_i {\bar{\psi}}_{+i}, \cr
\partial_+ \vartheta_i &=& 2 \left[-\rho_i^2 (\partial_+ \varphi_i + Q_i A_+)
+ {\bar{\psi}}_{+i} \psi_{+i} \right]. \eea
{}From the second relation, we see that
\be \label{M2}
\partial_- \vartheta_i = 2\rho_i^2 (\partial_- \varphi_i + Q_i A_-).
\ee
Note the
difference in the expressions for $\partial_+ \vartheta_i$ and
$\partial_- \vartheta_i$. Since vortices play a crucial role in the
construction of the superpotential, we begin by briefly reviewing
vortex instantons.


\subsection{A Review of Vortex Instantons}
\label{reviewinstantons} We briefly review the vortex instanton
solution of the two dimensional Abelian Higgs model. In order to
construct the one instanton solution, we wick rotate to Euclidean
space sending
$$y^0 \rr  -i y^2, \quad F_{01} \rr -iF_{12}.$$
The Euclideanized action for the Abelian Higgs model is
\be \label{eact} S= \int {\rm{d^2}} y \ \left[ \ \sum_i
\vert  \D_{i}\phi\vert^2 + \frac{1}{2e^2} F_{12}^2 + \frac{i\theta}{2\pi}
F_{12} + \frac{D^2}{2 e^2} \right], \ee
where $i=1,2$ and $D$ is given by
\be D= -e^2( Q\vert \phi\vert^2- r). \ee
In polar coordinates $(\rho, \theta)$,
the one-instanton configuration is given by
\be A_\rho =0, \quad  A_{\theta}= A(\rho), \quad
\phi = f(\rho) e^{i \theta}\ee
where for large $\rho$,
\be A(\rho) \sim \frac{1}{\rho} + {\rm{constant}} \times
\frac{e^{- {\sqrt r} \rho}}{\sqrt \rho},\ee
\be f(\rho) \sim {\sqrt r} + {\rm{constant}} \times  e^{-{\sqrt {2r}} \rho},\ee
and $A(0)=f(0)=0$. In writing the expression for $A(\rho)$ and
$f(\rho)$,
we have set $Q=e=1$.
The fields go to zero at the location of the instanton and also fall off
exponentially at spatial infinity. The Bogomol'nyi equations which
determine BPS instanton configurations are
\be \label{bone}(D_1 +i D_2)~\phi=0\ee
and
\be \label{btwo} D + F_{12}=0.\ee
On evaluating the instanton action~\C{eact}\ in this background, we
obtain $S=-2 \pi i t$, where $t=ir + \frac{\theta}{2\pi}$.
In the supersymmetric theories that we consider, there are fermion zero modes
in the instanton background which are crucial in our analysis of
non-perturbative corrections to the dual superpotential. We now turn
to the construction of the dual superpotential.


\subsection{$R$-charge Assignments}

We will restrict to the case where both chiral and Fermi
superfields are dualized, and where $E \neq 0$. 
So we proceed by constructing the dual theory in terms of the
neutral chiral superfields, $Y_i$, and the neutral Fermi
superfields, $F_a$. We recall from our previous analysis that the
relation between the original and dual Fermi superfields is a
local one where \be \label{fermiagain} F_a ={\bar{\Gamma}}_a {\cal
E} (\Phi_i). \ee Clearly, this definition is not unique and can be
subject to field redefinitions by gauge-invariant combinations of
the original superfields. This possibility will play a role when
we construct explicit examples. That the relation between the
original and dual Fermi superfields is a local one will make our
life easier in determining instanton corrections.

Recall that the component expansion for $\Sm$ takes the form \be
\label{Sg} \Sigma = \sigma + {\sqrt 2} \theta^+ {\bar{\lambda}}_+
-i\theta^+ {\bar{\theta}}^+ \partial_+ \sigma . \ee We need only
consider the case of $E_a = \Sigma g_a (\Phi_i)$ since the case
$E_a = f_a(\Phi_i)$ follows by giving $\Sm$ an expectation value,
$$ < \Sm > \neq 0. $$
The Lagrangian of the original \ZT\ theory given in~\C{originalLag}\
admits a classical $U(1)_R$ symmetry under which
\bea & \theta^+
\rightarrow & e^{-i\alpha} \theta^+, \cr
& \Upsilon \rightarrow & e^{-i\alpha} \Upsilon, \non \eea
while $\Phi_i$ and $\Gamma_a$ are left invariant.
In terms of component fields,  the non-trivial transformations are given by
\be \psi_{+i} \rightarrow e^{i\alpha} \psi_{+i}, \qquad \lambda_- \rightarrow
e^{-i\alpha} \lambda_-, \qquad E_a \rightarrow e^{-i\alpha} E_a, \ee
which means that $\sigma \rightarrow e^{-i\alpha} \sigma$. To avoid
confusion, we should note that $E_a$ has mass dimension $1$. The
dimensionful parameter in $E_a$ can either be absorbed in the definition of
$\Sm$, or inserted by hand. Either way, we call
this mass parameter $\sigma$, and it carries all the $R$-charge of $E_a$.
This classical $R$-symmetry is generally
anomalous, and leads to a shift of the theta angle given by
\be \theta \rightarrow \theta -\sum_i Q_i \, \alpha .\ee

How do the dual superfields transform under $U(1)_R$? In cases
where ${\cal E}_a$ is not zero, we see from \C{fermiagain}\ that
the corresponding $F_a$ is uncharged since the mass parameter
$\sigma$ does not appear in the relation. When $E_a=0$, the
relation is even simpler
$$ {\cal F}_a= \bar{\G}_a$$
and again the dual Fermi superfield is uncharged. In this case,
however, the dual Fermi field is charged under the gauge symmetry.

We also require the transformation
properties of $Y_i$ under the classical $R$-symmetry.
In order to find the transformation properties of $Y_i$, we follow the
procedure in~\cite{Hori:2000kt}. The classical $U(1)_R$ symmetry has a
conserved current given by
\be J_+^R = \sum_i {\bar{\psi}}_{i+} \psi_{i+} +\frac{i}{e^2}
\sigma \partial_+ \bar\sigma \ee
and
\be J_-^R = -\frac{1}{e^2} {\bar{\lambda}}_- \lambda_-
-\frac{i}{e^2} \bar\sigma \partial_- \sigma. \ee
Using these currents and the expressions for $\partial_+ \vartheta_i$
and $\partial_- \vartheta_i$ from \C{M1} and \C{M2}, we get that
\be   J_+^R (x) \partial_+ \vartheta_i (y) \sim  \frac{2}{(x^+ -
  y^+)^2}, \qquad
  J_{\pm}^R (x) \partial_- \vartheta_i (y) \sim  0, \qquad
  J_-^R (x) \partial_+ \vartheta_i (y) \sim 0,\ee
where we have dropped the regular terms in the operator product expansion.
This leads to the singularity structure
\be \label{OPE} J_+^R (x) \vartheta_i (y) \sim \frac{2}{(x^+ - y^+)}.\ee
Constructing the classically conserved charge $Q^R$ given by
\be Q^R = \frac{1}{2\pi} \int dx^1 (J_+^R + J_-^R),\ee
we obtain the relation
\be  \label{Yrcharge}
[Q^R, \vartheta_i (y)]= -i \quad \Rightarrow \quad [Q^R, Y_i (y)]= -1.
\ee
In evaluating the integral we have used the OPE \C{OPE} and also wick rotated
to Euclidean space. So we obtain the result
\be e^{i\alpha Q^R} Y_i (\theta^+,{\bar{\theta}}^+ ) e^{-i\alpha Q^R} = Y_i
(e^{-i\alpha} \theta^+,e^{i\alpha} {\bar{\theta}}^+) -i \alpha.\ee
Therefore the perturbative dual superpotential
\be  \widetilde{W} =
-\frac{i \Upsilon}{4}(\sum_i Q_i Y_i + it)
+ \frac{\Sigma}{\sqrt 2} \sum_a F_a\ee
yields the correct $U(1)_R$ anomaly under the shift of the $Y_i$
fields. From this we learn that the possible non-perturbative
corrections to $\widetilde{W}$, which we denote
${\widetilde{W}}_{non-pert}$,
must have $U(1)_R$ charge one.

\subsection{The Structure of Instanton Corrections}

The fermionic nature of the superpotential forces
non-perturbative corrections to be of the form
\be \Upsilon A + \sum_a B^a F_a \ee
where $A$ carries
no $R$-charge and $B^a$ has $R$-charge one.

First let us determine $A$. $A$ cannot be just a parameter since such
a term is ruled out by the perturbative non-renormalization theorem (note that
$\widetilde{W}$ already contains the term $\frac{t \Upsilon}{4}$).
Also, $A$ cannot depend solely on $\Sigma$ which has $R$-charge one.
Suppose $A$ is only a function of $Y_i$. Demanding that
the function be analytic in $Y_i$ allows us to expand
\be A = a_0 + \sum_i a_1^i Y_i + \sum_{ij} a_2^{ij} Y_i Y_j
+ \sum_{ijk} a_3^{ijk} Y_i Y_j Y_k +\ldots.\ee
Then $[Q^R,A]=0$ evaluated using \C{Yrcharge}\ implies that $A$ only
depends on the $Y_i$ in the combination
$$ \sum_i \alpha_i Y_i $$
where
\be \label{baddecay}
\sum_i \alpha_i=0.
\ee
Perturbative contributions to the superpotential are ruled out, so
we must look for single-valued terms of the form
$$
e^{ \left( \sum_i \alpha_i Y_i\right)}.
$$
However, because of condition \C{baddecay}, this kind of term
always grows as we make one or more of the $Y_i$ large. These
non-perturbative contributions are therefore ruled out, and we
conclude that $A$ cannot depend solely on the $Y_i$.

Suppose $A$ depends on both $\Sm$ and $Y_i$. Demanding regular
behaviour in $\Sigma$ allows us to expand $A$ in
the form
\be A = \Sigma f_1 (Y_j) +\Sigma^2 f_2 (Y_j) +\Sigma^3 f_3 (Y_j) +\ldots
= \sum_{k > 0} \Sigma^k f_k (Y_j),\ee
where $f_k (Y_i)$ has $R$-charge $-k$. We construct a solution in a
way similar to the prior case. Insisting that $f_k$ has $R$-charge
$-k$ tells us that
$$ \sum_i {\p f_k \over \p Y_i} =  k f_k. $$
A single-valued solution of this equation contains terms of the form
\be \label{growing} e^{\left( k \sum_i \alpha^k_i Y_i\right)} \ee
where, unlike the prior case,
$$ \sum_i \alpha^k_i =1.$$
Again, as some combination of $Y_i$ become large, terms of the form~\C{growing}\ must
diverge and are therefore ruled out. We conclude that $A=0$.

Next we proceed to constrain $B^a$ which must have $R$-charge $1$.
Clearly $B^a$ cannot depend only on $\Sigma$ since this would be a
perturbative term modifying the already present
$-\frac{\Sigma}{\sqrt 2} \sum_a F_a$ coupling. So we must consider
the possibility that $B^a$ depends on both $\Sm$ and $Y_i$.
Demanding regular behaviour in $\Sigma$ allows us to put $B^a$ in
the form \be B^a = f_0^a(Y_j) + \Sm f_1^a (Y_j)+ \Sigma^2 f_2^a
(Y_j)
 +\ldots
= \sum_{k} \left\{ \Sigma^k f_k^a (Y_j) \right\},\ee
where $f_k^a (Y_i)$ has $R$-charge $1-k$. From our prior discussion,
we know that each $f_k^a$ (for $k \neq 1$) contains terms of the form
\be \label{againgrowing} e^{\left( \{k-1\} \sum_i \alpha^k_i Y_i\right)}, \ee
where,
$$ \sum_i \alpha^k_i =1.$$
The case $k=1$ involves terms of the form
$e^{\left( \sum_i \alpha^1_i Y_i\right)}$, where
$$ \sum_i \alpha^1_i =0. $$
The only case that admits terms that decay as $ \sum_i Y_i \rightarrow \infty$
in all possible ways is $k=0$. Every other case is ruled out.
This leads to a possible non-perturbative superpotential
\be {\widetilde{W}}_{non-pert} =
\sum_{\mu a} \beta_{\mu a} F_a e^{-\sum_i \alpha_{\mu i} Y_i}.\ee
where $\sum_i \alpha_{\mu i} =1$ for each $\mu$.

\subsection{Constraining the Superpotential}

Let us now constrain ${\widetilde{W}}_{non-pert}$ further.
On integrating over the superspace variables, we see that
${\widetilde{W}}_{non-pert}$ leads to a term in the Lagrangian
\be L= \ldots+{\sqrt 2} \sum_{\mu a i} \beta_{\mu a} \alpha_{\mu i}
e^{-\sum_j \alpha_{\mu j} y_j} \eta_{-a} {\bar{\xi}}_{+i}.\ee
If such a term exists in the Lagrangian, then
$$\langle {\bar{\eta}}_{-a} \xi_{+i} \rangle \neq 0$$
for all $i$. It is instructive for us
to calculate this $2$-point function in the original theory. It can
only be non-vanishing in an instanton background.
Let use the duality map of~\C{M1},
$$\xi_{+i} =2 \phi_i {\bar{\psi}}_i,$$
from which we see that the $2$-point
function in the original theory involves a factor of $\phi_i$. If the
instanton is embedded in $\phi_m$, then $\phi_i =0$ for $i \neq m$. Hence,
only
$$\langle {\bar{\eta}}_{-a} \xi_{+m} \rangle$$
can possibly be non-zero
while all the other terms $\langle {\bar{\eta}}_{-a} \xi_{+i} \rangle$
for $i \neq m$ vanish trivially. For any instanton configuration,
only one term of this kind can possibly be
non-zero (this term may still vanish because of additional fermion
zero modes, as we shall see in later examples).

The structure of BPS instanton contributions tells us that
$B^a$ must be of the form,
\be B^a = \sum_i \beta_{i a} e^{-Y_i}, \ee
giving the non-perturbative superpotential
\be {\widetilde{W}}_{non-pert} =
\sum_{i a} \beta_{i a} F_a e^{-Y_i}.\ee
This can also be seen in a different way. Periodicity of $Y_i$ implies
that
$$ \alpha_{\mu i} \in \Z. $$
When combined with the constraint $\sum_i \alpha_{\mu i} =1$ and the
decay condition on ${\widetilde{W}}_{non-pert}$, we are
lead to the same conclusion: namely, that the exact dual superpotential is given by
 \be \label{ET} {\widetilde{W}}_{exact} =
-\frac{i \Upsilon}{4}(\sum_i Q_i Y_i + it)
+ \frac{\Sigma}{\sqrt 2} \sum_a  F_a
+ \mu \sum_{i a} \beta_{i a} F_a e^{-Y_i},\ee
where we have explicitly exhibited the mass scale $\mu$ in the
superpotential. What remains is the determination of the $\beta_{i a}$
parameters of the dual theory. Unlike the case of $(2,2)$
theories, these parameters depend on the particular theory under
consideration.

\section{The Vacuum Structure and Observables}
\label{yukawadiscussion}

We want to begin by studying the vacuum structure of these \ZT\
theories. In the absence of a superpotential, minimizing the bosonic
potential imposes the constraints
\be\label{hyperone}
E_a(\phi_i, \Sm)=0, \qquad \sum_i Q_i |\phi_i|^2 = r.
\ee
where $i=1, \ldots, N$, and each $E_a$ is associated to a left-moving
fermion, $\chi_{-a}$.
With a superpotential, there are additional holomorphic constraints
\be\label{hypertwo}
J^a(\phi_i, \Sm)=0.
\ee
Note that there need not be a $\Sm$ field in the theory. There are
typically multiple phases for these models, with $r>>0$ corresponding
to a geometric phase, while $r<<0$ corresponds to a Landau-Ginzburg
phase. With multiple $U(1)$ factors, hybrid phases are also
possible. There are a myriad of models that we could examine, but in
this effort, we will restrict to a few classes that we find
particularly interesting.

\subsection{Without a $\Sm$ field}
\label{nosigma}

There are really two distinct cases that we will
consider: let us first suppose that there is no $\Sm$ field. Each
$E_a$ depends only
on $\Phi_i$, and is a section of the line-bundle
\be {\cal O} (Q_a). \ee
Similarly, each $J^a$ is a section of ${\cal O} (-Q_a)$.
Minimizing the bosonic potential restricts us to the surface
$E_a=J^a=0$. Usually, we consider non-singular surfaces where
${\partial E_a \over \partial \phi_i} \neq 0$ and
${\partial J^a \over \partial \phi_i} \neq 0$ on the locus
$E_a=J^a=0$. This is not really a necessary condition for the physical
theory but it does simplify our analysis.

Suppose we have a single field $\chi_a$. The chirality condition $E
\cdot J =0 $ tells us that either $E$ or $J$ must be zero. If we have
more than a single left-moving field, there can be non-trivial
solutions to the chirality condition. However, if $(E_a, J^a)$ are
both non-zero for any $a$, the resulting surface is singular since
$$ dE_1 \wedge \cdots dE_{a_{max}} \wedge dJ^1 \wedge \cdots dJ^{a_{max}}
=0. $$
The linear sigma model is likely to be perfectly regular in this
case but again, for simplicity, we will restrict to non-singular
surfaces. For the moment, let us also take each $Q_i \geq 0$ so the
ambient space $\A$, defined by $\sum_i Q_i |\phi_i|^2 = r$,  is compact.
We will consider models with some negatively
charged fields later. Lastly, we note that $a_{max} \leq N$ for a
non-singular surface.

The last element of the low-energy description is the fermions. The
right-handed fermions are fixed by supersymmetry to be sections of the
tangent bundle to the hypersurface \C{hyperone}\ and
\C{hypertwo}\ regardless of whether there is or is not a $\Sm$ field.
It is worth seeing how this emerges directly from the
Yukawa couplings in this case since we will use the same techniques for the
left-moving fermions. The Yukawa couplings
are,
\be
- \left\{ i Q_i {\sqrt 2}
    {\bar{\phi}^i} \lambda_- \psi_{+i} + \bar{\chi}_{-a} {\partial E_a \over \partial
      \phi_i} \psi_{+i} +  \chi_{-a} \psi_{+i}  {\partial J^a \over
  \partial \phi_i} \right\} -  {\rm h.c.}
\ee
We want to determine which of the $\psi_{+i}$ fermions is
massless. Massless fermions satisfy the conditions
\be \label{Yukawacond}
\sum_i Q_i {\bar\phi}^i \psi_{+i}=0, \qquad \sum_i {\partial E_a \over \partial
      \phi_i} \psi_{+i}=0, \qquad \sum_i  {\partial J^a \over
  \partial \phi_i}\psi_{+i}=0,
\ee
{}for each $a$. Following~\cite{Witten:1993yc}, we interpret the first
condition as a gauge-fixing condition on the holomorphic equivalence
\be\label{gaugefix} \psi_{+i} \sim \psi_{+i} + \phi_i \psi. \ee
We encode this condition in a short sequence,
\be
0 \rightarrow {\cal O} {\buildrel \alpha \over \rightarrow}
\oplus_i \, {\cal O}(Q_i) \rightarrow 0,
\ee
where $\alpha$ is the map $\psi \rightarrow  \phi_i \psi.$ This
defines the tangent bundle to the ambient space, ${\cal A}$, defined by $\sum_i Q_i |\phi^i|^2
=r$ in terms of a quotient of line
bundles $ \oplus_i \,\, {\cal O}(Q_i)/ {\rm Im} (\alpha)$.

We can now impose the remaining conditions in turn. For example, for a
particular $E_a$, we consider the sequence
\be
0 \rightarrow T{\cal A} \,\, {\buildrel \alpha_E \over \rightarrow}
\,\, {\cal O}(Q_a) \rightarrow  0,
\ee
where
$$ \alpha_E: \, s_i \mapsto  \sum_i {\partial E_a \over \partial
      \phi_i} s_i $$
and $\{s_i\}$ is a section of $T\A$. This sequence simply defines the
restriction of $T\A$ to the hypersurface $E_a=0$. In a similar way, we
impose all the remaining Yukawa conditions \C{Yukawacond}. What we
learn (as expected from supersymmetry) is that the surviving light
$\psi_{+i}$ transform as sections of
the tangent bundle to the surface $E_a=J^a=0$.

More interesting are the left-moving fermions,
$\chi_{-a}$, with charge $Q_a$. These fermions satisfy the conditions
\be \label{Yukawacondtwo}
\sum_a {\partial \bar{E}_a \over \partial
      \bar{\phi}_i} {\chi}_{-a}=0, \qquad \sum_a  {\partial J^a \over
  \partial \phi_i}\chi_{-a}=0
\ee
{}for each $i$. The condition that the surface be non-singular
guarantees that for a given $a$, either $E_a$ or $J^a$ is non-zero but
not both.
The first condition of \C{Yukawacondtwo}\ is a gauge-fixing condition
for a holomorphic identification akin to
\C{gaugefix}
\be \label{chiyukawa}
\chi_{-a} \sim \chi_{-a} + \sum_i {\partial {E}_a \over \partial {\phi}_i} \chi^i
\ee
where $\chi^i$ has charge $Q_i$.

We must first dispense with
fermions, $\chi_{-a}$, for which both $E_a$ and $J^a$ are zero. These
fermions come along for the ride as we flow into the IR where they
transform as sections of ${\cal O} (Q_a)$ restricted to the
surface. They also contribute to the low-energy anomaly in a
straightforward way since,
$$ {\rm ch} ( \oplus_a {\cal O} (Q_a)) = \sum_a {\rm ch}( {\cal O}(Q_a)). $$

Any fermion for which $E_a$ or $J^a$ is non-trivial must
satisfy \C{chiyukawa}\ for each $i$. However, this imposes $N$
equations on $a_{max} \leq N$ variables so there are no surviving
left-moving fermions.

The low-energy theory is then a non-linear sigma model on the surface $\M$
obtained by setting
\be
E_a (\Phi) = J^a (\Phi)=0, \qquad  \sum_i Q_i |\phi_i|^2 = r.
\ee
The Chern classes of the surface can be computed using the adjunction
formula which tells us that
\be
c(T\M) = { \prod_i \left( 1+ Q_i J\right) \over \prod_{E_a \neq 0}
  \left( 1+ Q_a J\right) \, \prod_{J^a \neq 0} \left( 1 - Q_a J\right)}
\ee
{}from which we see that
\be
c_1(T\M) = \sum_i Q_i - \sum_{E_a \neq 0} Q_a + \sum_{J^a \neq 0} Q_a,
\qquad {\rm ch}_2 ( T\M) = {1\over 2}
c_1^2 -  c_2 = 0.
\ee
This low-energy theory is free of anomalies as we expect with no
left-moving fermions at all. It
is worth pointing out that we can even construct simple conformal models
$(c_1=0)$ of this kind.

\subsection{With a $\Sm$ field}
\label{withasm}
So far, our examples have given theories with no low-energy
left-moving fermions at all. To obtain interesting models with
left-movers, we need to include an uncharged field, $\Sm$. We consider
cases where
$$ E_a = \Sm {\cal E}_a(\Phi_i), \qquad J^a = J^a(\Phi_i).$$
Minimizing the bosonic potential gives two branches. If $ <\Sm > \neq
0$ then we must set $\E_a = J^a=0$, and the corresponding low-energy analysis
is exactly as before except there is an extra uncharged decoupled
chiral multiplet in the IR.

More interesting is the case where $\Sm =0$. This allows us to have
non-trivial $\E$ without the constraint $\E=0$. The analysis for the
right-moving fermions, $\psi_{+i}$, is as before. Again, we conclude
that they are tangent to the surface
$$   J^a(\phi_i)=0, \qquad \sum_i Q_i |\phi_i|^2 = r. $$
The only non-vanishing Yukawa couplings for the left-moving fermions
teach us that
\be \label{Yukawacondsig}
\sum_a \bar{\E}_a  {\chi}_{-a}=0, \qquad \sum_a  {\partial J^a \over
  \partial \phi_i}\chi_{-a}=0.
\ee Suppose there are no $J^a$ in the UV theory. The single
remaining constraint from \C{Yukawacondsig}\ is a gauge-fixing
condition on the equivalence,
$$\chi_{-a} \sim \chi_{-a} + {\E}_a \chi, $$
which tells us that the left-movers are sections of the quotient bundle  $
\oplus_a \,\, {\cal O}(Q_a)/ {\rm Im} (\alpha_\E)$ where
\be \label{leftseq}
0 \rightarrow  \cO \,\, {\buildrel \alpha_\E \over \rightarrow}
\,\, \oplus_a  {\cO}(Q_a) \rightarrow  0,
\ee
where
$$ \alpha_\E: \, \chi \mapsto  \E_a \chi.$$
This construction includes the special class of theories where for
each $\Phi_i$, we include one
$\chi_i$ ($a=i$) with charge $Q_i$ and
$$\E_i = \sqrt{2} Q_i \Phi_i.$$
{}For this particular choice, the left-movers are also sections
of the tangent bundle, and theory has enhanced $(2,2)$
supersymmetry. The target space is the ambient space, $\A$.

Now suppose that some $J^a$ are non-trivial in the UV. We are then confined
to the surface $J^a=0$ in $\A$. The second condition from
\C{Yukawacondsig}\ has no solutions for the partner $\chi_{-a}$
except the pure gauge solution,
$$ \chi_{-a} =  \E_a {\chi}, $$
which one can check is a solution on the surface using $\E \cdot
J=0$. Those $\chi_{-a}$ whose corresponding $J^a$ do vanish in the
UV survive. The bundle that appears in the IR can, however, now be
more interesting than a direct sum of line bundles. The
holomorphic bundle, $\V$, is defined by the cohomology of the
sequence \be 0 \rightarrow {\cal O} \,\, {\buildrel \alpha_\E
\over \rightarrow} \,\,  \oplus_a {\cal O}(Q_a)  \,\, {\buildrel
\beta_J \over \rightarrow} \,\, \oplus_i {\cal O} (-Q_i)
\rightarrow 0, \ee where \be \alpha_\E: \, \chi \mapsto \E_a \chi,
\qquad \quad \beta_J: \, \chi_{-a} \mapsto   \sum_a  {\partial J^a
\over
  \partial \phi_i} \chi_{-a}. \ee
The left-movers are therefore sections of $\V$ given by the $ {\rm
Ker} (\beta_J) / {\rm Im} (\alpha_\E)$. The rank of $\V$ is
$a_{max} - \{ \# (J^a \neq 0) + 1\}$. It is easy to generalize
this construction to cases where some $E_a, J^a$ depend on $\Sm$
while some do not.

\subsection{Vacua for Non-Linear Sigma Models}

In the geometric phase, the low-energy physics is captured by a
non-linear sigma model on the surface $\M$, with the left-moving
fermions taking values in the holomorphic bundle, $\V$, of rank
$r$. For corresponding $(2,2)$ models, the semi-classical ground
states of the sigma model are in one-to-one correspondence with
elements of de Rham cohomology, $H^*(\M, \R)$.

{}For \ZT\ theories, the situation is different. In a sector of
the Hilbert space with $m$ left-moving fermions excited, the
supercharge acts as the Dolbeault operator, $\bar{\p}_E$, twisted
in the holomorphic bundle $E=\wedge^m \V^\ast$. The semi-classical
ground states of the sigma model are therefore in correspondence
with the cohomology groups, \be H^{*} (\M,\wedge^m \V^\ast), \quad
m=0,\ldots, r-1\ee with dimension $h^{*} (\M,\wedge^m \V^\ast)$. Some
of these ground states might pair up and become massive but the
Witten index, \be {\rm Tr} (-1)^F = \sum_{p,m} (-1)^{p+m} \, h^{p}
(\M,\wedge^m \V^\ast), \ee should remain invariant. Lastly, we should
mention the existence of BPS solitons interpolating between these
vacua with mass gap. These are quite fascinating excitations that
merit further exploration, perhaps with the aim of generalizing
the structure of helices of coherent sheaves~\cite{Zaslow:1996nk},
and the attempt to classify massive N=2
theories~\cite{Cecotti:1993rm}.

\subsection{Moduli for Conformal Models}

In the case of conformal models where $\sum_i Q_i =0$, there are
particularly interesting operators that control the moduli of the
non-linear sigma model. The simplest to describe are the moduli
for the K\"ahler metric. Deformations of the K\"ahler and complex
structure correspond, respectively, to elements of
$$ H^{1}(\M, T^*\M), \quad H^1 (\M, T\M). $$
{}For models with a space-time interpretation, each cohomology element
gives rise to a space-time scalar field. Ignoring effects that are
non-perturbative in the string coupling, the potential for these
scalar fields has flat directions.

The last class of moduli parametrize continuous deformations
of the holomorphic bundle, $\V$, and correspond to elements of
$$ H^1(\M, {\rm End} \V). $$
Each of these deformations also gives rise to a space-time scalar.
Even in non-conformal models, these deformations are interesting
because they are relevant deformations.
For example, starting with the tangent bundle, $\V = T\M$, where
the theory is \TT, we can find families of \ZT\ theories by
deforming the bundle.

\subsection{Instanton Corrections}

The most natural set of observables to study both in massive and
conformal models are chiral operators. Both the vacua (via the
state-operator correspondence) and the moduli described above
correspond to particular chiral operators. A chiral operator,
$\cO$, satisfies
$$ \{ \bar{Q}_+, \cO \} =0.$$
Consider a correlator of chiral operators, \be < \cO_1(y_1) \cdots
\cO_n(y_n) >. \label{chiral}\ee Chirality ensures that the
correlator is independent of the insertion points, $y_i$, on the
world-sheet $\Sigma$.\footnote{We apologize for the multiple uses
of $\Sm$, but this notation for the world-sheet is conventional.}
The correlator must also depend on the parameters of the theory in
a holomorphic way, and so is protected from perturbative
corrections.

While there are no perturbative contributions to the correlation
function, there can be non-perturbative contributions arising from
instantons. In the linear sigma model, an instanton corresponds to
a BPS solution of the abelian Higgs model reviewed in
section~\ref{reviewinstantons}. In the IR non-linear sigma model,
these BPS instantons correspond to holomorphic maps
$$ \phi: \quad \Sm \rightarrow \M.$$ Each map is characterized by
winding number $n$, which is given by
$$ n = {1\over 2 \pi} \int_{\Sm} \phi^*(\omega) $$
where $\omega$ is the K\"ahler form of the target space $\M$.
Both in the linear and
non-linear sigma model, an $n$ instanton contribution to a
correlator function is suppresed by the instanton action (taking $n>0$),
$$S_{inst} \sim e^{2\pi int}, \qquad t = ir + {\theta\over 2\pi}. $$
However, the linear sigma model contains point-like instanton
contributions in addition to the usual smooth
instantons~\cite{Witten:1993yc}. The effect of these point-like
instantons is to renormalize $t$ as we flow from the UV to the IR.
The relation between $t$ in the linear and non-linear sigma models
has been computed for \TT\ theories in~\cite{Morrison:19 95fr},
where in some cases, the parameters were found to agree.

We can use symmetries to further constrain the correlation
functions. The main symmetry that we will consider is  the
right-moving $U(1)_R$ symmetry under which the right-moving
$\psi_+$ fermions have charge one. To obtain a selection rule, we
need to determine the number of right-moving fermion zero-modes in
a sector with instanton number $n$. On a genus $g$ world-sheet
$\Sm$, the count of fermion zero-modes follows from an index
theorem. In instanton sector $n$, there are
$$ {\rm dim}(\M)*(1-g)  + n\, c_1(\M)$$ right-moving zero
modes. We will primarily consider the plane (or equivalently a
genus $0$ world-sheet). For the perturbative sector where $n=0$ where 
we consider constant maps (the only holomorphic maps) from $\Sm \to
\M$, we learn that the
correlator~\C{chiral}\ is non-vanishing only when the product of
chiral operators, each associated to an element of twisted
Dolbeault cohomology, has anti-holomorphic degree ${\rm dim}
(\M)$, i.e., only when it is a top form. The semi-classical value
of the correlator~\C{chiral} then defines a map \be H^{*}(\M, E_1)
\times \ldots \times H^{*}(\M, E_m) \,\to\, \mC \ee where each
$E_i$ is a bundle of the form $\wedge^{*} \V^\ast$, and the total
anti-holomorphic form degree is ${\rm dim} (\M)$ or the correlator
vanishes. This is a kind of intersection form on
$\M$~\cite{Distler:1988ee}.

Let us consider the left-movers. To constrain the left-moving
fermions, we want to restrict to \ZT\ non-linear sigma models which are
the IR limits of GLSMs. In the UV GLSM, there is a classical $U(1)$
charge $Q_L$ where
\be
Q_L \sim  \int dx^1 \sum_a \bar\chi_{-a} \chi_{-a}. 
\ee 
In general, this is not a conserved charge like the $U(1)_R$
charge. However, the charge violation is proportional to the instanton
number. As we flow to the IR, some of the $\chi_-$ fermions become
massive. There is an index theorem that counts the net number of
$\chi_-$ zero modes, 
 $$ {\rm dim}(\V)*(1-g)  + n\, c_1(\V). $$
Absorbing these zero modes for $n=0, g=0$ gives a selection rule: the
correlator~\C{chiral}\ must contain ${\rm dim}(\V)$ left-moving
fermions. Note that ${\rm dim}(\V) =  {\rm rk}(\V)$ for these
holomorphic bundles so this constraint is again a statement that the
correlator be a top form. 

In non-conformal models, a combination of the $U(1)_L$ and $U(1)_R$
charges is conserved exactly in the UV. Both charges are individually  
violated by instantons. This permits a quantum deformation of the classical
geometric rings which satisfy the $n=0$ selection rules. In the \TT\
case, the instanton corrected ring is known as the 
quantum cohomology ring~\cite{Witten:1988xj, Witten:1990ig}. 
In the following section, we will find analogous
structures for \ZT\ theories.

The last issue we need to address is the coefficient of the
instanton corrections to a chiral correlator in a low-energy conformal
non-linear sigma model. Since the model is conformal, 
$U(1)_R$ is conserved. In a conformal model, $c_1(\M)=0$ so there are
no additional right-moving zero modes for $n>0$. This, combined with
the conservation of $U(1)_R$, implies that the only way that the
chiral ring
is modified quantum mechanically is via instanton corrections to the
classical ring coefficients. 

In the \TT\ case, this coefficient
`counted' the number of holomorphic curves in some suitable sense.
In the \ZT\ case, the basic picture is similar. Consider the
moduli space of instantons with charge $n$, which we denote
$\M_n$. There are subtle issues surrounding the compactification
of this space. We will take the physical compactification provided
by the linear sigma model. The zero-modes for the left-moving
$\chi_-$ fermions (which transform as a section of $\V$) in the
sector with instanton charge $n$ define a holomorphic bundle
$\V_n$ on $\M_n$. The effective theory of the instanton moduli is a sigma
model with target $\M_n$ and with a supercharge acting as the
$\bar{\p}$ operator twisted in the bundle $\oplus_m \wedge^m \V_n^\ast$. The
leading contribution of the path-integral over the moduli gives
instanton contributions \be <\cdots> \quad =\quad \sum_{n>0} \,
 \left( \sum_m (-1)^m {\rm Ind} (\bar{\p}_{\wedge^m \V_n^\ast}) \, 
\right) e^{- 2\pi i n t}.\ee 
More precisely, the path-integral computation gives the integral over
the index density over $\M_n$ which need not necessarily agree
with the index. When non-vanishing, these instanton contributions
modify the ring coefficients. In the \TT\ case, the coefficient of the
instanton correction reduces to $\chi(\M_n)$. In the \ZT\ case, we
find a natural generalization that depends on the choice of
holomorphic bundle, $\V$.

\section{Examples of Dual Pairs}

We now turn to the construction of specific \ZT\ dual pairs. There
are three broad classes of models. These classes are characterized
by whether the rank of the left-moving bundle, $\V$, is less than,
equal to, or greater than the rank of the tangent bundle $T\M$. As
we will see, the dual theory in the first case is quite different
from the latter two cases. Unlike the latter two cases, the dual
theory for ${\rm rk}(\V) < {\rm rk}(T\M)$ is typically a
non-linear sigma model so the duality relates two geometric
theories. In the remaining cases, the dual theory is typically a
\ZT\ Landau-Ginzburg theory with no flat directions in the
superpotential.

{}For brevity, in our subsequent discussion, we will not
explicitly write the gauge kinetic terms, the FI-terms, and the
$\theta$ terms in either the original or the dual theories. We
will always assume they are present. The first examples that we will
consider fall
in the category ${\rm rk}(\V) = {\rm rk}(T\M)$.

\subsection{One Chiral \& One Fermi Field}

We start with the simplest possible model containing one chiral
superfield, $\Phi$, and one Fermi superfield, $\Gamma$, both with
charge $Q$. The Lagrangian of the theory is given by \be L =
-\frac{i}{2} \int d^2 \theta ~{\bar{\Phi}} ({\cal{D}}_0 -
{\cal{D}}_1 ) \Phi
 -\frac{1}{2} \int d^2 \theta ~\bar{\Gamma} \Gamma ,
\ee In the definition of $\Gamma$, we have some freedom in our
choice of $E$. We consider two choices for $E$ below, and
construct the dual theories. In the first case, we find no
non-perturbative corrections to the dual superpotential, while in
the second case there is a correction.

\subsubsection{$E = i \alpha \Phi$}
{}For the first case, take $E = i \alpha \Phi$ so that $E$ is
itself a chiral superfield of charge $Q$ which satisfies
${\bar{\cal{D}}}_+ E=0$ for some parameter, $\alpha$. This theory
is free of anomalies.  Note that for this choice of $E$, this
theory is a \ZT\ theory that never has enhanced \TT\ supersymmetry
for any choice of $\alpha$. This is the case because there is no
$\Sigma$ superfield, and so no right-moving gauginos. Hence, the
left-moving fermions in the Fermi multiplet do not couple to the
gauginos at all. We could also equivalently start with a $\Sm$
field and the choice $E= \Sm \Phi$, and set
$$ < \Sm> = i \alpha$$
while setting the right-moving gauginos in $\Sm$ to zero.

Using the component field expansions for $\Phi$ and $\Gamma$, we
get that \bea L = (\partial_+ \rho)(\partial_- \rho) + \rho^2
(\partial_+ \varphi + Q A_+) (\partial_- \varphi + Q A_-) + i
{\bar{\psi}}_+ D_- \psi_+
- {\sqrt 2} iQ \bar{\phi} \lambda_- \psi_+ \\
+ {\sqrt 2} iQ \phi {\bar{\psi}}_+ {\bar{\lambda}}_- + QD \rho^2 +
i{\bar{\chi}}_- D_+ \chi_- - \vert \alpha \phi \vert^2 -i \alpha
{\bar{\chi}}_- \psi_+ + i \bar{\alpha} {\bar{\psi}}_+ \chi_-, \non
\eea where we have set $G=0$ by its classical equation of motion.
In the dual theory, we have a single neutral chiral superfield
$Y$, and a neutral Fermi superfield $F$. The relation between the
original and the dual Fermi fields follows from the component
expansion of the duality map~\C{fermiduality} \be ~{\bar{\eta}}_-
= -\bar{\phi} \chi_-, ~\eta_- = -\phi {\bar{\chi}}_- . \ee These
relations will be useful in determining the non-perturbative
corrections to the dual superpotential. The perturbative  dual
theory is given by the Lagrangian \bea \label{dual} \widetilde{L}
= \frac{1}{8} \int d^2 \theta ~ \left[\frac{i (Y - \bar{Y})}{Y +
\bar{Y}}
\partial_- (Y + \bar{Y})
 -8 \frac{\bar{F} F}{Y + \bar{Y}}
\right] \\ -\left[\frac{iQ}{4} \int d \theta^+ Y \Upsilon -
\frac{i \alpha }{\sqrt 2} \int d \theta^+ F + {\rm h.c.} \right].
\non \eea This dual description can be checked using the various
duality maps together with the identity (true up to total
derivatives), \be \frac{(\partial_+ \vartheta)(\partial_-
\vartheta)}{2 \varrho} - Q \vartheta F_{01} = \frac{(\partial_-
\vartheta)}{8 \varrho^2} \xi_+ {\bar{\xi}}_+ . \ee We also have to
integrate out the auxiliary field $H$ in the superfield $F$ using
its classical equation of motion to explicitly check the duality.

So in the dual theory, we find the perturbative superpotential \be
 \widetilde{W} = -\frac{i \Upsilon}{4}(QY + it)
+ \frac{i \alpha}{\sqrt 2} F. \ee Now we must consider the
possibility of non-perturbative corrections to the dual
superpotential: namely, is there an $F e^{-Y}$ addition to the
superpotential? We will argue that such a term does not arise. The
non-perturbative correction to the dual superpotential is
generated by instantons in the original theory. Because of the
$\vert \alpha \phi \vert^2$ term in the original action, there is
no BPS instanton because $\phi$ must be set to zero. For any
non-zero $\alpha$, there is no non-perturbative correction. The
perturbative dual superpotential is exact \be 
{\widetilde{W}}_{exact} = -\frac{i \Upsilon}{4}(QY + it) + \frac{i
\alpha}{\sqrt 2} F. \ee On integrating out $\Upsilon$, we find an
effective potential \be {\widetilde{W}}_{eff} = \frac{i
\alpha}{\sqrt 2} F, \ee with the constraint $QY =-it$. Note that
supersymmetry is spontaneously broken in both the original and
dual theories.

\subsubsection{A Vanishing Result for More General Cases}

We can extend the prior result to a more general setting.
Non-perturbative terms in the dual superpotential of the form $
\beta_{ia} F_a e^{-Y_i}$ lead to terms in the Lagrangian given by
\be L= \ldots +{\sqrt 2} \sum_{a i} \beta_{i a} e^{-y_i} \eta_{-a}
{\bar{\xi}}_{+i}.\ee The existence of these terms implies that the
correlator $\langle {\bar{\eta}}_{-a} \xi_{+i} \rangle$ must be
non-vanishing. Consider the case $E_a = f_a (\Phi_i)$ which is a
generalization of the case just considered. For this choice of
$E_a$, we see that the Lagrangian of the original theory contains
the term \be L =\ldots -\sum_i \vert f_a (\phi_i)\vert^2. \ee So
the condition for a BPS instanton solution is $f_a (\phi_i) =0$
for all $a$. From the duality map ${\bar{F}}_a = \Gamma_a
{\bar{\mathcal{E}}}_a$, we see that ${\bar{\eta}}_{-a} =
-\chi_{-a} {\bar{f}}_a ({\bar\phi}_i)$, which is zero for all $a$
using the BPS condition. Hence the two point function always
vanishes, and so do the non-perturbative corrections to the dual
superpotential. There is an apparent caveat to this argument; namely, the
kinetic terms for the $F_a$ superfields diverge like
$1/|f_a|^2$ since for an
instanton configuration $f_a=0$. However, in the dual theory, in terms
of $Y$ variables, $1/|f_a|^2$ is not holomorphic and so this
divergence should not affect the determination of the
superpotential.

\subsubsection{$E = c \Sigma \Phi$ }
\label{largesm}
Next we consider a case where, as we shall show, there is a
non-perturbative correction to the dual superpotential. We
consider the case where $E = c \Sigma \Phi$, where $c$
is a non-zero parameter. The key difference is the appearance of
$\Sm$ in $E$. This case can easily be generalized to a theory with $N$
chiral and Fermi superfields with charge $Q_i$ 
where $$E_i = c_i \Sigma \Phi_i.$$ These models are
deformations of theories with \TT\ supersymmetry which is restored at 
the point $c_i=\sqrt{2} Q_i$. However, this particular deformation is not a
relevant deformation although it does break supersymmetry. We can see this
from the low-energy perspective by considering the target space, $
W\mathbb{P}^N$. The left-moving bundle $\V$ is a deformation of the
tangent bundle specified by the sequence \C{leftseq}; however, the
bundles obtained from this deformation are all equivalent. We will see
this reflected in the low-energy physics of the dual description. Note,
however, that the bundle can degenerate by taking some $c_i
\rightarrow 0$.

\detail{Determining the $\beta_{ia}$ Coefficients}

While it is difficult to determine the $\beta_{ia}$ coefficients in
the superpotential for most models, in this case, we can explicitly
determine these parameters. The dual superpotential takes the form,
\be \label{Exacta} {\widetilde{W}}_{exact}
= -\frac{i \Upsilon}{4}(\sum_i Q_i Y_i + it)
+ \frac{\Sigma}{\sqrt 2} \sum_i c_i F_i
+ \mu \sum_{i j} \beta_{i j} F_i e^{-Y_j}.\ee
We have replaced $\beta_{ia}$ by $\beta_{ij}$ since we have an equal
number of chiral and Fermi superfields. We have also rescaled $F_i$ 
and $\beta_{i j}$ by a factor of $c_i$ in \C{NP3}\
and \C{ET}\ to get this form.

We shall see that we can determine $\beta_{i j}$ exactly. In the original
theory, we take $\sigma$, the lowest component field of
$\Sigma$, to be very large and slowly varying, and we give it a
specific expectation value.
Then from the terms in the Lagrangian given by
\be L=\ldots-\vert \sigma \vert^2 \sum_i \vert c_i \phi_i \vert^2
-\sigma \sum_i c_i {\bar{\chi}}_{-i} \psi_{+i}
-\bar\sigma \sum_i {\bar{c}}_i {\bar{\psi}}_{+i} \chi_{-i},\ee
we see that $\Phi_i$ and $\Gamma_i$ both get a large mass
of order $c_i \sigma$. We can therefore consider integrating out the massive
superfields, $\Phi_i$ and $\Gamma_i$, for a fixed value of $\sigma$,
together with the high frequency modes of $\Sigma$ (in the sense of
Wilsonian R.G.). This will
give us an effective superpotential, ${\widetilde{W}}_{eff} (\Upsilon, \Sigma)$,
{}for the remaining low energy degrees of freedom. We can also
integrate out the neutral superfields
$Y_i$ and $F_i$ in the dual theory to get another expression for
${\widetilde{W}}_{eff} (\Upsilon, \Sigma)$. Equating the two expressions
gives a constraint on the $\beta_{ij}$ coefficients.

First we focus on integrating out the massive superfields in the original
theory. The superpotential ${\widetilde{W}}_{eff} (\Upsilon, \Sigma)$, on
demanding analyticity in $\Upsilon$, is of the form
\be {\widetilde{W}}_{eff} (\Upsilon, \Sigma) = W_{eff}^0 (\Sigma)
+ \Upsilon  W_{eff} (\Sigma).\ee
The Grassmann odd nature of the superpotential forces $W_{eff}^0 (\Sigma) =0$,
leading to
\be {\widetilde{W}}_{eff} (\Upsilon, \Sigma)=\Upsilon  W_{eff} (\Sigma). \ee
This gives terms in the Lagrangian
\be \label{EffL}\frac{1}{4}
\int d \theta^+ ~{\widetilde{W}}_{eff} (\Upsilon, \Sigma)
+ {\rm h.c.} = -D ~{\rm Im} \{ W_{eff} (\sigma)\} +F_{01} ~{\rm Re} \{
W_{eff} (\sigma)\} +\ldots
\ee
where Im and Re are the imaginary and real parts of the complex
quantity. Therefore, in order to determine ${\widetilde{W}}_{eff}$, it is
enough to consider only the terms in the effective action that are linear
in $D$ and $F_{01}$. We need to evaluate
\be e^{i S_{eff} (\Upsilon, \Sigma)} =\int {\mathcal{D}} \Phi_i
{\mathcal{D}} {\bar{\Phi}}_i {\mathcal{D}} \Gamma_i {\mathcal{D}}
{\bar{\Gamma}}_i ~e^{i S(\Upsilon, \Sigma, \Phi_i,{\bar{\Phi}}_i,
\Gamma_i, {\bar{\Gamma}}_i)}.
\ee
Because each $E_i$ is linear in $\Phi_i$, we can exactly evaluate
the path integral and hence compute $S_{eff}$. In the limit of large $\sigma$,
the wick rotated Lagrangian in Euclidean space reduces to
\bea L^E &= \sum_i & \Big[ \, \vert D_{\alpha} \phi_i \vert^2 +i {\bar{\psi}}_{+i}
D_-^E \psi_{+i} -i {\bar{\chi}}_{-i} D_+^E \chi_{-i} -Q_i D \vert
\phi_i \vert^2 + \vert c_i \sigma \phi_i \vert^2
 \cr & & +\sigma c_i {\bar{\chi}}_{-i} \psi_{+i}
+\bar\sigma {\bar{c}}_i {\bar{\psi}}_{+i} \chi_{-i}\Big], \eea
where $D_{\pm}^E = D_1 \pm i D_2$.
Let us now extract the dependence of $L^E$ on the phase of $\sigma$
and the
$c_i$. We define $\sigma =\vert \sigma \vert e^{i\omega}$ and
$c_i = \vert c_i \vert e^{i\tau_i}$. Classically, these phases can be
absorbed by a phase rotation of the fermions given by
\be \psi_{+i} \rightarrow e^{-\frac{i}{2} (\omega + \tau_i)}\psi_{+i},
\qquad~\chi_{-i} \rightarrow
e^{\frac{i}{2} (\omega + \tau_i)}\chi_{-i}. \ee
However, this chiral rotation of the fermions is anomalous and shifts the
effective Lagrangian by
\be -i \sum_i Q_i (\omega + \tau_i) F_{12}.\ee
Hence,
\be L_{eff}^E (\sigma, c_i) =  L_{eff}^E (\vert \sigma \vert,
\vert c_i \vert) -i \sum_i Q_i (\omega + \tau_i) F_{12}.\ee
We calculate $L_{eff}^E (\vert \sigma \vert, \vert c_i \vert)$ finding
\be e^{-\int d^2 x L_{eff}^E (\vert \sigma \vert, \vert c_i \vert)}
= \prod_i \frac{{\rm det} \left(\begin{array}{cc} -\vert \sigma c_i \vert &
i D_+^E \\ iD_-^E & \vert \sigma c_i \vert \end{array} \right)}
{{\rm det}(-D_{\mu}^2 -Q_i D + \vert \sigma c_i \vert^2)}.\ee
The square of the Dirac operator in the numerator is
\be \left(\begin{array}{cc} -\vert \sigma c_i \vert &
i D_+^E \\ iD_-^E & \vert \sigma c_i \vert \end{array} \right)^2
= \left(\begin{array}{cc} -D_{\mu}^2 +Q_i F_{12} +\vert \sigma c_i \vert^2 &
0\\ 0& -D_{\mu}^2 -Q_i F_{12} +\vert \sigma c_i \vert^2 \end{array} \right)\ee
which gives an effective action
\bea  & \int d^2 x~L_{eff}^E (\vert \sigma \vert, \vert c_i \vert)
=  \sum_i \Big\{ {\rm log ~det} (-D_{\mu}^2 -Q_i D + \vert \sigma c_i \vert^2) \\
&  -\frac{1}{2} \,{\rm log ~det}
(-D_{\mu}^2 +Q_i F_{12} +\vert \sigma c_i \vert^2)
-\frac{1}{2} \, {\rm log ~det}
(-D_{\mu}^2 -Q_i F_{12} +\vert \sigma c_i \vert^2) \Big\}. \non 
\eea
It is easy to see that this gives no linear term in $F_{12}$. However it has a
term linear in $D$ given by
\be  \int d^2 x~L_{eff}^E (\vert \sigma \vert, \vert c_i \vert)
= -D \sum_i Q_i {\rm tr} (\frac{1}{-\partial_{\mu}^2 +
\vert \sigma c_i \vert^2 })+\ldots.
\ee
So we obtain an effective action
\be L_{eff}^E (\vert \sigma \vert, \vert c_i \vert)
= -\frac{D}{2}\sum_i Q_i {\rm ln}
(\frac{\Lambda_{UV}^2 +\vert \sigma c_i \vert^2}
{\vert \sigma c_i \vert^2 }) +\ldots\ee
which in the continuum limit
$\Lambda_{UV} \rightarrow \infty$ reduces to
\be L_{eff}^E (\vert \sigma \vert, \vert c_i \vert)
= -D\sum_i Q_i {\rm ln} (\frac{\Lambda_{UV}}{\vert \sigma c_i \vert })+\ldots. \ee
Putting together these results, we find that
\be L_{eff}^E (\sigma, c_i)
=  -D\sum_i Q_i {\rm ln} (\frac{\Lambda_{UV}}{\vert \sigma c_i \vert })
-iF_{12} \sum_i Q_i (\omega +\tau_i)+\ldots. \ee
Using \C{EffL}, we read off the effective superpotential
\be \widetilde{W}_{eff} (\Upsilon, \sigma, c_i)
= -\frac{i \Upsilon}{4}(\sum_i Q_i {\rm ln} (\frac{\Lambda_{UV}}{c_i \sigma})
+i t_0). \ee
Now we use the one-loop renormalization of $t$ given by
\be t(\mu) = i \sum_i Q_i {\rm ln} (\frac{\mu}{\Lambda}),\ee
where $\Lambda$ is the RG invariant dynamical scale of the theory given by
$\Lambda = \mu e^{it(\mu)/\sum_i Q_i}$, to obtain
\be \label{RGinv} \widetilde{W}_{eff} (\Upsilon, \sigma, c_i)
=-\frac{i \Upsilon}{4} \sum_i Q_i {\rm ln} \left(\frac{\Lambda}{c_i \sigma}\right).\ee

We will now argue that this is an exact result which receives no corrections from
integrating out the high frequency modes of $\Sigma$. Previously, we
described a classical $R$-symmetry under which $\sigma$
transforms as 
$$\sigma \rightarrow e^{-i\alpha}\sigma.$$ 
The RG invariant scale $\Lambda \rightarrow e^{-i\alpha} \Lambda$
under this classical symmetry,
so that $\widetilde{W}_{eff}$ remains invariant. Now for
$\frac{\sigma}{\Lambda} \rightarrow \infty$, $\widetilde{W}_{eff}$ must reduce
to \C{RGinv}. This constrains the form of the effective superpotential
\be \widetilde{W}_{eff} (\Upsilon, \sigma, c_i)
=-\frac{i \Upsilon}{4} \sum_i Q_i {\rm ln} (\frac{\Lambda}{c_i \sigma})
+ \sum_{n >0} a_n (\frac{\Lambda}{\sigma})^n .\ee
However, these corrections are non-perturbative in nature because of the
positive powers of $\Lambda$. We have obtained the result simply by
perturbatively integrating
out the high frequency modes so there should not be any non-perturbative
corrections to $\widetilde{W}_{eff}$. Hence all the $a_n$ vanish.
{}For $\Sigma$ large and slowly varying, we obtain the low-energy
effective superpotential of the original theory
\be \label{Oreff} \widetilde{W}_{eff} (\Upsilon, \Sigma)
= \frac{i \Upsilon}{4} \left\{\sum_i Q_i {\rm ln} \left(\frac{c_i \Sigma}{\mu}\right)
-i t(\mu) \right\}. \ee

Now consider the dual theory with the exact superpotential taking the form
\C{Exacta}. On taking $\sigma$ to be large and slowly varying, we see
that the neutral superfields, $Y_i$ and $F_i$, get masses
of order $\frac{c_i \sigma}{{\sqrt r}}$. We can therefore integrate out the
$Y_i$ and $F_i$ to get a low-energy effective superpotential
$\widetilde{W}_{eff} (\Upsilon, \Sigma)$. Integrating out $F_i$
teaches us that
\be \label{Eqn} \frac{\Sigma c_i}{\sqrt 2}
=-\mu \sum_{j} \beta_{i j} e^{-Y_j}.\ee
On substituting the value of $Y_i$ obtained from \C{Eqn}\ in the dual
superpotential,
we get $\widetilde{W}_{eff} (\Upsilon, \Sigma)$. However, in general,
we cannot solve \C{Eqn}\ exactly for $Y_i$.
Consider the case where the matrix $\mathcal{B}$ (with
entries $\beta_{i j}$) is
invertible ($\mathcal{B}^{-1}$ has entries $\beta^{i j}$). This is
actually  the case of interest in our example, but to show this
requires an instanton analysis that we will temporarily postpone. 
Using the invertibility of $\mathcal{B}$, we find that
\be \label{inv} Y_i =
-{\rm ln} \left\{ \frac{-\Sigma}{{\sqrt 2} \mu} \sum_{j} c_j \beta^{i j}\right\}.
\ee
This leads to the effective superpotential
\be \widetilde{W}_{eff} (\Upsilon, \Sigma)
= \frac{i \Upsilon}{4}\left\{\sum_i Q_i \, {\rm ln}
\left(\frac{-\Sigma}{\sqrt 2 \mu} \sum_{j} c_j \beta^{i j}\right) -i t(\mu)\right\}. \ee
On equating this with \C{Oreff}, we find a general constraint on the
$\beta_{i j}$ given by
\be  \label{Con}
\prod_i \left(\frac{-{\sqrt 2} c_i}{\sum_{j} c_j \beta^{i j}}\right)^{Q_i} =1.\ee

Let us now show that the matrix $\mathcal{B}$ is actually
diagonal. Consider a term $\beta_{ij} F_i e^{-Y_j}$ in the dual superpotential.
If this term is non-zero, then the two point function
$$\langle {\bar{\eta}}_{-i} \xi_{+j} \rangle$$ 
must be non-zero. Using the
duality maps, we obtain the relations,  ${\bar{\eta}}_{-i} =
-{\bar\phi}_i \chi_{-i}$ and $\xi_{+j} = 2 \phi_j  {\bar{\psi}}_{+i}$. We
evaluate this two point function in the instanton background in the original
theory. If $i \neq j$ then the correlator vanishes trivially since
either $\phi_i$  or $\phi_j$ is zero. 
We therefore define $\beta_{ij} = \delta_{ij} \frac{\beta_i}{\sqrt 2}$. From
our assumption that the matrix $\mathcal{B}$ is invertible, we see that all
the $\beta_i$ are non-vanishing. The constraint \C{Con}\ becomes
\be \label{Con2} \prod_i (-\beta_i)^{Q_i}=1.\ee
Actually, it is possible to obtain the $\beta_i$ from this
constraint. To do this, we use the result obtained
in~\cite{Hori:2000kt}\ for \TT\ theories.
For \TT\ theories, the dual theory has a non-perturbative
superpotential
\be {{\widetilde{W}}_{non-pert}}^{(2,2)} = \mu \sum_i e^{-\widetilde{Y}_i},
\ee
where $\widetilde{Y}_i$ is a neutral twisted chiral superfield satisfying
$\bar{D}_+ \widetilde{Y}_i=D_- \widetilde{Y}_i=0$. We simply reduce this
to \ZT\ form 
\be {\widetilde{W}}_{non-pert} = -\frac{\mu}{\sqrt 2}\sum_i F_i e^{-Y_i},
\ee
where $Y_i = \widetilde{Y}_i \vert_{\theta^- = {\bar\theta}^- =0}$ and
$-{\sqrt 2} F_i =
\bar{D}_- \widetilde{Y}_i \vert_{\theta^- = {\bar\theta}^- =0}$. We have scaled
$\mu$ suitably for notational convenience.  

Consider a specific $i$, say $i=m$, and take $E_m = c_m 
\Sigma \Phi_m$  where $c_m$
is arbitrary and non-zero. For all $i \neq m$, we take $E_i =
{\sqrt 2} Q_i \Sigma \Phi_i$, i.e., $c_i = \sqrt{2} Q_i$. Hence for all
$i \neq m$ we have $\beta_i =-1$. So the constraint \C{Con2}\ gives us that
\be (-\beta_m)^{Q_m}=1,\ee
leading to
\be \beta_m = - e^{\frac{2\pi i k}{Q_m}}\ee
where $k=0,1,\ldots, Q_m -1$. Note that $\beta_m$ is independent of
$c_m$, so we can determine it by considering any non-zero value of
$c_m$. For $c_m = {\sqrt 2} Q_m$, we know from the \TT\ 
result that $\beta_m =-1$ so this must be true for all values of
$c_m$; hence $k=0$. We can repeat this
analysis for each $\beta_i$ leading to the result
\be \label{Fix} \beta_i = -1. \ee
We therefore have the exact dual superpotential given by
\be \label{NPc} {\widetilde{W}}_{exact}
= -\frac{i \Upsilon}{4}(\sum_i Q_i Y_i + it)
+ \frac{\Sigma}{\sqrt 2} \sum_i c_i F_i
- \frac{\mu}{\sqrt 2} \sum_{i} F_i e^{-Y_i}.\ee

A few comments about the superpotential are in order. First, as mentioned
above, the non-perturbative corrections to the superpotential are independent
of $c_i$ for any non-zero $c_i$. We might ask what happens as we take
a particular
$c_m \rightarrow 0$. In the original theory, the bundle
degenerates. In the dual theory, this limit is singular because our
procedure for arriving at the effective superpotential involved integrating
out massive fields with masses of $O(c_m)$ in the original theory. In
the dual theory, we integrated out $Y_m$ and $F_m$ with masses 
of $O(\frac{c_m\s}{\sqrt r} )$. These fields become massless as $c_m
\rightarrow 0$ so the integration procedure leads to singularities in
the effective superpotential.

The effective superpotential \C{Oreff}\  gives us information
about the vacuum structure of the theory for large $\Sigma$. For large
$\Sigma$, the charged heavy fields $\Phi_i$ and
$\Gamma_i$ are frozen at zero vacuum expectation value. As is
standard, the potential energy of the theory is then given by
\be U = \frac{e^2 r^2}{2} + \frac{e^2}{2}(\frac{\tilde\theta}{2\pi})^2 =
 \frac{e^2}{2} \vert \tilde{t} \vert^2,    \ee
where $(\frac{\tilde\theta}{2\pi})^2$ is the minimum value of
$(\frac{\theta}{2\pi} -n)^2$ for $n \in \Z$~\cite{Coleman:1976uz}. In the
expression for $U$, the first term comes from the $D$-term while the second
term comes from the energy density generated by 
the $\theta$-term. Here, $\tilde{t}$ (defined with appropriate shifts in
$\frac{\theta}{2\pi}$ by integer amounts) is basically due to the FI-term in
the Lagrangian $${t\over 4}\int d\tp~ \U
\vert_{{\bar{\theta}}^+ =0} + {\rm h.c}.$$ 
In the calculation above for the
effective superpotential, we allowed $\Phi_i$ and $\Gamma_i$ to
fluctuate about their classical zero expectation values to take quantum
effects into account. From \C{Oreff}, we see that this leads to a
renormalization of $t$
\be U = \frac{e_{eff}^2}{2}  \vert t_{eff} (\sigma)\vert^2 ,\ee
where
\be  t_{eff} (\sigma) = t(\mu) +i\sum_i Q_i {\rm ln}
(\frac{c_i \sigma}{\mu}).\ee
This can also be determined from the one-loop renormalization of
$t$. The supersymmetric ground states of the theory for large $\Sigma$
are then given by
$t_{eff} (\sigma) =0$ which has solutions,
\be \sigma^{\sum_i Q_i} =
\frac{\mu^{\sum_i Q_i} e^{it(\mu)}}{\prod_i c_i^{Q_i}} =
\frac{\Lambda^{\sum_i Q_i}}{\prod_i c_i^{Q_i}}.
\ee
Hence for large $\Sigma$, there are $\vert \sum_i Q_i \vert$ vacuum states
labelled by
\be \sigma =
\frac{\mu e^{{it\over \sum_i Q_i}}}{(\prod_i c_i^{Q_i})^{ {1\over\sum_i Q_i}}}
\times e^{{2\pi ik\over\sum_i Q_i}},\ee
for $k=0,1,\ldots, \sum_i Q_i -1$. For \TT\ theories where $c_i =
{\sqrt 2} Q_i$ for all $i$, we recover the relation 
\be \sigma^{\sum_i Q_i} =
\frac{\Lambda^{\sum_i Q_i}}{\prod_i ({\sqrt 2Q_i})^{Q_i}}\ee
which is indeed true.

Let us turn to the \ZT\ $\mathbb{P}^{N-1}$ model. For generic choices of
$c_i$ where we only have \ZT\ supersymmetry
in the UV, we find the relation
\be \label{Qcohom} \sigma^N = \frac{\mu^N e^{it}}{\prod_i c_i}
= \frac{\Lambda^N}{\prod_i c_i}, \ee
which shows us that quantum cohomology ring is unchanged by the
deformation modulo a numerical scaling. This is in accord with our
expectation that this deformation is not a relevant one. The number of
vacua is also unchanged with $N$ vacua given by 
\be  \sigma  = \frac{\mu e^{{it}/ N}}{(\prod_i c_i)^{1/N}}
\times e^{2\pi ik/N} \ee
{}for $k=0,\ldots,N-1$.

\detail{A Direct Computation via Instantons}

So far, we determined the superpotential by using symmetries, the
effective superpotential, and the
known \TT\ result. For the case $\sqrt{2} Q\vert c \vert =1$ but a
non-trivial phase, we can do better. In this case, the fermion zero
modes can be explicitly constructed in a one instanton background, and
a non-perturbative correction to the dual superpotential can be
directly exhibited. 

Let us return to the original case of one chiral and one Fermi field.
Consider
the Lagrangian \be L =  -\frac{i}{2} \int d^2 \theta ~{\bar{\Phi}}
({\cal{D}}_0 - {\cal{D}}_1 ) \Phi
 -\frac{1}{2} \int d^2 \theta ~\bar{\Gamma} \Gamma. \ee
Using the duality map for the Fermi superfield, we see that 
\be
{\bar{\eta}}_- = -\frac{1}{\sqrt 2} \bar\phi \chi_-, ~\eta_-
=-\frac{1}{\sqrt 2} \phi {\bar{\chi}}_- .
\ee 
The dual theory has a
perturbative superpotential given by \be 
\widetilde{W} = -\frac{i \Upsilon}{4}(QY + it) + cQ \Sigma F. \ee
Is there an $F e^{-Y}$ correction to the superpotential? As
before, this implies that $\langle {\bar{\eta}}_- \xi_+ \rangle$
should be non-zero, which we now argue directly is the case. 

The
Euclidean action of the original theory has vortex instantons for
$\sigma =0$. There are two fermion zero modes in this instanton
background. The first is given by, 
\be \label{fzm} \mu^0 = \pmatrix{ {\bar\psi}_{+}^0 \cr \lambda_-^0\cr} =
\pmatrix{
  -{\sqrt 2} ({\bar{D}}_1  + i{\bar{D}}_2){\bar\phi} \cr D-F_{12}\cr}\ee
and the second is,
\be \nu^0 = \pmatrix{ \chi_-^0
\cr {\bar{\lambda}}_+^0\cr} = \pmatrix{
  -2Q c (D_1  - i D_2) \phi \cr D-F_{12}\cr}.\ee
The fact that $\vert c \vert =1$ is necessary to show that the
$\nu^0$ zero mode is annihilated by the Dirac-Higgs operator. So,
\be \langle {\bar{\eta}}_- \xi_+ \rangle \sim c \int d^2 x_0 e^{-2
\pi i t} \vert \phi (D_1 - i D_2) \phi \vert^2 \ee which is
clearly non-zero. Hence the exact superpotential is \be
{\widetilde{W}}_{exact} = -\frac{i \Upsilon}{4}(QY + it) +cQ
\Sigma F +\beta \mu F e^{-Y}, \ee where $\mu$ is the energy scale
of the theory and $\beta$ is a non-zero constant. Using our prior
discussion, we see that $\beta$ is independent of $c$ and is given by
$\beta =-\frac{1}{\sqrt 2}$, which leads to the exact result 
\be \label{Exactc}
{\widetilde{W}}_{exact} = -\frac{i \Upsilon}{4}(QY + it) + cQ
\Sigma F - \frac{\mu}{\sqrt 2} F e^{-Y}. \ee 

\detail{The Vacuum Structure}

We can now directly analyze the vacuum structure of the \ZT\ $\mathbb{P}^{N-1}$ model
with $E_i = c_i \Sigma \Phi_i$. Earlier from the large $\Sm$ analysis,
we obtained $N$ vacuua and the chiral ring relation \C{Qcohom}. Using the dual
theory, we show that these conclusions are indeed correct. 

{}For the
$\mathbb{P}^{N-1}$ model, $Q_i=1$ for all $i$. The exact
superpotential is given by \C{NPc}. We will determine the vacua of
this superpotential. Integrating out $\Upsilon$ gives the constraint
$$\sum_i Y_i =-it $$ 
which is solved by setting $Y_i =-\Theta_i$ (for
$i=1,\ldots,N-1$) and $Y_N =-it + \sum_{i =1}^{N-1} \Theta_i$.  Each 
$\Theta_i$ is a periodic variable with period $2\pi$. Integrating
out $\Sm$ gives the 
constraint $$\sum_i c_i F_i =0$$ which is solved by  $F_i = \frac{1}{c_i}
{\mathcal{G}}_i$ (for $i=1,\ldots,N-1$) and  $F_N = -\frac{1}{c_N}
\sum_{i =1}^{N-1} {\mathcal{G}}_i$. 
Finally, defining  $X_i =
e^{\Theta_i}$, we exhibit an effective superpotential \be {\widetilde{W}}_{eff} = -
\frac{\mu}{\sqrt 2}  \sum_{i =1}^{N-1} {\mathcal{G}}_i
\left(\frac{X_i}{c_i} -\frac{e^{it}}{c_N X_1 \cdots X_{N-1}} \right). \ee 
We obtain the supersymmetric ground states by solving
${\partial{\widetilde{W}}_{eff} \over \partial {\mathcal{G}}_i}=0$ 
for all $i$. This gives us
\be \label{Vaceqn} \frac{X_1}{c_1}
=  \frac{X_2}{c_2} =\ldots= \frac{X_{N-1}}{c_{N-1}}
=\frac{e^{it}}{c_N X_1 \cdots X_{N-1}}.\ee Also the linearity of
${\widetilde{W}}_{eff}$ in ${\mathcal{G}}_i$ sets
${\widetilde{W}}_{eff} =0$. Setting $\frac{X_i}{c_i} =
\frac{x}{\mu}$, we see that 
\be x^N = \frac{\mu^N
e^{it}}{\prod_i c_i} = \frac{\Lambda^N}{\prod_i c_i},
\ee 
which is the quantum cohomology ring (or chiral ring) relation for this theory.
The vacuum states are given by \be x  = \frac{\mu e^{{it}/
N}}{(\prod_i c_i)^{1/N}} \times e^{2\pi ik/N},\ee for
$k=0,1,\ldots,N-1$. There are indeed $N$ supersymmetric vacua,
which confirms that the large $\Sigma$ analysis did capture all the
vacuum states. 

\subsubsection{The Case of Equal and Opposite Charges}
\label{equalandopp}

Next we consider a theory with one chiral superfield $\Phi$ of charge
$Q$, and one Fermi superfield
$\Gamma$ of charge $-Q$. With these charge assignments, this theory is
never \TT, but it is a consistent \ZT\ theory. Because $\Gamma$
carries charge $-Q$, we see that $E$ has to be zero in the
theory. This is because the only possibility consistent with
chirality and the charge assignments is $E \sim
\frac{1}{\Phi}$ which is singular. So the theory described
has the Lagrangian \be L =  -\frac{i}{2} \int d^2 \theta
~{\bar{\Phi}} ({\cal{D}}_0 - {\cal{D}}_1 ) \Phi
 -\frac{1}{2} \int d^2 \theta ~\bar{\Gamma} \Gamma .
\ee

The case of $E=0$ is problematic for us since it corresponds to a
singular choice of section. This model is simple enough that we can
postulate a reasonable dual description as follows. We dualize only
the chiral superfield, initially leaving
the Fermi superfield untouched. In the dual theory, we find
a neutral chiral superfield, $Y$, and a charged Fermi
superfield $\Gamma$. 

However, as we discussed earlier, it is difficult to study (and
perhaps even define) the dual theory in terms of $Y$ and
$\Gamma$. So we proceed by constructing the dual in terms of $Y$, and a
neutral Fermi superfield $F$. We will define $F$ by 
$$F =\Phi \Gamma$$  so that
\be {{\eta}}_- ={\phi} {{\chi}}_-. \ee
Now the dual Lagrangian is 
\bea \widetilde{L}
=  \frac{1}{8} \int d^2 \theta ~ \Big[\frac{i(Y -\bar{Y})}{Y +
\bar{Y}} \partial_- (Y + \bar{Y})
 -8\frac{\bar{F} F}{Y + \bar{Y}} \Big]
  -\Big[\frac{iQ}{4}
\int d \theta^+ Y \Upsilon + {\rm h.c.} \Big]. \eea 
The perturbative dual
superpotential is given by \be  \widetilde{W} =
-\frac{i \Upsilon}{4}(QY + it) . \ee 
We now consider the
possibility of non-perturbative corrections to the dual
superpotential of the usual form $F e^{-Y}$.
We can check if there is such a term by computing,
\be \langle {\bar{\eta}}_- \xi_+ \rangle \sim
\int d^2 x_0 \vert \phi \vert^2 \phi
({\bar{D}}_1 + i {\bar{D}}_2) \bar{\phi}. \ee
To obtain this expression, we have used the $\bar{\psi}_+$ zero mode given by \C{fzm},
and the $\bar{\chi}_{-}$ zero mode given by 
$$ \bar{\chi}_{-}^0 = \phi.$$
However, the integral vanishes using the identity (in Euclidean space)
\be \label{iden} 2i \phi ({\bar{D}}_1 + i{\bar{D}}_2) \bar{\phi} +
(\partial_1 +
i\partial_2) (D -F_{12}) =0.\ee
So this non-perturbative correction is absent. 
Our conjectured dual superpotential is therefore
\be {\widetilde{W}}_{exact} = -\frac{i \Upsilon}{4}(QY + it), \ee
leading to $ {\widetilde{W}}_{eff} =0$ with the constraint $QY
=-it$. This is consistent with the original theory where there is a
single supersymmetric vacuum with mass gap.

\subsection{Relevant Deformations of $\IP^1 \times \IP^1$}
\label{reldef}
We now want to construct the dual of
a theory that admits non-trivial bundle deformations. As a
particularly simple example, we
take $\M=\IP^1 \times \IP^1$.\footnote{It is our pleasure to thank Sheldon Katz for
suggesting this example, and describing the following computation of  
$H^1 (\M, {\rm End }(T\M))$.} 
Deformations of the tangent bundle are
parametrized by $H^1 (\M, {\rm End }(T\M))$. In this case, the tangent
bundle is a sum of line-bundles over each $\IP^1$ which we denote 
$$ T\M = \cO(2,0) \oplus \cO(0,2).$$
The cohomology of ${\rm End}( T\M) = \cO\oplus\cO\oplus \cO(-2,2)\oplus
\cO(2,-2)$ can be computed easily by using the Kunneth formula and the
relations  
\be
H^1(\IP^1, \cO(-2))=\IC, \qquad H^0(\IP^1, \cO(2)) = \IC^3.
\ee
Therefore $H^1 (\M, {\rm End }(T\M))=\IC^6.$ We want to both realize these
$6$ deformations in a GLSM, and explicitly construct the dual
description. This will allow us to solve for the instanton corrected
chiral ring of the IR sigma model. 

\detail{The Original Theory}

To realize $\IP^1\times \IP^1$, we need a GLSM with a $U(1)_1\times
U(1)_2$ gauge symmetry. The fields are 
$$\Phi_1,\Phi_2, \widetilde\Phi_1,
\widetilde\Phi_2, \Gamma_1, \Gamma_2, \widetilde\Gamma_1 ,
\widetilde\Gamma_2, \Sigma, \widetilde\Sigma.$$ 
The fields with charge $1$ under $U(1)_1$ are $\Phi_1, \Phi_2, \Gamma_1$ and
$\Gamma_2$, while the fields with charge $1$ under $U(1)_2$ are $\widetilde\Phi_1,
\widetilde\Phi_2, \widetilde\Gamma_1$ and $\widetilde\Gamma_2$. 
Both $\Sigma$ and $\widetilde\Sigma$ are
neutral under both $U(1)$ factors. We take the following choices for
$E$ and $\widetilde{E}$ 
\bea \label{Es} E_1 ={\sqrt 2} \{\Phi_1
\Sigma + \widetilde\Sigma (\alpha_1 \Phi_1
+ \alpha_2 \Phi_2)\},\nonumber \\
E_2 ={\sqrt 2} \{\Phi_2 \Sigma + \widetilde\Sigma (\alpha_1 '
\Phi_1
+ \alpha_2 ' \Phi_2)\}, \\
\widetilde{E_1} ={\sqrt 2} \{ \widetilde\Phi_1 \widetilde\Sigma +
\Sigma (\beta_1 \widetilde\Phi_1
+ \beta_2 \widetilde\Phi_2)\}, \nonumber \\
\widetilde{E_2} ={\sqrt 2} \{ \widetilde\Phi_2 \widetilde\Sigma +
\Sigma (\beta_1 ' \widetilde\Phi_1 + \beta_2 ' \widetilde\Phi_2)\}.
\nonumber \eea 
Here $\alpha_i, \alpha_i ', \beta_i, \beta_i '$ are
complex parameters. Not all of these parameters correspond to
independent deformations. Rescaling  $\alpha_i,
\alpha_i'$ and $ \beta_i, \beta_i'$ independently by any non-zero complex
number correspond to trivial deformations. These projective 
identifications leave us with the six
degrees of freedom parametrizing deformations of $T\M$. Intuitively,
these deformations couple the tangent bundles of each $\IP^1.$ 
Note that when the deformation parameters are taken to zero, we
recover a \TT\ GLSM.

 The vacuum solution of the GLSM is given by \be \sum_i \vert
\phi_i \vert^2 =  r_1,~~ \sum_i \vert \widetilde\phi_i \vert^2 =
r_2, \ee i.e., the product of $\IP^1 \times \IP^1$ with K\"ahler
classes $r_1$ and $r_2$ respectively, and \be E_i =
{\widetilde{E}}_i =0.\ee Generically, $ E_i = {\widetilde{E}}_i
=0$ has a solution $\sigma = \widetilde\sigma =0$. However there
do exist vacuum solutions with $\sigma \neq 0$ and
$\widetilde\sigma \neq 0$. These correspond to new branches in the
moduli space of solutions. Typically, where these branches meet is
extremely interesting since there is usually a singularity at the
intersection locus which should be resolved in the full
two-dimensional field theory. In this case, such a singularity must be a
kind of bundle degeneration. 

{}For example, let us construct a vacuum
solution with 
$$(\phi_1 = {\sqrt r_1},  \phi_2 = 0) \qquad (\widetilde\phi_1
= {\sqrt r_2}, \widetilde\phi_2 = 0).$$ 
Now we can have a solution
with $\sigma \neq 0$ and $\widetilde\sigma \neq 0$ given by 
$$ \alpha_1 ' =  \beta_1 ' =0, \qquad {\sigma} =
-\alpha_1\widetilde\sigma, \qquad \alpha_1 \beta_1 =1.$$ 
In this case, we see
that $\Sigma$ is proportional to $\widetilde\Sigma$, and from the
analysis of the left-moving Yukawa couplings (which we described in
section~\ref{yukawadiscussion}), we see that the rank of the bundle decreases by $1$
instead of decreasing by $2$ when $\Sigma$ and $\widetilde\Sigma$
are linearly independent. This is in accord with our general
expectations. Although these degeneration locii are fascinating, we
will continue by considering the generic vacuum solution where
$\Sigma = \widetilde\Sigma =0$.

We now consider the massless fermionic degrees of freedom of the
low-energy theory. Let the $U(1)_1\times U(1)_2$ gauginos be $\lambda_{-1}$ and
$\lambda_{-2}$, respectively. From the Yukawa couplings for
$\lambda_{-1}$, we see that the massless right-moving fermions
satisfy 
\be \sum_i \bar\phi_i \psi_{+i} =0,\ee 
which we can interpret as a gauge fixing constraint as before. From the Yukawa
couplings for $\lambda_{-2}$, we see that the massless right-moving
fermions satisfy 
\be \sum_i \bar{\widetilde\phi}_i
\widetilde\psi_{+i} =0, \ee 
which we again interpret as a gauge fixing constraint.

Let us denote the fermionic component field of $\Sigma$ and
$\widetilde\Sigma$ by $\bar\lambda_+$ and
$\bar{\widetilde\lambda}_+$, respectively. From their Yukawa
couplings, we see that the left-moving massless fermions satisfy
\be \sum_i \bar\phi_i \chi_{-i} + \widetilde\chi_{-1} \sum_i
\bar\beta_i  \bar{\widetilde\phi}_i + \widetilde\chi_{-2} \sum_i
\bar\beta_i '  \bar{\widetilde\phi}_i =0 \ee and, \be \sum_i
\bar{\widetilde\phi}_i \widetilde\chi_{-i} + \chi_{-1} \sum_i
\bar\alpha_i  \bar\phi_i + \chi_{-2} \sum_i \bar\alpha_i '
\bar\phi_i =0. \ee 
These are again interpretable as gauge fixing 
constraints.

\detail{The Dual Description}

Let us analyse the dual theory. The dual classical
Lagrangian is given by 
\bea \widetilde{L} =  \frac{i}{8} \sum_i
\int d^2 \theta ~ \frac{Y_i - {\bar{Y}}_i}{Y_i + {\bar{Y}}_i}
\partial_- (Y_i + {\bar{Y}}_i)
+ \frac{i}{8} \sum_i \int d^2 \theta ~ \frac{{\widetilde{Y}}_i -
\bar{\widetilde{Y}}_i}{{\widetilde{Y}}_i + \bar{\widetilde{Y}}_i}
\partial_- ({\widetilde{Y}}_i + \bar{\widetilde{Y}}_i)
\nonumber \\
-\frac{1}{2} \sum_i \int d^2 \theta~{\bar{\mathcal{F}}}_i
{\mathcal{F}}_i -\frac{1}{2} \sum_i \int d^2
\theta~\bar{\widetilde{\mathcal{F}}}_i \widetilde{\mathcal{F}}_i
+ \int d \theta^+ ~\widetilde{W} + {\rm h.c.} \eea where \be
\widetilde{W} = - \frac{i\Upsilon_1}{4} (\sum_i  Y_i + it_1)
 - \frac{i\Upsilon_2}{4} (\sum_i  {\widetilde{Y}}_i + it_2)
+ \frac{1}{\sqrt 2} \sum_i E_i {\mathcal{F}}_i + \frac{1}{\sqrt 2}
\sum_i {\widetilde{E}}_i \widetilde{\mathcal{F}}_i. \ee 
Here the ${\mathcal{F}}_i, \widetilde{\mathcal{F}}_i$ are charged Fermi
superfields. The duality maps (modulo fermion bilinears) for the
bosonic superfields are 
\bea  \label{bilin1}\bar{\Phi}_i \Phi_i
~=~\frac{1}{2} (Y_i+\bar{Y}_i),\qquad
 \bar{\Phi}_i ( {\buildrel \leftrightarrow \over \partial_-} +i V_1) \Phi_i
=-\frac{1}{4} \partial_- (Y_i -\bar{Y}_i), \eea
\bea \label{bilin2}\bar{\widetilde\Phi}_i \widetilde\Phi_i ~=~ \frac{1}{2}
({\widetilde{Y}}_i+\bar{\widetilde{Y}_i}),\qquad
\bar{\widetilde{\Phi}}_i ( {\buildrel \leftrightarrow \over
\partial_-} +i V_2) \widetilde{\Phi}_i =-\frac{1}{4} \partial_-
({\widetilde{Y}}_i -\bar{\widetilde{Y}}_i), \eea
while the fermionic superfields map according to,
$$ \bar{\Gamma}_i=
{\mathcal{F}}_i,~~~\bar{\widetilde\Gamma}_i=
\widetilde{\mathcal{F}}_i. $$
 The dual
{}Fermi superpotential term in the action can be written as 
\be \label{fermsuper}
\int d \theta^+ ~\Sigma (F_1 + F_2) + \int d \theta^+
~\widetilde\Sigma ({\widetilde{F}}_1 + {\widetilde{F}}_2) + {\rm
h.c.} \ee 
where 
\bea \label{redefineF} F_1 = \Phi_1 {\mathcal{F}}_1 + (\beta_1
\widetilde\Phi_1 +
\beta_2 \widetilde\Phi_2) \widetilde{\mathcal{F}}_1, \nonumber \\
F_2 = \Phi_2 {\mathcal{F}}_2 + (\beta_1 ' \widetilde\Phi_1 +
\beta_2 ' \widetilde\Phi_2) \widetilde{\mathcal{F}}_2, \\
\widetilde{F_1} = \widetilde\Phi_1 \widetilde{\mathcal{F}}_1 +
(\alpha_1 \Phi_1 +
\alpha_2 \Phi_2) {\mathcal{F}}_1 \nonumber,\\
\widetilde{F_2} = \widetilde\Phi_2 \widetilde{\mathcal{F}}_2 +
(\alpha_1 ' \Phi_1 + \alpha_2  '\Phi_2) {\mathcal{F}}_2 ,\nonumber
\eea 
where $F_i, {\widetilde{F}}_i$ are neutral Fermi superfields. Note
that there is no unique way of defining $F$ in terms of
${\mathcal{F}}$, but there is a natural choice given in
\C{redefineF}. With this choice, $\Sm$ only couples to $F$ while
$\widetilde\Sigma$ only couples to $\widetilde{F}$ in the
superpotential \C{fermsuper}. 

It is worth noting that the kinetic terms for the dual neutral Fermi
superfields are not singular, even for field configurations that
correspond to instantons in the original theory. To see this, we
consider generic deformations of the left-moving bundle given
in~\C{Es}. We can solve for $\F, \widetilde{\F}$ in terms of $F,
\widetilde{F}$ and $\Phi, \widetilde{\Phi}$
\bea & {\mathcal{F}}_1 =
\frac{\widetilde{\Phi}_1 F_1 - A \widetilde{F}_1}{
\widetilde\Phi_1 \Phi_1 -AC},\qquad & {\mathcal{F}}_2 =
\frac{\widetilde{\Phi}_2 F_2 -B \widetilde{F}_2}{
\widetilde\Phi_2 \Phi_2 -BD},\non \\
& \widetilde{\mathcal{F}}_1 = \frac{F_1 - \Phi_1
{\mathcal{F}}_1}{A},\qquad & \widetilde{\mathcal{F}}_2 = \frac{F_2 -
\Phi_2 {\mathcal{F}}_2}{B} , \non\eea 
where $A = \beta_1
\widetilde\Phi_1 + \beta_2  \widetilde\Phi_2, ~B = \beta_1 '
\widetilde\Phi_1 + \beta_2 ' \widetilde\Phi_2,~C= \alpha_1 \Phi_1
+ \alpha_2\Phi_2$ and $D= \alpha_1 ' \Phi_1 + \alpha_2 ' \Phi_2$.
So for generic choices of the parameters, all the denominators are
non-vanishing, even in instanton backgrounds. Consider embedding an
instanton in  $\phi_1$ (or $\phi_2$) and
$\widetilde\phi_1$ (or $\widetilde\phi_2$), then it is easy to see
that $A,B,C,D$ are each non-vanishing.

Note that on the degeneration locus described before
where $\Sigma$ is proportional to $\widetilde\Sigma$, we find that
\be
\widetilde\Phi_1 \Phi_1 -AC=0, \qquad B= \widetilde\Phi_2 \Phi_2=0.
\ee
Only $A$ is non-zero and  equal to $\beta_1
\widetilde\Phi_1$. This leads to singular kinetic energy terms which
is natural for a singular locus.

We therefore obtain the exact dual superpotential 
\bea \label{bW} \widetilde{W} &=&
-\frac{i\Upsilon_1}{4} (\sum_i  Y_i + it_1) -\frac{i\Upsilon_2}{4}
(\sum_i  \widetilde{Y}_i + it_2)
+\Sigma \sum_i F_i +\widetilde\Sigma \sum_i \widetilde{F}_i \cr
&& +\mu \sum_{ij} \left(\beta_{ij} F_i e^{-Y_j} + \beta_{\tilde{\imath}
\tilde{\jmath}} \widetilde{F}_i e^{-\widetilde{Y}_j} +\beta_{i
\tilde{\jmath}} F_i e^{-\widetilde{Y}_j} + \beta_{\tilde{\imath} j}
\widetilde{F}_i e^{-Y_j}\right). \eea 
The duality map for the Fermi
superfields is given by 
\bea F_1 = \Phi_1 \bar{\Gamma}_1 +
(\beta_1 \widetilde\Phi_1 +
\beta_2 \widetilde\Phi_2) \bar{\widetilde{\Gamma}}_1, \nonumber\\
F_2 = \Phi_2 \bar{\Gamma}_2 + (\beta_1 ' \widetilde\Phi_1 +
\beta_2 ' \widetilde\Phi_2) \bar{\widetilde{\Gamma}}_2, \\
\widetilde{F_1} = \widetilde\Phi_1 \bar{\widetilde{\Gamma}}_1 +
(\alpha_1 \Phi_1 +
\alpha_2 \Phi_2) \bar{\Gamma}_1, \nonumber\\
\widetilde{F_2} = \widetilde\Phi_2 \bar{\widetilde{\Gamma}}_2 +
(\alpha_1 ' \Phi_1 + \alpha_2  '\Phi_2) \bar{\Gamma}_2 .\nonumber
\eea 
Our task is to relate the $\beta$ parameters to the
original bundle deformation parameters given in~\C{Es}. The difficulty
in determining this map is easy to explain. The $\beta$ parameters are
determined by instanton computations in the original theory. In an
instanton background, the right-moving fermion zero modes can be
determined exactly. However, the left-moving zero modes depend
sensitively on the choice of $E, \widetilde{E}$ given in~\C{Es}. To
determine the $\beta$ parameters, we need to be able to evaluate
exactly instanton corrections to various two point functions in the
original theory. This is a hard task so we will need to be more clever. 

\detail{The Vacuum Structure}

Before determining the parameter map, let us examine the general
vacuum structure for the dual theory.
Integrating out the massive field strength multiplets, $\Upsilon, 
\widetilde{\Upsilon}$,  we obtain the constraint
\be Y_1 + Y_2 =-it_1, \qquad {\widetilde{Y}}_1 + {\widetilde{Y}}_2
=-it_2.
\ee
On integrating out the massive $\Sigma $ and
$\widetilde\Sigma$ fields we find 
\be F_1 + F_2 =0,\qquad
{\widetilde{F}}_1 + {\widetilde{F}}_2 =0. 
\ee 
We solve these constraints by setting
\be \label{con1} Y_1 =Y,\qquad Y_2 = -Y -it_1,\qquad F_1 = -F_2 = F, \ee and
\be \label{con2} \widetilde{Y}_1 =\widetilde{Y},\qquad\widetilde{Y}_2 =
-\widetilde{Y} -it_2, \qquad \widetilde{F}_1 = -\widetilde{F}_2 =
\widetilde{F}.\ee 
Recall that the imaginary parts of the $Y, \widetilde{Y}$ variables
are periodic. Let us define the low-energy theory in terms of
single-valued degrees of freedom
$X$ and $\widetilde{X}$ where
$$X= e^{-Y}, \qquad \widetilde{X} = e^{-\widetilde{Y}}.$$ 
In terms of these variables, 
\bea && \mu^{-1}
  {\widetilde{W}}_{eff}  =  F \left[X(\beta_{11} -\beta_{21}) +
\frac{e^{it_1}}{X} (\beta_{12} -\beta_{22} ) + \widetilde{X}
(\beta_{1 \tilde{1}} - \beta_{2 \tilde{1}})  +
\frac{e^{it_2}}{\widetilde{X}} (\beta_{1 \tilde{2}} - \beta_{2
\tilde{2}}) \right] 
\cr
&& + \widetilde{F} \left[\widetilde{X} (\beta_{\tilde{1} \tilde{1}} -
\beta_{\tilde{2} \tilde{1}}) + \frac{e^{it_2}}{\widetilde{X}}
(\beta_{\tilde{1} \tilde{2}} - \beta_{\tilde{2} \tilde{2}}) + X
(\beta_{\tilde{1} 1} -\beta_{\tilde{2} 1}) + \frac{e^{it_1}}{X}
(\beta_{\tilde{1} 2} -\beta_{\tilde{2} 2} ) \right]. 
\eea 
Because we deformed the bundle for the left-movers, the chiral ring of
the IR (or low-energy) theory is deformed. This will define our
analogue of the usual quantum cohomology ring of \TT\ theories.    

In order to construct the chiral ring, we set 
$$\frac{\partial {\widetilde{W}}_{eff}}{\partial F}
= \frac{\partial {\widetilde{W}}_{eff}}{\partial \widetilde{F}}
=0$$ 
{}from which we obtain the deformed chiral ring relations 
\be X + p
\frac{e^{it_1}}{X} +  q \widetilde{X} + s
\frac{e^{it_2}}{\widetilde{X}} =0,    \ee and \be \widetilde{X} +
\widetilde{p} \frac{e^{it_2}}{\widetilde{X}} + \widetilde{q} X +
\widetilde{s} \frac{e^{it_1}}{X} =0. 
\ee 
In these equations, 
\be
p= \frac{\beta_{12} -\beta_{22}}{\beta_{11} -\beta_{21}} , \quad q=
\frac{\beta_{1 \tilde{1}} - \beta_{2 \tilde{1}}}{\beta_{11}
-\beta_{21}} , \quad s= \frac{\beta_{1 \tilde{2}} - \beta_{2
\tilde{2}}}{\beta_{11} -\beta_{21}} ,\ee and \be \widetilde{p}
=\frac{\beta_{\tilde{1} \tilde{2}} - \beta_{\tilde{2}
\tilde{2}}}{\beta_{\tilde{1} \tilde{1}} - \beta_{\tilde{2}
\tilde{1}}}, \quad\widetilde{q} =\frac{\beta_{\tilde{1} 1} -
\beta_{\tilde{2} 1}}{\beta_{\tilde{1} \tilde{1}} -
\beta_{\tilde{2} \tilde{1}}}, \quad \widetilde{s}
=\frac{\beta_{\tilde{1} 2} - \beta_{\tilde{2} 2}}{\beta_{\tilde{1}
\tilde{1}} - \beta_{\tilde{2} \tilde{1}}}. 
\ee 
So the \ZT\ chiral ring relations mix the generators of the chiral
ring for each $\mP^1$; these generators correspond to the K\"ahler
classes of each $\mP^1$. The mixing occurs because we have deformed
the bundle for the left-movers away from the
tangent bundle (in a holomorphic way).

In the limit in which
the bundle deformations vanish, we should recover two
decoupled chiral rings; one for each $\mP^1$. It is easy to
see that this is true. As the bundle deformations vanish, we recover
\TT\ supersymmetry and only the diagonal $\beta$ parameters survive
giving
\be p= -1,\quad
q=s=0,\qquad {\widetilde{p}} =-1,\quad {\widetilde{q}} ={\widetilde{s}}
=0.
\ee 
Therefore, we find a decoupled ring 
\be X^2 =
e^{it_1},\qquad {\widetilde{X}}^2 = e^{it_2} ,\ee 
for each $\mP^1$ as we expect. 

\detail{Determining the Exact Parameter Map}

We now want to solve this theory completely by determining the exact
parameter map. We want to know how the $\beta$ parameters depend on 
${\alpha_i, \alpha'_i, \beta_i, \beta'_i}$. Our tools for this task
will be global $U(1)$ symmetries and a large $\Sm, \widetilde{\Sm}$
analysis of the kind described in section~\ref{largesm}. 
The strategy in constructing a
$U(1)$ global symmetry is to assign suitable $U(1)$ charges to
the various superfields as well as to the deformation parameters.
This $U(1)$ is, in general, anomalous. In the dual theory, the $U(1)$
acts by shifting the $Y_i,\widetilde{Y}_i$ fields, and the anomaly is
realized by a non-invariant term in the perturbative dual
superpotential. This is exactly analogous to the case of the
$R$-symmetry. If the $\beta$ parameters are charged under the global
$U(1)$, we can use the symmetry to constrain their dependence on the
deformation parameters.  

However, we now show that unless some of the deformation parameters are
set to zero, no choice of $U(1)$ symmetry will help us fix the 
$\beta$ parameters. To see this, let us go back to the definitions
given in \C{Es}\ to make charge assignments. Assign
the superfields the following charges 
$$ (\Phi_1, p_1),\quad (\Phi_2, p_2), \quad (\widetilde\Phi_1,
\widetilde{p}_1), \quad  (\widetilde\Phi_2, \widetilde{p}_2), \quad
(\Sigma, k), \quad (\widetilde\Sigma, \widetilde{k})$$ 
where, for example, $\Phi_1$ has charge $p_1$. 
We then see that the deformation parameters have 
the following charges: 
$$ (\alpha_1, \alpha_2', k-\widetilde{k}),\quad
(\alpha_2, k -\widetilde{k} +p_1 -p_2), \quad (\alpha_1', k
-\widetilde{k}-p_1 +p_2), $$ $$ (\beta_1, \beta_2', \widetilde{k}
-k),\quad  (\beta_2, \widetilde{k} -k
+\widetilde{p}_1 -\widetilde{p}_2), \quad (\beta_1',  
\widetilde{k} -k -\widetilde{p}_1 +\widetilde{p}_2).$$
So in particular, we see that arbitrary powers of 
$$\alpha_1 \beta_1,\quad  \alpha_1 
\beta_2',\quad \alpha_2' \beta_1, \quad \alpha_2' \beta_2'$$
carry zero charge. The $\beta$ parameters could depend on these
combinations in arbitrary ways.  

Let us therefore set some deformation parameters to zero in order to
usefully employ global $U(1)$ symmetries. 
In \C{Es}, we take $\alpha_1 =\epsilon_1$ and 
$\alpha_2' =\epsilon_2$ and set all other deformation parameters to zero. 
Thus we start with 
\bea E_1 ={\sqrt 2} (\Phi_1 \Sigma + \epsilon_1 \widetilde\Sigma \Phi_1),
\qquad\widetilde{E_1} ={\sqrt 2} \widetilde\Phi_1 \widetilde\Sigma, 
\cr
E_2 ={\sqrt 2} (\Phi_2 \Sigma + \epsilon_2 \widetilde\Sigma \Phi_2), 
\qquad\widetilde{E_2} ={\sqrt 2} \widetilde\Phi_2 \widetilde\Sigma. \eea    
This choice gives the following expressions for the dual fermions
\bea F_1 = \Phi_1 \bar{\Gamma}_1,\qquad \widetilde{F}_1 = \widetilde\Phi_1 
\bar{\widetilde{\Gamma}}_1 +\epsilon_1 \Phi_1 \bar{\Gamma}_1,\cr
F_2 = \Phi_2 \bar{\Gamma}_2,\qquad\widetilde{F}_2 = \widetilde\Phi_2 
\bar{\widetilde{\Gamma}}_2 +\epsilon_2 \Phi_2 \bar{\Gamma}_2 . 
\eea
The exact dual superpotential is given by \C{bW}. We assign global $U(1)$ 
charges as discussed above. (Note that $\epsilon_1$ and $\epsilon_2$ have the 
same charge $k-\widetilde{k}$.) So the terms $\Sigma F_i$ and 
$\widetilde\Sigma \widetilde{F}_i$ in the dual perturbative superpotential
are charge zero. However, this $U(1)$ symmetry is anomalous: the
$U(1)_1$ gauge symmetry shifts $\sum_i Y_i$ by $-2k$, while the
$U(1)_2$ gauge symmetry
shifts $\sum_i \widetilde{Y}_i$ by $-2\widetilde{k}$. However,  this does 
not tell us the amount by which each individual $Y_i$ or $\widetilde{Y}_i$ shifts 
under the anomaly. The individual shifts can be determined from the
duality maps if we know the complete maps including the fermion
bilinear terms in~\C{bilin1}\ and~\C{bilin2}. 

In the limit in which the deformations vanish, we have a \TT\ theory 
with 
\be\label{ttbeta}\beta_{11} =\beta_{22} = \beta_{\tilde{1}\tilde{1}} =
\beta_{\tilde{2}\tilde{2}} =-\frac{1}{\sqrt 2}\ee
with all other $\beta$ parameters vanishing. From the $U(1)$
invariance of 
$$F_1 e^{-Y_1}, \quad F_2 e^{-Y_2}, \quad \widetilde{F}_1
e^{-\widetilde{Y}_1}, \quad \widetilde{F}_2 
e^{-\widetilde{Y}_2},$$ 
we see that $e^{-Y_i}$ has $U(1)$ charge $k$ while
$e^{-\widetilde{Y}_i}$ has $U(1)$ charge $\widetilde{k}$. In this way,
we determine the individual shifts of the $Y_i, \widetilde{Y}_i$
fields without knowing the fermion bilinear terms in the duality map. 

We can now determine the $U(1)$ charges for the remaining $\beta$
parameters. The parameter $\beta_{i \tilde{\jmath}}$ has charge
$k-\widetilde{k}$ while $\beta_{\tilde{\imath} j}$ has charge $\widetilde{k} -k$.
Because the $\beta$ parameters depend smoothly on the deformation
parameters, we conclude that  $\beta_{i \tilde{\jmath}}$ is
proportional to  $\epsilon_1$ or
$\epsilon_2$, while  $\beta_{\tilde{\imath} j}$ is zero. This is also
fixes the diagonal $\beta$ parameters at their \TT\ value given in~\C{ttbeta}.
We therefore find, 
\bea \widetilde{W}_{non-pert} &=& -\frac{\mu}{\sqrt 2} \sum_{i} (F_i e^{-Y_i} + 
\widetilde{F}_i e^{-\widetilde{Y}_i}) 
 -\frac{\epsilon_1 \mu}{\sqrt 2} (c_1 F_1 e^{-\widetilde{Y}_1} +c_2 F_1 
e^{-\widetilde{Y}_2} +c_3 F_2 e^{-\widetilde{Y}_1} +c_4 F_2 
e^{-\widetilde{Y}_2}) \cr
&& -\frac{\epsilon_2 \mu}{\sqrt 2} (d_1 F_1 e^{-\widetilde{Y}_1} +d_2 F_1 
e^{-\widetilde{Y}_2} +d_3 F_2 e^{-\widetilde{Y}_1} +d_4 F_2 
e^{-\widetilde{Y}_2}). \eea         
The particular deformation we are considering does not distinguish
between $\widetilde{Y}_1$ and $\widetilde{Y}_2$. There is also an obvious
$\Z_2$ symmetry exchanging $\e_1$ and $\e_2$, and all the $1$ and $2$
fields. Together, these symmetries imply 
$$c_1 = c_2 =d_3 =d_4 \equiv \frac{a}{2}, \qquad c_3 = c_4 =d_1 =d_2
\equiv \frac{b}{2}.
$$  
Thus, 
\bea \widetilde{W}_{non-pert} &=& -\frac{\mu}{\sqrt 2} \sum_{i} (F_i e^{-Y_i} + 
\widetilde{F}_i e^{-\widetilde{Y}_i}) \cr &&
-\frac{\mu}{2 {\sqrt 2}} \left[ \epsilon_1 (a F_1 +b F_2) +\epsilon_2
  (b F_1 +aF_2)
\right]
\left(e^{-\widetilde{Y}_1} +e^{-\widetilde{Y}_2}\right) . \eea
Here $a$ and $b$ are numbers which we now evaluate. 
These numbers can be evaluated using the large $\Sigma, \widetilde\Sigma$ 
approach along the lines of section~\ref{largesm}, so we shall be brief. In the 
original theory, take $\Sigma, \widetilde\Sigma$ to be large and slowly 
varying. Integrate out the chiral and Fermi superfields
exactly; since the Lagrangian is quadratic, we can do this exactly
giving an effective action
\be \widetilde{W}_{eff} (\Sigma, \widetilde\Sigma, \Upsilon_1, \Upsilon_2) =
\Upsilon_1 W_1 (\Sigma, \widetilde\Sigma) + \Upsilon_2 W_2 (\Sigma, 
\widetilde\Sigma).  \ee   
This superpotential gives terms in the action
\bea  \frac{1}{4} \int d\theta^+ \widetilde{W}_{eff} (\Sigma, \widetilde\Sigma, 
\Upsilon_1, \Upsilon_2) + {\rm h.c.} = -D_1 {\rm Im} \{ W_1 (\sigma, 
\widetilde\sigma) \} -D_2 {\rm Im} \{ W_2 (\sigma, \widetilde\sigma)\} \cr
 + F_{01} {\rm Re}\{ W_1 (\sigma, \widetilde\sigma)\} + \widetilde{F}_{01} 
{\rm Re} \{ W_2 (\sigma, \widetilde\sigma) \} +\ldots, \eea
where $D_1, D_2$ ($F_{01}, \widetilde{F}_{01}$) are the $D$ terms (field 
strengths) for $U(1)_1$ and $U(1)_2$, respectively. We 
have only included terms that are linear in the $D_i$ fields, and in the field 
strengths. In order to determine $W_1 (\Sigma, \widetilde\Sigma)$ and  
 $W_2 (\Sigma, \widetilde\Sigma)$, we only need to retain
terms linear in the $D_i$ fields and the field strengths. 
It turns out that there are no 
terms linear in the field strengths so the entire contribution comes 
from terms linear in the $D_i$ fields. The calculation is very similar
to the one in section~\ref{largesm}, giving the result
\be \label{origeqn} \widetilde{W}_{eff} (\Sigma, \widetilde\Sigma, 
\Upsilon_1, \Upsilon_2) =
\frac{i\Upsilon_1}{4} \Big\{\sum_i {\rm ln} \left[ 
\frac{{\sqrt 2} (\Sigma +\epsilon_i
\widetilde\Sigma)}{\mu}\right] -it_1 \Big\}
+\frac{i\Upsilon_2}{4} \Big\{ 2{\rm ln} \left[ 
\frac{{\sqrt 2}\widetilde\Sigma}{\mu} \right]
-it_2\Big\}.\ee
Now we proceed to the dual theory and integrate out $F_i$ and 
$\widetilde{F}_i$. It is easy to solve for $Y_i, \widetilde{Y}_i$ from
the four resulting equations 
\be  Y_1 = -{\rm ln} \left[ \frac{{\sqrt 2}(\Sigma -(a\epsilon_1 + b\epsilon_2)
\widetilde\Sigma)}{\mu}\right] , \qquad
Y_2 = -{\rm ln} \left[ \frac{{\sqrt 2}(\Sigma -(b\epsilon_1 + a\epsilon_2)
\widetilde\Sigma)}{\mu}\right] ,  \ee 
$$ \widetilde{Y}_1 = \widetilde{Y}_2 =-{\rm ln} \left[ \frac{{\sqrt 2} 
\widetilde\Sigma}{\mu}\right]. $$    
Thus in the dual theory, we get 
\bea \label{dualeqn} \widetilde{W}_{eff} (\Sigma, \widetilde\Sigma, 
\Upsilon_1, \Upsilon_2) &=&
\frac{i\Upsilon_1}{4} \Big\{ {\rm ln} \left[ 
\frac{{\sqrt 2}(\Sigma -(a\epsilon_1 + 
b\epsilon_2)\widetilde\Sigma)}{\mu} \right]
 +{\rm ln} \left[ \frac{{\sqrt 2}(\Sigma -
(b\epsilon_1 + a\epsilon_2) \widetilde\Sigma)}{\mu}\right] -it_1 \Big\} \cr && 
+\frac{i\Upsilon_2}{4}\Big\{2{\rm ln} \left[ \frac{{\sqrt
	2}\widetilde\Sigma}{\mu} \right]
-it_2\Big\}.\eea
Equating coefficients in \C{origeqn} and \C{dualeqn} gives the relation
\be (\Sigma +\epsilon_1\widetilde\Sigma) (\Sigma +\epsilon_2\widetilde\Sigma)
=(\Sigma -\{a\epsilon_1 +b\epsilon_2\}\widetilde\Sigma) (\Sigma 
-\{b\epsilon_1 +a\epsilon_2\}\widetilde\Sigma). \ee 
Equating the coefficients of $\Sigma \widetilde\Sigma$ and
$\widetilde\Sigma^2$ gives two equations from which we determine $a$
and $b$. There are two 
solutions given by (i) $a=0, b=-1$ and (ii) $a=-1, b=0$. In the first case, 
\be \label{caseone} \widetilde{W}_{non-pert} = -\frac{\mu}{\sqrt 2} \sum_i
(F_i e^{-Y_i} + \widetilde{F}_i e^{-\widetilde{Y}_i}) +\frac{\mu}{2 
{\sqrt 2}} (\epsilon_1 F_2 +\epsilon_2 F_1) (e^{-\widetilde{Y}_1} 
+e^{-\widetilde{Y}_2}), \ee
while in the second case,
\be \label{casetwo} \widetilde{W}_{non-pert} = -\frac{\mu}{\sqrt 2} \sum_i
(F_i e^{-Y_i} + \widetilde{F}_i e^{-\widetilde{Y}_i}) +\frac{\mu}{2 
{\sqrt 2}} (\epsilon_1 F_1 +\epsilon_2 F_2) (e^{-\widetilde{Y}_1} 
+e^{-\widetilde{Y}_2}). \ee
Note that the two superpotentials explicitly exhibit the symmetry of the 
theory under interchange of $\epsilon_1$ and $\epsilon_2$. Using
\C{con1}\ and \C{con2}, we obtain the chiral ring relations 
\be \widetilde{X} = \frac{e^{it_2}}{\widetilde{X}} ,\qquad X -\frac{e^{it_1}}{X}
\pm (\epsilon_1 -\epsilon_2) \widetilde{X} =0,\ee
where the $\pm$ is corresponds to \C{caseone}\ and \C{casetwo},
respectively. Note that this sign ambiguity in the ring relation has
no physical meaning because $(\e_1, \e_2)$ are projective coordinates,
and can be freely rescaled by any non-zero complex number.  

Since $\widetilde{E}_1$ and $\widetilde{E}_2$ are at their \TT\ values, the
chiral ring relation for the $\mP^1$ corresponding to $U(1)_2$ is undeformed. 
The other ring for the $\mP^1$ corresponding to $U(1)_1$ is deformed
because $E_1$ and $E_2$ involve  $\widetilde\Sigma$ couplings. This is
an example of a non-trivial bundle deformation where we have
explicitly solved for the dual superpotential, and determined the
chiral ring. It should be possible to directly compute this ring by
studying instantons in the IR \ZT\ non-linear sigma model along the
lines described in section~\ref{yukawadiscussion}. Lastly, note that
for $\epsilon_1 =\epsilon_2$, the ring relations remain undeformed and
correspond to two decoupled $\mP^1$ spaces.  

\subsection{Examples of Conformal Models}

Next we consider conformal cases where the total space
is a non-compact Calabi-Yau manifold. The two examples that we 
consider are the total spaces of bundles over $\mP^1 \times \mP^1$,
with the bundles suitably chosen so that the models are
conformal. We continue to use the same notation of
section~\ref{reldef}\ for the fields of the $\mP^1 \times \mP^1$
GLSM. 
In our first example,
the dual IR theory is a $\Z_2$ Landau-Ginzburg (LG)
orbifold, while in our second example, the dual is a $(\Z_2)^2$ LG
orbifold.

\subsubsection{A Uniquely \ZT\ Example}
To the fields of the $\mP^1 \times \mP^1$ GLSM described in the
last section, we add
a chiral superfield $P$ and a Fermi superfield $\Gamma$. Both
$P$ and $\Gamma$ carry charge $-2$ under both $U(1)_1$ and $U(1)_2$. 
Since the sum of the charges for the right-movers is zero, the model
is conformal: the IR theory is a non-linear sigma model on a
non-compact Calabi-Yau space. The target space is the total space of
the line-bundle $\cO(-2,-2)$ over $\mP^1 \times \mP^1$.  

We keep the same choice of $E_i,~\widetilde{E}_i$ as in~\C{Es}. For
the $E$ associated to $\G$, we take the choice 
  \be E = -2{\sqrt 2} (\Sigma +
\widetilde\Sigma) P. \ee
Note that with this choice of $E$, this model never enjoys \TT\
supersymmetry; hence the
title of this section. The vacuum solution of the GLSM is given
by 
\be \sum_i \vert \phi_i \vert^2  -2 \vert p \vert^2 =  r_1,\qquad
\sum_i \vert \widetilde\phi_i \vert^2  -2 \vert p \vert^2 =  r_2,
\ee 
and 
\be E_i = {\widetilde{E}}_i = E =0.\ee 
Once again, we
choose the generic vacuum solution $\Sigma =\widetilde\Sigma =0$.
Now because of the presence of the superfield $P$, the two $D$
term equations for the vacuum solution are no longer decoupled from each
other.

Let us define 
$$ P = p + \sqrt{2} \theta^+ \psi_+ + \ldots, \qquad\qquad \G = \chi_- + \ldots.$$
{}From the various
Yukawa couplings, we see that the massless fermionic degrees of
freedom of the low-energy theory satisfy 
\bea && \sum_i \bar\phi_i
\psi_{+i} -2\bar{p} \psi_+ =0,\qquad
\sum_i \bar{\widetilde\phi}_i \widetilde\psi_{+i} -2\bar{p} \psi_+ =0,\cr
&& \sum_i \bar\phi_i \chi_{-i} + \widetilde\chi_{-1} \sum_i
\bar\beta_i  \bar{\widetilde\phi}_i + \widetilde\chi_{-2}
\sum_i \bar\beta_i '  \bar{\widetilde\phi}_i -2\bar{p} \chi_- =0 \cr
&& \sum_i \bar{\widetilde\phi}_i \widetilde\chi_{-i} + \chi_{-1}
\sum_i \bar\alpha_i  \bar\phi_i + \chi_{-2} \sum_i \bar\alpha_i'
\bar\phi_i  -2\bar{p} \chi_-=0. \eea 
In the dual
theory, the classical Lagrangian has the K\"ahler terms given in the
$\mP^1 \times \mP^1$ example along with the following additional terms 
\be
\widetilde{L} =\ldots  +\frac{i}{8} \int d^2 \theta ~ \frac{Y -
\bar{Y}}{Y + \bar{Y}}
\partial_- (Y + \bar{Y})
- \frac{1}{2} \int d^2 \theta~{\bar{\mathcal{F}}} {\mathcal{F}},
\ee 
where we have the duality map (again, modulo fermion bilinears) 
\be  \bar{P} P ~=~\frac{1}{2}
(Y+\bar{Y}),\qquad
 \bar{P} ( {\buildrel \leftrightarrow \over \partial_-} -2i V_1 -2i V_2) P
=-\frac{1}{4} \partial_- (Y -\bar{Y}),
\ee 
and 
\be \bar\Gamma
=\mathcal{F}.
\ee 
The classical dual superpotential is given by \be
\widetilde{W} = - \frac{i\Upsilon_1}{4} (\sum_i  Y_i -2Y + it_1)
 - \frac{i\Upsilon_2}{4} (\sum_i  {\widetilde{Y}}_i -2Y + it_2)
- \frac{1}{\sqrt 2} (\sum_i E_i {\mathcal{F}}_i
 +\sum_i {\widetilde{E}}_i \widetilde{\mathcal{F}}_i +E {\mathcal{F}}).
\ee 
The last term can be written in the form
\be -\int d \theta^+ ~\Sigma (F_1 +
F_2 -2F) -\int d \theta^+ ~\widetilde\Sigma ({\widetilde{F}}_1 +
{\widetilde{F}}_2 -2F), \ee 
where $F = P{\mathcal{F}}$. The
exact dual superpotential is therefore given by 
\bea \widetilde{W} &=&
-\frac{i\Upsilon_1}{4} (\sum_i  Y_i -2Y + it_1)
-\frac{i\Upsilon_2}{4} (\sum_i  \widetilde{Y}_i -2Y + it_2)
-\Sigma (\sum_i F_i -2F) \cr
&&  -\widetilde\Sigma (\sum_i \widetilde{F}_i -2F)
+\mu (\sum_{ij} \beta_{ij} F_i e^{-Y_j} + \beta_{\tilde{i}
\tilde{j}} \widetilde{F}_i e^{-\widetilde{Y}_j} +\beta_{i
\tilde{j}} F_i e^{-\widetilde{Y}_j}
+ \beta_{\tilde{i} j} \widetilde{F}_i e^{-Y_j}) \cr
&& +2\mu F(\omega e^{-Y} + \sum_i \omega_i  e^{-Y_i} + \sum_i
\widetilde\omega_i  e^{-\widetilde{Y}_i}) +\mu \sum_i (\nu_i F_i +
\widetilde\nu_i \widetilde{F}_i) e^{-Y}.\eea 

We now
analyse the vacuum solutions of this theory for generic $\beta,
\omega$ and $\nu$ parameters. The vacuum solution
is determined by solving 
\be Y_1 + Y_2 -2Y=-it_1,~~{\widetilde{Y}}_1 +
{\widetilde{Y}}_2 -2Y=-it_2,\ee and \be F_1 + F_2 -2F=0, \qquad
{\widetilde{F}}_1 + {\widetilde{F}}_2  -2F=0. 
\ee 
We construct solutions where
\be X_1= e^{-Y_1 /2}, \qquad X_2=e^{-Y_2 /2},
\qquad e^{-Y} =e^{-it_1/2} X_1 X_2, \ee
and
\be X_3 = e^{-\widetilde{Y}_1}, \qquad e^{-\widetilde{Y}_2} =
e^{i(t_2 -t_1)} \frac{(X_1 X_2)^2}{X_3},\ee 
for the Bose
superfields while 
\be G_1= F_1,\qquad G_2= F_2,\qquad F= \frac{G_1
+G_2}{2},\qquad G_3= \widetilde{F}_1,\qquad \widetilde{F}_2 = G_1 +G_2
-G_3,
\ee 
{}for the Fermi superfields. Note that by definition, $(X_1,X_2)$ are
not  single-valued and, as we shall soon see, the low-energy
Landau-Ginzburg theory is an orbifold conformal field
theory.

After some straight forward algebra, the effective superpotential
of the low-energy theory turns out to be 
\bea \mu^{-1}
{\widetilde{W}}_{eff} &=& A G_1 (X_1^2 + p X_2^2 +q X_3 +s
\frac{(X_1 X_2)^2}{X_3} +u X_1 X_2)+  \cr
&& B G_2 (X_2^2 + p' X_1^2 +q' X_3 +s' \frac{(X_1 X_2)^2}{X_3} +u' X_1 X_2)+  \\
&& C G_3 (X_3 + p'' X_1^2 +q'' X_2^2 +s'' \frac{(X_1 X_2)^2}{X_3}
+u'' X_1 X_2), \nonumber \eea 
where 
\bea
&A& = \beta_{11} +\beta_{\tilde{2} 1} +\omega_1, \cr
&B&= \beta_{22} +\beta_{\tilde{2} 2} +\omega_2, \\
&C&= \beta_{\tilde{1} \tilde{1}} -\beta_{\tilde{2}
\tilde{1}},\nonumber 
\eea 
and 
\bea 
&  p=(\beta_{12} +\beta_{\tilde{2}
2}  +\omega_2)/A,  \qquad   q=(\beta_{1 \tilde{1}} +\beta_{\tilde{2}
\tilde{1}} +\widetilde\omega)/A, 
\qquad s= e^{i(t_2 -t_1)}(\beta_{1 \tilde{2}} +\beta_{\tilde{2}
\tilde{2}} + \widetilde\omega_2)/A, & \cr
&  u= e^{-it_1 /2}(\nu_1
+\widetilde{\nu}_2 +\omega)/A,  \qquad
 p'=(\beta_{21} +\beta_{\tilde{2} 1} +\omega_1)/B, & \cr
& q'=(\beta_{\tilde{2}  \tilde{1}} +
\beta_{2 \tilde{1}} +\widetilde\omega_1)/B,\qquad  
s' =e^{i(t_2 -t_1)}(\beta_{2 \tilde{2}} +\beta_{\tilde{2}
\tilde{2}} +\widetilde\omega_2)/B, & \\
&  u'= e^{-it_1 /2}(\nu_2 +\widetilde{\nu}_2 +\omega)/B, \qquad
p''=(\beta_{\tilde{1} 1} - \beta_{ \tilde{2}
1})/C, & \cr 
&  q''=(\beta_{\tilde{1} 2}-
\beta_{\tilde{2} 2})/C,\qquad s''=e^{i(t_2 -t_1)}(\beta_{\tilde{1} \tilde{2}} -
\beta_{\tilde{2} \tilde{2}})/C, \qquad  u''= e^{-it_1
/2}(\widetilde{\nu}_1 -\widetilde{\nu}_2)/C. &\nonumber \eea 
We see that the effective superpotential is invariant under the
diagonal $\Z_2$ which sends
$$ X_1 \rightarrow \pm X_1, \qquad X_2 \rightarrow \pm X_2$$ 
while keeping $X_1 X_2$ invariant. The low-energy theory is therefore a
well-defined $\Z_2$ orbifold of the low-energy Landau-Ginzburg theory.  
Lastly, the chiral ring relations are given by
\bea
&& X_1^2 +p X_2^2 + q X_3 +s \frac{(X_1 X_2)^2}{X_3} +u X_1 X_2=0, \cr
&& X_2^2 +p' X_1^2 + q' X_3 +s' \frac{(X_1 X_2)^2}{X_3} +u' X_1 X_2=0, \\
&& X_3 +p'' X_1^2 + q'' X_2^2 +s'' \frac{(X_1 X_2)^2}{X_3}+u'' X_1
X_2 =0. \nonumber \eea

\subsubsection{A \TT\ Deformation}

Now we start with our base $\mP^1\times \mP^1$ GLSM, and add a 
chiral superfield $P$ and a
Fermi superfield $\Gamma$ carrying charge $-2$ only under $U(1)_1$, and a 
chiral superfield
$\widetilde{P}$ and a Fermi superfield $\widetilde\Gamma$ carrying
charge $-2$ only under $U(1)_2$. The model is again conformal, but the
bundle is quite different from the prior case. In this case, the
target space for the low-energy theory is the total space of
$\cO(-2)\oplus \cO(-2)$ over $\mP^1\times \mP^1$. 
We will see the
difference between the two cases reflected in the dual description.  

We take as our choice of $E$ in the definition of $\G$
\be E \equiv E_{\Gamma} = -2{\sqrt
2}(\Sigma +\widetilde\epsilon \widetilde\Sigma)P, \ee 
while in defining $\widetilde\Gamma$ we take \be \widetilde{E} \equiv
E_{\widetilde{\Gamma}} = -2{\sqrt 2}(\widetilde\Sigma + \epsilon
\Sigma)\widetilde{P}. \ee 
The vacuum solution of the GLSM is \be
\sum_i \vert \phi_i \vert^2 -2 \vert p \vert^2=  r_1,\qquad \sum_i
\vert \widetilde\phi_i \vert^2 -2 \vert \widetilde{p} \vert^2  =
r_2. \ee 
The generic vacuum has $\Sigma= \widetilde\Sigma =0$.
Now, unlike the previous example, the $D$ term equations decouple. 

As before, let us define 
$$ P = p + \sqrt{2} \theta^+ \psi_+ + \ldots, \qquad\qquad \G = \chi_-
+ \ldots,$$
and,
$$ \widetilde{P} = \widetilde{p} + \sqrt{2} \theta^+
\widetilde{\psi}_+ + \ldots, \qquad\qquad \widetilde\G = \widetilde{\chi}_-
+ \ldots. $$
{}From the various
Yukawa couplings, we see that the massless fermionic degrees of
freedom of the low-energy theory satisfy 
\bea &&
\sum_i \bar\phi_i \psi_{+i} -2\bar{p} \psi_+ =0,\qquad
\sum_i \bar{\widetilde\phi}_i \widetilde\psi_{+i}
-2\bar{\widetilde{p}}
\widetilde\psi_+ =0,\cr
&& \sum_i \bar\phi_i \chi_{-i} + \widetilde\chi_{-1} \sum_i
\bar\beta_i  \bar{\widetilde\phi}_i + \widetilde\chi_{-2} \sum_i
\bar\beta_i '  \bar{\widetilde\phi}_i -2\bar{p} \chi_-
-2\bar\epsilon \bar{\widetilde{p}} \widetilde\chi_-=0, \\
&& \sum_i \bar{\widetilde\phi}_i \widetilde\chi_{-i} + \chi_{-1}
\sum_i \bar\alpha_i  \bar\phi_i + \chi_{-2} \sum_i \bar\alpha_i '
\bar\phi_i  -2 \bar{\widetilde\epsilon} \bar{p} \chi_- -2
\bar{\widetilde{p}}\widetilde\chi_- =0. \nonumber \eea 
The dual theory has a classical Lagrangian with the K\"ahler terms given in
the $\mP^1 \times \mP^1$ example along with the additional terms \bea
\widetilde{L} &=& \ldots  +\frac{i}{8} \int d^2 \theta ~ \frac{Y -
\bar{Y}}{Y + \bar{Y}}
\partial_- (Y + \bar{Y})
+\frac{i}{8} \int d^2 \theta~ \frac{\widetilde{Y} -
\bar{\widetilde{Y}}}{\widetilde{Y} + \bar{\widetilde{Y}}}
\partial_- (\widetilde{Y} + \bar{\widetilde{Y}}) \\
&& - \frac{1}{2} \int d^2 \theta~{\bar{\mathcal{F}}} {\mathcal{F}} -
\frac{1}{2} \int d^2 \theta~{\bar{\mathcal{\widetilde{F}}}}
{\mathcal{\widetilde{F}}}, \nonumber 
\eea 
where the duality map is (again, modulo fermion bilinears)
\bea  
&& \bar{P} P ~=~\frac{1}{2} (Y+\bar{Y}),\qquad
 \bar{P} ( {\buildrel \leftrightarrow \over \partial_-} -2i V_1 )
P =-\frac{1}{4} \partial_- (Y -\bar{Y}), \cr
&& \bar{\widetilde{P}} \widetilde{P} ~=~\frac{1}{2} (\widetilde{Y}+
\bar{\widetilde{Y}}),\qquad
 \bar{\widetilde{P}} ( {\buildrel \leftrightarrow \over \partial_-} -2i V_2)
\widetilde{P} =-\frac{1}{4} \partial_- (\widetilde{Y}
-\bar{\widetilde{Y}}), \nonumber 
\eea 
and 
\be \bar\Gamma
=\mathcal{F},\qquad \bar{\widetilde\Gamma} =\mathcal{\widetilde{F}}.
\ee 
The perturbative dual superpotential is given by 
\bea
\widetilde{W} &=& - \frac{i\Upsilon_1}{4} (\sum_i  Y_i -2Y + it_1)
 - \frac{i\Upsilon_2}{4} (\sum_i  {\widetilde{Y}}_i -2\widetilde{Y} + it_2)
\cr
&& - \frac{1}{\sqrt 2} (\sum_i E_i {\mathcal{F}}_i
 +\sum_i {\widetilde{E}}_i \widetilde{\mathcal{F}}_i +E {\mathcal{F}}
+\widetilde{E} {\mathcal{\widetilde{F}}}), \eea 
where again we write the last term in the form
\be -\int d \theta^+\Sigma (F_1 + F_2 -2F -2 \epsilon \widetilde{F})
-\int d \theta^+\widetilde\Sigma ({\widetilde{F}}_1 + {\widetilde{F}}_2
-2\widetilde{F} -2 \widetilde\epsilon F), 
\ee 
where $F = P{\mathcal{F}}$ and $\widetilde{F} = \widetilde{P}
{\mathcal{\widetilde{F}}}$.

The exact dual superpotential is then given by the lengthy expression 
\bea \label{evenbiggersup}\widetilde{W}
&=& -\frac{i\Upsilon_1}{4} (\sum_i  Y_i -2Y + it_1)
-\frac{i\Upsilon_2}{4} (\sum_i  \widetilde{Y}_i -2\widetilde{Y} +
it_2)
-\Sigma (\sum_i F_i -2F -2 \epsilon \widetilde{F}) \cr
&&-\widetilde\Sigma (\sum_i {\widetilde{F}}_i -2\widetilde{F} -2
\widetilde\epsilon F) +\mu (\sum_{ij} \beta_{ij} F_i e^{-Y_j} +
\beta_{\tilde{i} \tilde{j}} \widetilde{F}_i e^{-\widetilde{Y}_j}
+\beta_{i \tilde{j}} F_i e^{-\widetilde{Y}_j}
+ \beta_{\tilde{i} j} \widetilde{F}_i e^{-Y_j}) \cr
&& +2\mu F(\omega e^{-Y} + \widetilde\omega e^{-\widetilde{Y}} +
\sum_i \omega_i  e^{-Y_i} +\sum_i \widetilde\omega_i
e^{-\widetilde{Y}_i}) +\mu \sum_i (\nu_i F_i + \widetilde\nu_i
\widetilde{F}_i) e^{-Y}
\cr
&& +2\mu \widetilde{F}(\omega' e^{-Y} + \widetilde\omega'
e^{-\widetilde{Y}} + \sum_i \omega_i'  e^{-Y_i} +\sum_i
\widetilde\omega_i'  e^{-\widetilde{Y}_i}) +\mu \sum_i (\nu_i' F_i
+ \widetilde\nu_i' \widetilde{F}_i) e^{-\widetilde{Y}}. 
\eea 
The
vacuum solution is given by \be Y_1 + Y_2
-2Y=-it_1,\qquad {\widetilde{Y}}_1 + {\widetilde{Y}}_2
-2\widetilde{Y}=-it_2, \ee 
and 
\be F_1 + F_2 -2F -2 \epsilon
\widetilde{F}=0,\qquad {\widetilde{F}}_1 + {\widetilde{F}}_2
-2\widetilde{F} -2 \widetilde\epsilon F =0. 
\ee 
We solve these constraints in the following way 
\bea X_1= e^{-Y_1 /2}, \qquad X_2=e^{-Y_2 /2}, \qquad e^{-Y} =e^{-it_1
/2} X_1 X_2,
\cr
\widetilde{X}_1=e^{-\widetilde{Y}_1 /2}, \qquad \widetilde{X}_2=  
e^{-\widetilde{Y}_2 /2}, \qquad e^{-\widetilde{Y}} =e^{-it_2 /2} \widetilde{X}_1
\widetilde{X}_2, \eea 
for the bosonic superfields. For the fermionic superfields, we define
\bea G_1= F_1, \qquad G_2= F_2, \qquad F= \frac{1}{2(1- \epsilon \widetilde\epsilon)}
(G_1 +G_2 -\epsilon(\widetilde{G}_1 +\widetilde{G}_2)),
\cr
\widetilde{G}_1 = \widetilde{F}_1,\qquad \widetilde{G}_2 =
\widetilde{F}_2,\qquad  \widetilde{F} = \frac{1}{2(1- \epsilon
\widetilde\epsilon)} (\widetilde{G}_1 +\widetilde{G}_2
-\widetilde\epsilon (G_1 +G_2)). \eea
Again,  $(X_1,X_2,\widetilde{X}_1,\widetilde{X}_2)$ are not
single-valued, and the low-energy theory will be an orbifold.

So the low-energy theory has the effective superpotential 
\bea\label{bigpot}
\mu^{-1} {\widetilde{W}}_{eff} &=& A G_1 (X_1^2 + p X_2^2 +q
\widetilde{X}_1^2 +s \widetilde{X}_2^2
+u X_1 X_2 +v \widetilde{X}_1 \widetilde{X}_2)+  \cr
&& B G_2 (X_2^2 + p' X_1^2 +q' \widetilde{X}_1^2 +s'
\widetilde{X}_2^2
+u' X_1 X_2 + v' \widetilde{X}_1 \widetilde{X}_2)+  \cr
&& \widetilde{A} \widetilde{G}_1 (\widetilde{X}_1^2 + \widetilde{p}
X_1^2 +\widetilde{q} X_2^2 +\widetilde{s} \widetilde{X}_2^2
+\widetilde{u} X_1 X_2
+\widetilde{v} \widetilde{X}_1 \widetilde{X}_2) +\cr
&& \widetilde{B} \widetilde{G}_2 (\widetilde{X}_2^2 + \widetilde{p}'
X_1^2 +\widetilde{q}' X_2^2 +\widetilde{s}' \widetilde{X}_1^2
+\widetilde{u}' X_1 X_2 +\widetilde{v}' \widetilde{X}_1
\widetilde{X}_2) \eea 
where 
\bea A= \beta_{11}
+\kappa\omega_1 -\widetilde\epsilon \kappa\omega_1', \nonumber
\\
B= \beta_{22} +\kappa\omega_2 -\widetilde\epsilon \kappa\omega_2', \\
\widetilde{A}= \beta_{\tilde{1} \tilde{1}} +\kappa
\widetilde\omega_1' -\epsilon \kappa \widetilde\omega_1,
\nonumber \\
\widetilde{B}= \beta_{\tilde{2} \tilde{2}} +\kappa
\widetilde\omega_2' -\epsilon \kappa \widetilde\omega_2, \nonumber
\eea 
and $\kappa = 1/ (1- \epsilon \widetilde\epsilon)$. All the remaining
parameters appearing in~\C{bigpot}\ can be
expressed in terms of the parameters appearing
in~\C{evenbiggersup}. For example, 
$$p =(\beta_{12} +\kappa\omega_2  -\widetilde\epsilon
\kappa\omega_2')/A. $$  
We will not list the remaining lengthy expressions since they are not
particularly enlightening. 

The
effective superpotential is invariant under a $\Z_2\times \Z_2$
symmetry sending
$$X_1 \rightarrow \pm X_1, \qquad X_2 \rightarrow \pm X_2$$ 
holding the product $X_1 X_2$ invariant, and also sending 
$$\widetilde{X}_1 \rightarrow \pm
\widetilde{X}_1, \qquad \widetilde{X}_2 \rightarrow \pm
\widetilde{X}_2$$ holding the product $\widetilde{X}_1
\widetilde{X}_2$ invariant. Hence the IR theory is a $\Z_2\times \Z_2$
orbifold of the Landau-Ginzburg theory. This is quite different from the
previous example.

Finally, the chiral ring relations are given by
\bea && X_1^2 + p X_2^2 +q \widetilde{X}_1^2 +s \widetilde{X}_2^2
+u X_1 X_2 +v \widetilde{X}_1 \widetilde{X}_2 =0,\cr
&& X_2^2 + p' X_1^2 +q' \widetilde{X}_1^2 +s' \widetilde{X}_2^2
+u' X_1 X_2 + v' \widetilde{X}_1 \widetilde{X}_2 =0, \cr
&& \widetilde{X}_1^2 + \widetilde{p} X_1^2 +\widetilde{q} X_2^2
+\widetilde{s} \widetilde{X}_2^2 +\widetilde{u} X_1 X_2
+\widetilde{v} \widetilde{X}_1 \widetilde{X}_2 =0,\cr
&& \widetilde{X}_2^2 + \widetilde{p}' X_1^2 +\widetilde{q}' X_2^2
+\widetilde{s}' \widetilde{X}_1^2 +\widetilde{u}' X_1 X_2
+\widetilde{v}' \widetilde{X}_1 \widetilde{X}_2 =0. \eea

\subsection{Models With ${\rm rk}(\V)> {\rm rk}(T\M)$}

So far in all our examples, we have considered cases where we have an
equal number of Fermi and chiral superfields. At special loci in their
parameter spaces, many of these models enjoy enhanced \TT\
supersymmetry. These models flow in the IR to non-linear sigma models
with ${\rm rk}(\V)= {\rm rk}(T\M)$. We now turn to cases where the
number of Fermi superfields is greater than the number of chiral
superfields; in the IR sigma model, the bundles satisfy 
${\rm rk}(\V)> {\rm rk}(T\M)$. 

On general grounds, we expect the low-energy dual theory to be quite
different from the previous examples. As in our prior discussion, to
find the  low-energy theory,
we need to solve the constraints
$$\sum_{i=1}^N Q_i Y_i =-it, \qquad \sum_{a=1}^M Q_a F_a =0, $$
where now $M>N$. We are left with $N-1$ $Y$ variables, and $M-1$ Fermi
superfields. A generic non-perturbative superpotential  of the form $\mu \sum_{ia}
\beta_{ia} F_a e^{-Y_i}$ imposes a further $M-1$ constraints on the
$Y$ fields -- one for each light Fermi superfield. Since $M>N$,
generically the only solution is $Y_i \rightarrow\infty$ for all
$i$.\footnote{If we were to consider a superpotential, the situation
is likely to be quite different. There should then be many examples
with ${\rm rk}(\V)> {\rm rk}(T\M)$ where the dual theory flows to an
interacting SCFT. This illustrates some of the subtleties we expect to
encounter when attempting to dualize with a tree-level
superpotential.} This is clearly quite different from the  ${\rm rk}(\V)= {\rm
rk}(T\M)$ cases.

However, there can be interesting non-generic cases where we get non-trivial vacuum
solutions of the theory. This happens when some of the  
vacuum solution equations are linearly dependent. There can then be
solutions for finite values of the $Y$ fields, even when the rank of
the left-moving vector bundle is 
greater than the rank of the tangent bundle! We now consider two
examples which illustrate two possible situations: in the first, the
vacuum manifold consists of isolated points, while in the second, the
vacuum manifold is a geometric surface.

\subsubsection{A Model With Isolated Vacua}
 
Let us describe an example where generically we find isolated points as the 
vacua of the theory. In the GLSM, we take $3$ chiral superfields 
$\Phi_1, \Phi_2$ and $\Phi_3$ carrying gauge charges $1,1$ and $-2$,  
respectively under a single $U(1)$ gauge group. 
This model is conformal and flows in the IR to a NLSM with a target
space given by the total space of $\cO(-2)$ over $\mP^1$. 

We also take  $6$ Fermi superfields, 
$\Gamma_1,\ldots,\Gamma_6$, with gauge charges $(1,1,1,-1,-1,-1)$,      
respectively. To each $\Gamma_i$, there is a concomitant $E$ given by
\bea && E_1 = E_3 = {\sqrt 2} \Sigma \Phi_1, \qquad E_2 = {\sqrt 2} \Sigma 
\Phi_2,  \cr
&& E_4 = E_5 = E_6 = -{\sqrt 2} \Sigma \Phi_3 (\Phi_1 + \Phi_2). 
\eea         
However our analysis goes through for any (generic) $E_4, E_5, E_6$ 
satisfying $E_4 =E_5 =E_6$. The only constraint
on the choice of $E_4$ comes from demanding charge conservation and
non-singularity. Our choice of $E_4, E_5, E_6$ is just a 
particular one chosen to illustrate the general vacuum structure.    
   
The vacuum solution of the GLSM requires solving
\be \vert \phi_1 \vert^2  +\vert \phi_2 \vert^2 -2 \vert \phi_3 \vert^2 =  
r,\ee
while the analysis of the massless fermions follows straightforwardly from 
the Yukawa couplings as done in the previous examples. The perturbative
dual theory is given by
\be \widetilde{L} = \frac{i}{8} \sum_{i=1}^3 \int d^2 \theta
~ \frac{Y_i - \bar{Y}_i}{Y_i + \bar{Y}_i}
\partial_- (Y_i + \bar{Y}_i) 
- \frac{1}{2} \sum_{a=1}^6 \int d^2 \theta~\frac{\bar{F}_a F_a}
{\vert Y_{\mathcal{E}_a} + \bar{Y}_{\mathcal{E}_a} \vert^2} + 
\int d \theta^+ \widetilde{W} + {\rm h.c.},
\ee
where 
$${\mathcal{E}}_1= {\mathcal{E}}_3 = \Phi_1, \qquad {\mathcal{E}}_2=
\Phi_2, \qquad {\mathcal{E}}_4= {\mathcal{E}}_5 = {\mathcal{E}}_6 =
\Phi_3 (\Phi_1 + \Phi_2),$$ 
and
\be  \widetilde{W} = - \frac{i\Upsilon_1}{4} \left(Y_1 +Y_2 -2Y_3 + it\right) 
- \frac{\Sigma}{\sqrt 2} \left(\sum_{i=1}^3 F_i -\sum_{i=4}^6 F_i\right). 
\ee
The duality maps are the standard ones, and have not been written down for brevity. 
The non-perturbative dual superpotential is given by
\be \widetilde{W}_{non-pert} = \mu \sum_{ia} \beta_{ia} F_a e^{-Y_i}.\ee   

The $\beta$ parameters are highly constrained because of our symmetric
choice of $E_a$. These symmetries imply that
\bea \label{paramlist} & \beta_{11} =\beta_{22} 
=\beta_{13} \equiv a, \qquad &\beta_{12} =\beta_{21} =\beta_{23} \equiv b, \cr
& \beta_{31}=\beta_{33} \equiv c, \qquad &
\beta_{14} =\beta_{15} =\beta_{16} \equiv p, \cr
& \beta_{24} =\beta_{25} =\beta_{26} \equiv q, \qquad 
& \beta_{34} =\beta_{35} =\beta_{36} \equiv s. \eea 
We also set $d=\beta_{32}$.

Now we can determine the vacuum structure. We solve the 
constraint $Y_1 + Y_2 -2Y_3 =-it$ by setting 
$$ X_1= e^{-Y_1 /2}, \qquad  X_2= e^{-Y_2 /2}$$ 
so that 
$$e^{-Y_3} = e^{-it /2} X_1 X_2.$$ 
The other constraint yields $F_1 +F_2 +F_3 =F_4 +F_5 +F_6$. This gives the effective 
superpotential       
\bea \label{redun}
\mu^{-1} \widetilde{W}_{eff} &=& (F_1+F_3) \left[(a+p)X_1^2 +(b+q)X_2^2 +e^{-it /2} 
(c+s)X_1 X_2\right]+ \cr
&& F_2 \left[(b+p)X_1^2 +(a+q)X_2^2 +e^{-it /2}(d+s)X_1 X_2\right]. \eea
Note that only $F_1,F_2$ and $F_3$ are required to specify the effective 
superpotential, because of the symmetries of the $E_a$. We also
clearly see from~\C{redun}\ that the vacuum equations for $F_1,F_3$ 
are dependent. 

So the vacua are given by the solutions of
\bea (a+p)X_1^2 +(b+q)X_2^2 +e^{-it /2}(c+s)X_1 X_2 =0,\\
(b+p)X_1^2 +(a+q)X_2^2 +e^{-it /2}(d+s)X_1 X_2 =0. \nonumber \eea    
We have two equations for the two independent variables $X_1, X_2$. For
generic choices of $E_a$, we get isolated vacua. Also, the 
low-energy theory is invariant under
$$X_1 \rightarrow \pm X_1, \qquad X_2 \rightarrow \pm X_2$$ 
with $X_1 X_2$ held invariant. The low-energy theory is again a $\Z_2$ orbifold 
SCFT. We should stress that we assumed that the parameters of
\C{paramlist}\ are generically non-zero (but subject to symmetry
constraints). This is actually a worse case scenario; if some of the
parameters vanish, we would find additional vacua.

\subsubsection{A Model With a Continuum of Vacua}

Now we consider an example where we get a geometric surface, and not isolated
points, as the vacuuum manifold. The field content of the GLSM is exactly 
as in the previous example,  but now we take
\be E_1 =E_2 =E_3 ={\sqrt 2}\Sigma (\Phi_1 + \Phi_2), \qquad E_4 =E_5 =E_6 
=-{\sqrt 2}\Sigma \Phi_3 (\Phi_1 + \Phi_2). \ee 
Again, the analysis of the vacuum structure really only relies on the
relation 
$$E_1 =E_2 =E_3, \qquad E_4 =E_5 =E_6, $$
and we have just made a special choice. 

We go directly to the analysis of the non-perturbative superpotential
\be \widetilde{W}_{non-pert} = \mu \sum_{ia} \beta_{ia} F_a e^{-Y_i}.\ee 
{}From the symmetries of the $E_a$, we obtain the effective 
superpotential  
\be \mu^{-1} \widetilde{W}_{eff} = (F_1 +F_2 + F_3) \left[(\widetilde{a}+ 
\widetilde{p}) X_1^2 +(\widetilde{b}+\widetilde{q}) X_2^2 +e^{-it /2} 
(\widetilde{c}+\widetilde{s}) X_1 X_2\right]. \ee
So the vacuum is given by the solution of
\be \label{vaceq} (\widetilde{a}+ 
\widetilde{p}) X_1^2 +(\widetilde{b}+\widetilde{q}) X_2^2 +e^{-it /2} 
(\widetilde{c}+\widetilde{s}) X_1 X_2 =0.\ee
Thus there is only one equation constraining the two independent variables, 
$X_1$ and $X_2$. The vacuum is a one (complex) dimensional
surface~\C{vaceq}\ in $(X_1,X_2)$ space. The effective field theory is a $\Z_2$
orbifold SCFT as before. However, the low-energy theory is itself a
non-linear sigma model. There is an issue we have not yet addressed
in this model; namely, the kinetic terms become singular in this model,
and all models where the effective potential has flat directions. We
now turn to this issue in the context of ${\rm rk}(\V)<{\rm rk}(T\M)$
models for which this situation is generic.

\subsection{Models With ${\rm rk}(\V)< {\rm rk}(T\M)$}

The last class of examples have ${\rm rk}(\V)< {\rm rk}(T\M)$. The
dual descriptions are generically quite different from any of the
prior cases. The reason is a matter of counting constraints. The
vacuum is determined by solving   
the constraints
$$\sum_{i=1}^N Q_i Y_i =-it, \qquad \sum_{a=1}^M Q_a F_a =0, $$
where now $N>M$. We are left with $N-1$ $Y$ variables, and $M-1$ Fermi
superfields. A generic non-perturbative superpotential  of the 
form $\mu \sum_{ia} \beta_{ia} F_a e^{-Y_i}$ imposes a further 
$M-1$ constraints on the
$Y$ fields, as before. However, this potential must have flat
directions corresponding to excitations of the remaining $N-M$ light
$Y$ fields. The low-energy theory is not a Landau-Ginzburg theory, but
a \ZT\ non-linear sigma model with the vacuum manifold as a target space. 

We need to examine the metric on this target space. After dualizing a
single charged chiral field, we see from~\C{kinetic}\ that the dual
theory, parametrized by $Y$, has a K\"ahler metric with K\"ahler
potential 
\be K(Y,\bar{Y}) = (Y+\bar{Y}) \ln (Y+\bar{Y}) \quad \Rightarrow \quad
g_{y\bar{y}} = {dy d\bar{y} \over (y+\bar{y})}. 
\ee
Recall that ${\rm Re} (Y) \geq 0$ so the metric singularity at $Y=0$
is at finite distance. How is this singularity resolved?  

The situation is actually quite similar to string theory on the
two-dimensional black-hole solution found
in~\cite{Witten:1991yr}. We expect this metric to be accompanied by a
non-trivial dilaton diverging at $Y=0$. To see that this is the case,
we recall that under T-duality, the dilaton is usually shifted by a
metric 
factor $g_{\varphi\varphi}$ where $\varphi$ is
the isometry direction~\cite{Buscher:1988qj}. 

In our case, the metric factor is $\ln(y+\bar{y})$ but there is a
subtlety involving the gauge field. To see how this works, consider
the first order action
\be \label{dileqn} S = \int d^2 \xi~\left[ -\frac{1}{4\rho^2}
{\sqrt \gamma} \gamma^{\mu \nu}
B_{\mu} B_{\nu} +\epsilon^{\mu \nu} B_{\mu}(\partial_{\nu} \varphi +A_{\nu})
+{\sqrt \gamma} R^{(2)} \Phi\right]\ee  
where $B\m$ is a 1-form, and $\gamma_{\m\n}$ is the world-sheet
metric. Integrating out $B_\m$ generates the dilaton shift~\cite{Buscher:1988qj}
\be \Phi \rightarrow \Phi -\frac{1}{2} {\rm ln} (-g_{\varphi\varphi})
= \Phi +\frac{1}{2} {\rm ln}(4\rho^2).\ee
If we integrate out $A$, we expect an analogous shift of the
dilaton but with respect to the dual metric $g_{\vartheta\vartheta}  
= 1/g_{\varphi\varphi}$, 
\be
\Phi \rightarrow \Phi - \frac{1}{2} {\rm ln}(4\rho^2).
\ee 
These two shifts should cancel for this model as also argued
in~\cite{Hori:2000kt}. 

In the general case where we have many chiral fields with charge
$Q_i$, it appears that the shift is given by
\be\label{shifted}
\Phi \rightarrow \Phi  -\frac{1}{2} \sum_i {\rm ln}
(-g_{\varphi_i\varphi_i}) 
-\frac{1}{2} {\rm ln} (- \sum_i {Q_i \over g_{\varphi_i\varphi_i}}). 
\ee
With many $U(1)$ factors, there are  just more terms like the last one
appearing in~\C{shifted}. 
This also makes sense from the low-energy
target space perspective: We are T-dualizing  one phase for each chiral
superfield but each gauged $U(1)$ kills one combination of chiral
superfields, reducing the overall dilaton shift. 

Therefore, whenever we have a non-trivial vacuum manifold in the dual
description, we expect a corresponding dilaton diverging at the
location of the metric singularitites. This is a \ZT\ generalization
of the duality between minimal models and a sigma model
dual with diverging dilaton (see, for example,~\cite{Giveon:1994fu}).  

\subsubsection{A Surface in $\mP^3$}

To conclude our discussion, we will examine two models based on the
examples of~\cite{Basu:2003bq}. For the first
case, the target space geometry is a hypersurface in $\mP^3$. Our
basic GLSM has $4$ superfields of charge $1$ under a single $U(1)$
gauge symmetry. We take $1$ Fermi superfield $\G$ with charge
$2$. Associated to $\G$ is a choice of $E$, and we consider the case
\be
E= \alpha^{ij} \Phi_i \Phi_j. 
\ee
Note there is no $\Sm$ in $E$ so the constraint $E=0$ restricts us to a
hypersurface, $\M$,  in $\mP^3$. 

The low-energy theory is quite beautiful. There are no left-moving
fermions at all, but ${\rm ch}_2(\V)=0$ as described in
section~\ref{nosigma}. Whether supersymmetry is broken in the
low-energy theory can be tested by computing 
${\rm Ind} (\bar\p)$ which counts (with sign) the number of
supersymmetric ground states. First we note that the hypersurface $\M$
has Chern classes, 
$$ c_1(\M) = 2, \qquad c_2(\M)=2. $$
The index is given by
\bea  {\rm Ind} (\bar\p) &=& \int_{\M} {\rm td}(\M), \cr
&=& \int_{\M} \left( {c_1^2+c_2\over 12}\right) 
=  {1\over 2}\int_{\mP^3} J^2 \wedge 2J = 1, \eea
where $J$ is the K\"ahler form of $\mP^3.$ So supersymmetry is
unbroken, and we generically expect a single vacuum state with mass
gap.

Now we turn to the dual description. We want to determine whether
there are non-perturbative corrections to the dual superpotential. Let
us take a particular choice of $E$, say $E =\Phi_4^2$.\footnote{This
  is actually a degenerate section of $\cO(2)$ since $E = dE=0$ has a
  solution. Fortunately, this will not affect the subsequent
  analysis since we can always perturb $E$ by a small amount with no
  real change in the analysis.} 
To perform an
instanton zero mode analysis, we need the following relevant terms in
the action,
\be \label{somerel}i {\bar{\chi}}_- D_+ \chi_- - \vert \phi_4^2 \vert^2
-2 (\phi_4 {\bar{\chi}}_- \psi_{+4}
+ {\bar{\phi}}_4 {\bar{\psi}}_{+4} \chi_-) + \ldots. \ee 
A BPS instanton requires setting $\phi_4=0$. We must embed the
instanton in some other $\phi$, say $\phi_1$. In this (and any other
BPS configuration), all the potential terms in~\C{somerel} vanish and
we can exactly determine the fermion zero modes: there are $4$
right-moving zero modes. For $\psi_{+1}$, the zero mode is given by  
$$\mu^0 = \pmatrix{ {\bar\psi}_{+1}^0 \cr \lambda_-^0\cr} =
\pmatrix{-{\sqrt 2} ({\bar{D}}_1  + i{\bar{D}}_2){\bar\phi_1} \cr
  D-F_{12}\cr}, $$
while $\psi_{+i}^0 = \bar\phi_1$ for $i=2,3,4$. For the left-mover,
there is a single zero mode $\chi_{-}^0 = \phi_1^2$. Any two-point
function can only absorb two zero modes. Quantum effects could, in
principle, lift zero modes but since the remaining $3$ zero modes are
right-moving, they must remain massless. These zero modes kill the
correlation function. We conclude that there are no non-perturbative
corrections to the dual superpotential. This is very similar to the
argument in~\cite{Basu:2003bq}. 

The exact dual Lagrangian is therefore given by
\bea \widetilde{L} &=&   \frac{i}{8} \int d^2 \theta
~ \sum_i \frac{Y_i -{\bar{Y}}_i}{Y_i + {\bar{Y}}_i}
\partial_- (Y_i + {\bar{Y}}_i)
 - 2\int d^2 \theta~\frac{\bar{F} F} {(Y_4 + {\bar{Y}}_4)^2} \\
&& - \left[\frac{i}{4} \sum_i \int d \theta^+ Y_i \Upsilon
- \frac{1}{\sqrt 2} \int d\theta^+ F + {\rm h.c.}\right]. \non \eea
The vacuum solution is obtained by setting
\be \sum_i Y_i = -it. \ee
Integrating out $F$ generates a potential for $Y_4$ of the form
\be\label{bospot} V \sim  |y_4+\bar{y}_4|^2.  \ee
To find the vacuum manifold, we must set $Y_4=0$. The low-energy
theory is therefore a  non-linear sigma model on a two-dimensional
target space with metric determined by solving these constraints. There are
no left-moving fermions at all, and the space has metric singularities
at loci where the dilaton diverges. From our analysis of the original
model, we can predict that supersymmetry is unbroken and that the
index is $1$. It should be possible to verify these predictions
directly in the low-energy dual theory. It may also be possible to
relate the dual theory to a construction involving \ZT\ gauged WZW models.

Next we consider a special case where the potential term $|E|^2$
itself has flat directions. A simple specific choice  is 
$E = \Phi_1 \Phi_2$. The relevant terms in the action are, 
\be i {\bar{\chi}}_- D_+ \chi_- - \vert  \phi_1 \phi_2 \vert^2
-( \phi_2 {\bar{\chi}}_- \psi_{+1}
+ \phi_1 {\bar{\chi}}_- \psi_{+2}
+ {\bar{\phi}}_2 {\bar{\psi}}_{+1} \chi_-
+{\bar{\phi}}_1 {\bar{\psi}}_{+2} \chi_-)+\ldots. \ee
We argue that there are no non-perturbative corrections to the dual
superpotential in the following way: perturb $E$ by an infinitessimal
amount so the resulting section of $\cO(2)$ is generic. By our
previous analysis, the non-perturbative superpotential must
vanish. Since the dual superpotential varies holomorphically with
the deformation parameter, it cannot depend on the parameter at
all. Therefore, there are no corrections for this case. The only
difference from the prior case is that on integrating out $F$, we
obtain a potential 
$$ V \sim |(y_1 + \bar{y}_1)(y_2 + \bar{y}_2)| $$
which has a different structure from~\C{bospot}.

\subsubsection{A Bundle Over $\mP^3$}

Let us take the same model just discussed but consider a different
choice for $E$ where
\be E = \Sm \E = \Sm \alpha^{ij} \Phi_i \Phi_j. \ee
Because of the appearance of $\Sm$ in $E$, we expect the low-energy
theory to contain a left-moving fermion which is a section of $\cO(2)$
over $\mP^3$. There is a subtlety here worth explaining: the Yukawa
couplings described in section~\ref{withasm}\ would seem to give mass
to the single $\chi_{-}$ fermion in the UV. How can there be a
low-energy left-moving fermion at all? The resolution of this puzzle
goes as follows. The $\Sm$ superfield becomes massive when $\E \neq
0$, and can be integrated out. However, on performing this
integration, we see that $\chi_{-}$ does not pick up a mass but picks
up a derivative coupling. It therefore survives as a light degree of
freedom as required by consistency of the low-energy theory. 

We count the number of supersymmetric vacua in this theory (weighted by
signs) by evaluating the Witten index, 
\be {\rm Tr} (-1)^F = \sum_{p,m} (-1)^{p+m} \, h^{p}
(\M,\wedge^m \V), \ee
where $\V = \cO(2)$. This is easily done. For the sector with no
excited left-moving fermion ($m=0$), the only contribution comes from
$\bar\p$-cohomology which consists only of constant functions so
$h^0(\mP^3)=1$. For the other case where $m=1$, the only contribution
comes from $h^0(\mP^3, \cO(2)) = 10$. In total, there are a net $9$
fermionic ground states. Supersymmetry is unbroken. 

Now we turn to the dual theory.  
It is easy to see that there are no non-perturbative corrections to
the superpotential. In any instanton background, there are always $3$
right-moving zero modes that cannot be paired. These zero modes kill
any instanton contributions. The resulting superpotential is,
\be \label {Ex2} {\widetilde{W}}_{exact} =
-\frac{i \Upsilon}{4}(\sum_i Y_i + it)
+ \frac{1}{\sqrt 2} \Sigma F. \ee
The only resulting constraint is $\sum_i Y_i =-it$. The $\Sm F$
coupling gives a mass to $\Sm$ so the low-energy
theory is again a  non-linear sigma model with no effective
superpotential. We predict that
supersymmetry is unbroken in this theory.

\section*{Acknowledgements}

It is our pleasure to thank P.~Aspinwall, J.~Distler, J.~Harvey,
K.~Hori, D.~Kutasov,
R.~Plesser, and M.~Stern for helpful discussions.
We would particularly like to thank S.~Katz for useful suggestions.

A.~A. sends profound thanks to the wonderful Theory Group at TIFR
for spectacular hospitality in Bombay, as well as to the Math. Dept.
at TIFR, the Theory Group at HRI in Allahabad, and Gokul in Colaba.
A.~A. and S.~S. would also like to thank the Aspen Center for Physics 
for hospitality while this work was completed.

The work of A.~A. was supported by a Junior Fellowship from the
Harvard Society of Fellows, as well as a Visiting Fellowship at the Tata
Institute of Fundamental Research.
The work of A.~B. is supported in part by  NSF Grant No.
PHY-0204608.
The work of S.~S. is supported in part by NSF CAREER Grant No.
PHY-0094328, and by the Alfred P. Sloan Foundation.

\appendix
\section{Expressing $(2,2)$ Theories in $(0,2)$ Notation}
\label{reduction}

In this Appendix, we express a $(2,2)$ GLSM in terms of \ZT\
fields. Our starting point is the $(2,2)$ Lagrangian describing a
chiral field, $\Phi$,  and the gauge field $V$ with
field strength $\Sm$,
\be L = \int d^4 \theta ~\bar{\Phi} e^{2 Q V} \Phi
- \frac{1}{4 e^2} \int d^4 \theta ~\bar{\Sigma} \Sigma -
\left( \frac{it}{2 {\sqrt 2}} \int d^2 \widetilde{\theta} ~\Sigma + \rm{h.c.}\right).\ee
The gauge coupling constant is given by $e$, while $t= ir + \frac{\theta}{2\pi}$ is the
complexified Fayet-Iliopoulos parameter. We also use the short hand,
$$ d^2 \widetilde{\theta}= d \theta^+ d {\bar{\theta}}^-.  $$
To obtain a \ZT\ Lagrangian, we just need to integrate out $\t^-, \tb^-$
which we can do by noting
\bea  L_\Phi &=& \int d^4 \theta ~\bar{\Phi} e^{2 Q V} \Phi \\
&=& - \int d^2 \theta ~{\bar{\D}}_- \D_- \left( \bar{\Phi} e^{2 Q V}
\Phi \right).
\eea
Next we reduce this expression to a \ZT\ Lagrangian by explicitly
applying the supercovariant derivatives,
\bea
L_\Phi = - \int d^2 \theta & ~\Big[  & 2Q ~({\bar{\D}}_- {\bar{\Phi}}) ( \D_- V)
e^{2 Q V} \Phi + 2Q ~{\bar{\Phi}} ( {\bar{\D}}_-  \D_- V) e^{2 Q V} \Phi \non \\
& & -4 Q^2 ~{\bar{\Phi}} ( \D_- V) ( {\bar{\D}}_- V) e^{2 Q V} \Phi
+ ({\bar{\D}}_- {\bar{\Phi}}) e^{2 Q V} (\D_- \Phi) \\
& & + 2Q ~{\bar{\Phi}} ( {\bar{\D}}_- V) e^{2 Q V} (\D_- \Phi) +
{\bar{\Phi}} e^{2 Q V} ( {\bar{\D}}_-  \D_- \Phi) ~\Big]
\vert_{\theta^- = {\bar{\theta}}^- =0}. \non
\eea
{}From now on for brevity, we will not explicitly write
$\theta^- = {\bar{\theta}}^- =0$. This final reduction will always be
implied. Let us reduce term by term to \ZT\ superspace.

\detail{Term 1}

The first term to consider is
\be  - 2Q \int d^2 \theta~({\bar{\D}}_- {\bar{\Phi}}) ( \D_- V)
e^{2 Q V} \Phi \ee
We use that the results of section~\ref{ttsusy}\ to write
\be {\bar{\D}}_- {\bar{\Phi}} = {\sqrt 2} ({\bar{\psi}}_- -
 {\sqrt 2} \bar{\theta}^+
 \bar{F} + i \theta^+ {\bar{\theta}}^+  \partial_+  {\bar{\psi}}_- )
= {\sqrt 2} e^{- Q \Psi} {\bar{\Gamma}} + 2 \theta^+ \bar{E}  \ee
where $\Gamma$ is a charged \ZT\ Fermi superfield satisfying
$${\bar{\cal{D}}}_+\Gamma =
{\sqrt 2} E.$$
Also $\Psi = \theta^+ {\bar{\theta}}^+ A_+ $. We also recall that
\be \D_- V = - {\sqrt 2} {\bar{\theta}}^+ \Sigma_0, \ee
where
$\Sigma_0 = \Sigma \vert_{\theta^- = {\bar{\theta}}^- =0}
$.\footnote{In section~\ref{ttsusy}, we used the notation $\Sm^{(0,2)}$
for the $\theta^- = {\bar{\theta}}^- =0$ component of the $(2,2)$
chiral field $\Sm$. For notational simplicity, here we just use $\Sm_0$.}
Finally note that the uncharged field $\Phi\vert_{\theta^- =
  {\bar{\theta}}^- =0}$ satisfies
$${\DB}_+ \Phi\vert_{\theta^- = {\bar{\theta}}^- =0} =0,$$ and is given by
\be \Phi\vert_{\theta^- = {\bar{\theta}}^- =0} =
\phi + {\sqrt 2} \theta^+ \psi_+ -i \theta^+ {\bar{\theta}}^+
\partial_+ \phi = e^{-Q \Psi} \Phi_0\ee
where $\Phi_0$ satisfies ${\bar{\cal{D}}}_+ \Phi_0 =0$, and is a \ZT\
charged chiral superfield.
Therefore, in terms of \ZT\ superfields, we express term 1 as
\be  -4Q \int d^2 \theta ~{\bar{\theta}}^+ \bar{\Gamma} \Sigma_0 \Phi_0
 + 4 {\sqrt 2} Q \int d^2 \theta ~\theta^+ {\bar{\theta}}^+
\bar{E} \Sigma_0 \Phi_0. \ee

\detail{Term 2}

The second term to consider is
\be - 2Q \int d^2 \theta ~ {\bar{\Phi}} ( {\DB}_-  \D_- V)
 e^{2 Q V} \Phi. \ee
Recall from section~\ref{ttsusy}\ that
\be ( \DB_-  \D_- V) = -V_0 + i \partial_- \Psi \non \ee
where  $V_0$ is given by
\be V_0 = A_- -2i\theta^+ {\bar{\lambda}}_- -2i {\bar{\theta}}^+ \lambda_-
+2\theta^+ {\bar{\theta}}^+ D.\ee
Term 2 is therefore
\be  -2Q \int d^2 \theta ~{\bar{\Phi}}_0 (-V_0 + i \partial_- \Psi)
\Phi_0. \ee

\detail{Term 3}

The next term is immediately reduced
\be  4 Q^2 \int d^2 \theta ~{\bar{\Phi}} ( \D_- V)
( {\DB}_- V) e^{2 Q V} \Phi = 8 Q^2 \int d^2 \theta
~{\bar{\theta}}^+ \theta^+ \vert \Phi_0 \Sigma_0 \vert^2.  \ee

\detail{Term 4}

Similarly for term 4,
\bea  -\int d^2 \theta ~({\DB}_- {\bar{\Phi}}) e^{2 Q V} (\D_- \Phi)\non
&=& -2 \int d^2 \theta ~\bar{\Gamma} \Gamma + 2 {\sqrt 2}
\int d^2 \theta ~{\bar{\theta}}^+ \bar{\Gamma} E \non \\
& & -  2 {\sqrt 2} \int d^2 \theta ~\theta^+ \Gamma
 \bar{E} +4 \int d^2 \theta
~{\bar{\theta}}^+ \theta^+ \vert E \vert^2.  \eea

\detail{Term 5}

We see that term 5,
\bea  -2 Q \int d^2 \theta ~{\bar{\Phi}} ( {\DB}_- V)
e^{2 Q V} (\D_- \Phi)
 = 4Q \int d^2 \theta ~\theta^+ \Gamma {\bar{\Sigma}}_0 {\bar{\Phi}}_0
 + 4 {\sqrt 2} Q \int d^2 \theta ~\theta^+ {\bar{\theta}}^+
E {\bar{\Sigma}}_0 {\bar{\Phi}}_0,  \eea
is just the conjugate of term 1.

\detail{Term 6}

Lastly, we come to
\be  -\int d^2 \theta ~{\bar{\Phi}} e^{2 Q V} ( {\DB}_-  \D_- \Phi)
= -2i \int d^2 \theta~{\bar{\Phi}}_0 (\partial_- \Phi_0 -
Q \Phi_0 \partial_- \Psi).
\ee

\detail{Some Simplifications}

Consider adding terms 2 and 6. The sum gives the gauge covariant combination
\be  -2i \int d^2 \theta ~{\bar{\Phi}}_0 ({\cal{D}}_0 -
{\cal{D}}_1 ) \Phi_0 \ee
where $({\cal{D}}_0 - {\cal{D}}_1) = \partial_- + i Q V_0 $.
On summing all terms, we find
\bea \label{bigeqn}
L_\Phi &=&  -2i \int d^2 \theta ~{\bar{\Phi}}_0 ({\cal{D}}_0 - {\cal{D}}_1 ) \Phi_0
 -2 \int d^2 \theta ~\bar{\Gamma} \Gamma
-  2 {\sqrt 2} \int d^2 \theta ~\theta^+ \Gamma \bar{E}
 + 4  Q \int d^2 \theta ~\theta^+
\Gamma {\bar{\Sigma}}_0 {\bar{\Phi}}_0 \non \\
& & + 2 {\sqrt 2} \int d^2 \theta ~{\bar{\theta}}^+ \bar{\Gamma} E
 - 4 Q \int d^2 \theta ~{\bar{\theta}}^+
\bar{\Gamma} \Sigma_0 \Phi_0
 + 8 Q^2 \int d^2 \theta
~{\bar{\theta}}^+ \theta^+ \vert \Phi_0 \Sigma_0 \vert^2
+4 \int d^2 \theta
~{\bar{\theta}}^+ \theta^+ \vert E \vert^2 \non \\
& & + 4 {\sqrt 2} Q \int d^2 \theta ~\theta^+ {\bar{\theta}}^+
\bar{E} \Sigma_0 \Phi_0
+ 4 {\sqrt 2} Q \int d^2 \theta ~\theta^+ {\bar{\theta}}^+
E {\bar{\Sigma}}_0 {\bar{\Phi}}_0.
\eea
This is the $(2,2)$ theory reduced to \ZT\ variables. Finally for a $(2,2)$
theory reduced this way,
$$E = {\sqrt 2} Q \Sigma_0  \Phi_0.$$
Substituting this explicit expression leads to a large number of
cancellations. When the dust settles, we are left with the simple Lagrangian
\be
L_\Phi =  -2i \int d^2 \theta ~{\bar{\Phi}} ({\cal{D}}_0 - {\cal{D}}_1 ) \Phi
 -2 \int d^2 \theta ~\bar{\Gamma} \Gamma ,
\ee
where ${\bar{\cal{D}}}_+ \Phi = 0$ and ${\bar{\cal{D}}}_+ \Gamma=
{\sqrt 2} E$. We have also dropped the subscript in the definition of the
chiral superfield. When rescaled by a factor of $1/4$, this is the standard \ZT\
Lagrangian.
The only remaining terms involve $\Sm$ and they reduce
straightforwardly to give,
\be \label{LSm}
L_\Sm = {i\over 2 e^2} \int d^2 \theta
 ~{\bar{\Sm}_0} \p_- \Sm_0  + {1\over 8e^2}\int d^2\t ~\bar{\U}\U \non  + \left\{
{ t\over 4}\int d\tp~ \U
\vert_{{\bar{\theta}}^+ =0} + {\rm h.c.} \right\},
\ee
where $\U$ is the field strength for the \ZT\ vector multiplet.

\detail{The Dual Description}

The dual Lagrangian is given in terms of the $(2,2)$ field strength
$\Sm$ and an uncharged chiral multiplet $Y$
\be \widetilde{L} = L_\Sm - \frac{1}{8} \int d^4 \theta
~(Y + \bar{Y}) {\rm{ln}} (Y + \bar{Y})
 - \left( {\frac{Q}{2 {\sqrt 2}}} \int d^2 \widetilde{\theta}~\Sigma
Y + {\rm h.c.} \right).\ee
The first term is given in \C{LSm}\ so need only
consider the remaining terms. We start with the twisted
superpotential. As before, we want to
reduce it to \ZT\ superspace,
\be \widetilde{L} = \ldots +  \left( {\frac{Q}{2 {\sqrt 2}}}
\int d \theta^+ [ ({\DB}_- \Sigma) Y + \Sigma ({\DB}_- Y)]
 \vert_{\theta^- = {\bar{\theta}}^- =0} + {\rm h.c.} \right). \ee
Using the results ${\DB}_- \Sigma = -\frac{i}{\sqrt 2} \Upsilon$,
where
\be \Upsilon = -2 \lambda_- +2 i \theta^+ (D -i F_{01})
+2i \theta^+ {\bar{\theta}}^+ \partial_+ \lambda_- ,\ee
and ${\DB}_- Y = -{\sqrt 2} F$, where
${\DB}_+ F = 0$, we get that
\be \widetilde{L} = \ldots - ( \frac{Q}{2}
\int d \theta^+  [ \Sigma_0 F + \frac{i}{2} Y_0 \Upsilon ]
+ {\rm h.c.}).\ee
where $Y_0 = Y \vert_{\theta^- = {\bar{\theta}}^- =0}$. Note that
${\DB}_+ Y_0 = 0$ and $Y_0$ is a neutral \ZT\ chiral superfield.
Next consider the kinetic term
\be \widetilde{L} = \frac{1}{8} \int d^2 \theta
 \, {\DB}_- \D_- (Y + \bar{Y}) {\rm{ln}} (Y + \bar{Y}) + \ldots  \ee
Up to a total derivative, this gives us
\bea \widetilde{L}
=\frac{1}{8} \int d^2 \theta
 ~\left[i\frac{ Y_0 -{\bar{Y}}_0}{Y_0 + {\bar{Y}}_0} \partial_- (Y_0 + {\bar{Y}}_0)
 - 2 \frac{\bar{F} F}{Y_0 + {\bar{Y}}_0} \right] + \ldots.
\eea
Excluding the terms involving only $\Sm$ given in \C{LSm},
we obtain the dual Lagrangian
\bea \widetilde{L} &=&  \frac{i}{8} \int d^2 \theta
~\left[\frac{Y-\bar{Y}}{Y + \bar{Y}} \partial_- (Y + \bar{Y})
 +2i\frac{\bar{F} F}{Y + \bar{Y}} \right] \\ & & - \left( \frac{Q}{2}
\int d \theta^+  \left[ \Sigma F + \frac{i}{2} Y \Upsilon \right]
+ {\rm h.c.} \right)+\ldots, \non \eea
where ${\DB}_+ Y = {\DB}_+ F = 0$ and we have dropped the subscript
on the neutral chiral superfield $Y$.

\detail{The Duality Map}

The $(2,2)$ duality map is given by
\be \label{22M} \bar{\Phi} ~e^{2 Q V} \Phi = {1\over 2} \left( Y +
\bar{Y} \right).\ee
To express this map \ZT\ language, we will make use of the relations
\be \Phi = e^{-Q \Psi} \Phi_0 + \theta^-
( {\sqrt 2} e^{-Q \Psi} \Gamma + 2 {\bar{\theta}}^+ E)
-i \theta^- {\bar{\theta}}^- \partial_- (e^{-Q \Psi} \Phi_0), \ee
\be Y = Y_0 +{\sqrt 2} {\bar{\theta}}^- F +i \theta^- {\bar{\theta}}^-
\partial_- Y_0 ,\ee
and
\be V = \Psi - {\sqrt 2} \theta^- {\bar{\theta}}^+ \Sigma_0
- {\sqrt 2} \theta^+ {\bar{\theta}}^- {\bar{\Sigma}}_0
+ \theta^- {\bar{\theta}}^- V_0. \ee
Substituting these expressions into \C{22M}, we obtain the \ZT\ duality map.
Equating terms independent of the fermionic superspace coordinates, we get
\be \bar\Phi \Phi = \frac{1}{2} (Y + \bar{Y}).\ee
Again, here we have dropped the subscript on the
\ZT\ fields for brevity.
Equating terms proportional to $\theta^- {\bar{\theta}}^-$, we get
\be  -i \bar{\Phi} ({\buildrel \leftrightarrow \over
   \partial_-} +iQ V) \Phi + \bar\Gamma \Gamma
=\frac{i}{4} \partial_- (Y-\bar{Y}).\ee
Finally on equating terms proportional to $\theta^-$, we get the relation
\be \frac{1}{2} \bar{F} = \bar\Phi \Gamma.\ee
and its conjugate from terms proportional to
${\bar{\theta}}^-$.



\endgroup

\end{document}